Imperial College London
Department of Mathematics

# Analysis of dissipative dynamics on noncommutative spaces and statistical inference of continuous time network stochastic processes

by

Shreya Mehta

A thesis presented for the degree of Doctor of Philosophy
in Mathematics of Imperial College London, August 2024

# Declaration

I certify that this thesis, and the research to which it refers, are the product of my own work, and that any ideas or quotations from the work of other people, published or otherwise, are fully acknowledged in accordance with the standard referencing practices of the discipline.

<div style="text-align:right">Shreya Mehta</div>

# Copyright

# Abstract

In this thesis, we analyse the generalisations of the Ornstein-Uhlenbeck (OU) semigroup and study them in both quantum and classical setups.

In the first three chapters, we analyse the dissipative dynamics on noncommutative/quantum spaces, in particular, the systems with multiparticle interactions associated to CCR algebras. We provide various models where the dissipative dynamics are constructed using noncommutative Dirichlet forms. Some of our models decay to equilibrium algebraically and the Poincaré inequality does not hold.

Using the classical representation of generators of nilpotent Lie algebras, we provide the noncommutative representations of Lie algebras in terms of creation and annihilation operators and discuss the construction of corresponding Dirichlet forms. This introduces the opportunity to explore quantum stochastic processes related to Lie algebras and nilpotent Lie algebras. Additionally, these representations enable the investigation of the noncommutative analogue of hypoellipticity.

In another direction, we explore the potential for introducing statistical models within a quantum framework. In this thesis, however, we present a classical statistical model of multivariate Graph superposition of OU (Gr supOU) process which allows for long(er) memory in the modelling of sparse graphs. We estimate these processes using generalised method of moments and show that it yields consistent estimators. We demonstrate the asymptotic normality of the moment estimators and validate these estimators through a simulation study.

# Acknowledgements

I would like to thank my supervisor, Prof. Boguslaw Zegarlinski, for his patience, kindness, and immense support over the past four years. Working and studying under his guidance has been an invaluable learning experience, and without his mentorship and insightful discussions, this thesis would not have been possible.

I am also thankful to my secondary supervisor, Prof. Almut Veraart, for her willingness to collaborate with me and her invaluable assistance in exploring the area of statistical modeling. I am immensely thankful for the countless coffee talks we had and for her close guidance on my work.

I would also like to thank Prof. Michel Hilsum, Prof. A. Swaminathan, Prof. Anil Kumar Karn, Mrs. Uma Versha Kakkar and Mr. Yograj Singh for their support and inspiration throughout my academic journey, encouraging me to delve into and appreciate advanced mathematics.

I am also grateful to all the members of the Imperial Analysis and Statistics Group for fostering a stimulating yet welcoming atmosphere. I am especially thankful to all the good friends I made at Imperial: Yaozhong, Esther, Kamilla, Mengchun, Shubham and Xianfeng. I would also like to thank our administrator, Aga, for her help in various funding matters, as well as the donors of the Schrödinger and Roth scholarships for funding my research.

I am grateful to all my friends and alumni from IIT and Lady Shri Ram in London for tons of laughter, good times, and always being there.

Above all, I am grateful to my sister, Shruti, for being my unwavering source of strength through the toughest times and for always cheering me on. Lastly, I am deeply thankful to my mom and dad, Simmi and Narinder, for being the coolest parents and for their constant encouragement and support. I could not have undertaken this journey without their love.

'Fall in love with some activity, and do it! Nobody ever figures out what life is all about, and it doesn't matter. Explore the world. Nearly everything is really interesting if you go into it deeply enough. Work as hard and as much as you want to on the things you like to do the best. Don't think about what you want to be, but what you want to do.'

*Richard Feynman*

# Contents

# List of Publications

1. S. Mehta, B. Zegarlinski, 'Dissipative dynamics in infinite lattice systems', Infinite Dimensional Analysis, Quantum Probability and Related Topics, `https://doi.org/10.1142/S0219025723500303`

   **Abstract**: We study dissipative dynamics constructed by means of non-commutative Dirichlet forms for various lattice systems with multiparticle interactions associated to CCR algebras. We give a number of explicit examples of such models. Using an idea of quasi-invariance of a state, we show how one can construct unitary representations of various groups. Moreover in models with locally conserved quantities associated to an infinite lattice we show that there is no spectral gap and the corresponding dissipative dynamics decay to equilibrium polynomially in time.

2. S. Mehta, B. Zegarlinski, Quantum Dissipative Systems in Infinite Dimensions. In: Chatzakou, M., Ruzhansky, M., Stoeva, D. (eds) Women in Analysis and PDE. GFOW APDEGS 2022 2021. Trends in Mathematics(), vol 5. Birkhäuser, Cham. `https://doi.org/10.1007/978-3-031-57005-6_26`

   **Abstract**: We analyze infinite-dimensional dissipative dynamics constructed by means of non-commutative Dirichlet forms associated to Gibbs states for various interactions and discuss their decay to equilibrium.

3. S. Mehta, A.E.D. Veraart, Statistical inference for Levy-driven graph supOU processes: From short- to long-memory in high-dimensional time series. `https://doi.org/10.48550/arXiv.2502.08838`

   **Abstract**: This article introduces Levy-driven graph supOU processes, offering a parsimonious parametrisation for high-dimensional time-series, where dependencies between the individual components are governed via a graph structure. Specifically, we propose a model specification that allows for a smooth transition between short- and long-memory settings while accommodating a wide range of marginal distributions. We further develop an inference procedure based on the generalised method of moments, establish its asymptotic properties and demonstrate its strong finite sample performance through a simulation study. Finally, we illustrate the practical relevance of our new model and estimation method in an empirical study of wind capacity factors in an European electricity network context.

# Notations

| | |
|---|---|
| $\mathbb{N}$ | set of natural numbers |
| $\mathbb{Z}$ | set of integers |
| $[a, b]$ | commutator of operators $a$ and $b$ |
| $a^*$ | adjoint of an operator $a$ |
| $\omega$ | state of the system |
| $\alpha_t$ | modular dynamics |
| $\rho$ | density matrix |
| $\mathbb{L}_p(\omega)$ | noncommutative $\mathbb{L}_p$ spaces with respect to state $\omega$ |
| $\mathcal{E}(f)$ | Dirichlet form |
| $P_t$ | Markov semigroup |

# Chapter 1

# Introduction

The central idea of this thesis is to investigate various generalisations of Ornstein–Uhlenbeck semigroup and process, in quantum and classical setup respectively. A Lévy driven Ornstein–Uhlenbeck process $(\mathbb{X}_t)_{t\in\mathbb{R}}$ is the solution of the stochastic differential equation,

$$d\mathbb{X}_t = a\mathbb{X}_t dt + d\mathbb{L}_t, \tag{1.1}$$

where $a \in \mathbb{R}$ and $(\mathbb{L}_t)_{t\in\mathbb{R}}$ is a Lévy process. The corresponding Ornstein–Uhlenbeck semigroup $(P_t)_{t\in\mathbb{R}}$ is given by the relation

$$P_t f(x) = \mathrm{E}_x(f(\mathbb{X}_t)),$$

where $x \in \mathbb{R}$, $f$ is a bounded uniformly continuous function and $E_x$ denotes the expectation of the process starting at $x$. An important example of a semigroup of this type is explicitly given as follows

$$P_t f = e^{t\mathcal{L}} f,$$

$$\mathcal{L} = \Delta - x\nabla$$

In $\mathbb{L}_2$ space with $n$-dimensional Gaussian measure $d\gamma = \frac{1}{(2\pi)^{n/2}} e^{\frac{-1}{2}x^2} d_n x$, the quadratic form of the operator $-\mathcal{L}$ is given the following Dirichlet form

$$\mathcal{E}(f) = \langle \nabla f, \nabla f \rangle_\gamma = \int |\nabla f|^2 d\gamma$$

which means on a dense domain we have

$$\langle f, -\mathcal{L}f\rangle_\gamma = \mathcal{E}(f).$$

By this definition, $\mathcal{L}$ is densely defined, closed and symmetric operator $\mathbb{L}_2(\gamma)$. By Beurling-Deny theorem, $P_t$ is positivity and unit preserving, symmetric and contractive in $\mathbb{L}_2(\gamma)$.

In the quantum setup, we replace the probability measure by state on a noncommutative algebra. Then, we have many possibilities to introduce a scalar product associated to a state $\omega$. There is a natural notion of positivity in the algebra, however it may be different than the natural choice of positive cone in $\mathbb{L}_2(\omega)$ with the chosen scalar product. The other problem is how to combine the positivity preservation with the symmetry in the given $\mathbb{L}_2(\omega)$ space. For description of the corresponding problems, see e.g. [Ali76],[Cip97], [GZ02] and references therein.

In the first three chapters, we construct and analyse the dissipative dynamics of various generalisations of the quantum Ornstein–Uhlenbeck process [CFL00, KP04]. The theory of quantum dissipative systems has significantly advanced in recent decades (see, for example, [CZ24] and the references cited therein). Such systems can be described by an equation of the following form

$$\frac{\partial}{\partial t}P_t f = \mathfrak{L}P_t f, \; P_0 = id, \tag{1.2}$$

where $\mathfrak{L}$ is a Markov generator and $P_t \equiv e^{t\mathfrak{L}}$ is the associated Markov semigroup in a suitable noncommutative $\mathbb{L}_2(\omega)$ space associated to a state on noncommutative algebra to which operator $f$ belongs. The semigroup $P_t$ we consider has both positivity preserving and unit preserving properties. A challenging issue in the quantum case is achieving both symmetry in a Hilbert space associated with a state and positivity preservation of the generated semigroup simultaneously. One approach to address this issue is by proving the closability and Markov property of the pre-Dirichlet form. The theory of Dirichlet forms was introduced in [AHK77] for the trace state and was later fully developed in [Cip97]. For further advancements, see the references in [Cip08, Par00, Zeg02, CZ24]. The generator $\mathfrak{L}$ of the semigroup is constructed using Quantum Dirichlet form given by

$$\mathcal{E}(f) \equiv \sum_{j\in\mathcal{J}} \int_\mathbb{R} \left(\nu_j \langle \delta_{\alpha_t(X_j)}(f), \delta_{\alpha_t(X_j)}(f)\rangle_\omega + \mu_j \langle \delta_{\alpha_t(X_j^*)}(f), \delta_{\alpha_t(X_j^*)}(f)\rangle_\omega\right)\eta(t)dt, \tag{1.3}$$

for some constants $\nu_j, \mu_j$ and $\mathcal{J}$ is an index set, with a scalar product

$$\langle f, g \rangle_\omega = Tr(\rho^{1/2} f^* \rho^{1/2} g)$$

where $\omega$ is a state corresponding to a density matrix $\rho$, and the modular dynamics corresponding to $\omega$ is defined by

$$\alpha_t(B) = \rho^{it} B \rho^{-it}. \tag{1.4}$$

The function $\eta(t)$ is an admissible function 2.2 in the sense of Park, [Par00], the operators $X_j$'s are chosen appropriately later for specific cases and $\delta_X$ denotes a derivation associated to an operator $X$. According to Beurling-Deny theory ([BD59]), along with its noncommutative generalization ([Cip97, Cip08]), there is a one-to-one correspondence between Markov semigroups and Dirichlet forms. This correspondence allows the problem of analysing the generator of Markov semigroups to be translated into the study of their corresponding Dirichlet forms. We say that the semigroup $P_t$ converges to equilibrium if $\lim_{t \to \infty} P_t(f) = \omega(f)$, where the limit is taken with respect to noncommutative $\mathbb{L}_p$ spaces or the norm of the algebra. Often, the decay to equilibrium in $\mathbb{L}_2$ and $\mathbb{L}_p$ spaces can be examined using coercive inequalities of the Poincaré and Log-Sobolev type, as studied in [OZ99, CM15, CS07] and the references therein.

For the construction of Dirichlet form of models, we start with the the quantum harmonic oscillator [CFL00]. Given a Hilbert space, say $\mathfrak{h}$ with $\{e_n\}$ as the orthonormal basis, one defines the creation and annihilation operators by

$$A^* e_n = \sqrt{n+1} e_{n+1}, \tag{1.5}$$

$$A e_n = \sqrt{n} e_{n-1}, \tag{1.6}$$

and the associated particle number operator

$$N e_n \equiv A^* A e_n = n e_n,$$

with dense domains $D(A) = D(A^*) = D(\sqrt{N})$. For $U = N$, we define a density matrix $\rho = \frac{1}{Z} e^{-\beta U}$, with a normalization constant $Z \in (0, \infty)$. In commutative analysis, Ornstein-Uhlenbeck (OU) semigroups are defined by the Dirichlet generator associated with a Gaussian measure. The noncommutative generalization of the OU semigroup is introduced in [CFL00], [KP04] with the generator ([CM17]) given by an extension of Alicki's theorem [Ali76],

$$\mathfrak{L}f = \sum_j \left(-e^{-\frac{\beta}{2}}[V_j, f]V_j^* + e^{\frac{\beta}{2}}V_j^*[V_j, f]\right), \tag{1.7}$$

where $V_j$ are the eigenvectors of the modular operator (3.3). Quantum Markov semigroups generated by generators of the form (2.18) were recently studied in [CM17] within a finite-dimensional setting. The authors also derived some entropic dissipation inequalities for the necessarily infinite-dimensional Bose OU semigroup.

In the first strand of this thesis, we discuss the infinite dimensional setting on the lattice $\mathbb{Z}^d$ and analyse the associated interacting particle system. In the context of infinite systems with interactions, defining Dirichlet forms and generators on dense domains requires the finite speed of propagation of information for the Hamiltonian dynamics. This condition, for quantum spin systems with bounded multi-particle interactions, was established in [LR72]. We extend this framework to systems with multi-particle interactions involving unbounded operators.

Dissipative dynamics for quantum spin systems on a lattice were previously discussed in [MZ96], [GM91], [Zeg02], where the existence of dynamics with an exponential decay to equilibrium in the high-temperature region was demonstrated. General systems with quadratic interactions were studied in [OZ08], [BKP03], and [FQ05]. The presence of unbounded operators in the Markov generators complicates these issues significantly.

In our work, we construct models of interacting dissipative systems on the infinite dimensional setting with finite-range interactions. We consider multi-particle interactions and generalize the setup of quantum spin systems as used in [MZ96, LR72, BR87]. For some of our models with locally conserved quantities, we provide a detailed analysis showing that the Poincaré inequality cannot be satisfied, and the system converges to equilibrium at a polynomial rate. This extends the commutative case considered in [INZ12]. A quantum Brownian motion model provided in [CFL00] has no spectral gap and no equilibrium state. Some of our models are more general, featuring no spectral gap at the bottom of the spectrum of the Markov generator, while still having an equilibrium state.

In the work that follows, we turn our discussion towards analysing the dissipative dynamics for the noncommutative analogue of the Hörmander type operators of the form

$$\mathfrak{L} = \sum_{j \in \mathcal{J}} X_j^2 \tag{1.8}$$

where $\{X_j : j \in \mathcal{J}\}$ is a family of noncommuting vector fields on an algebra $\mathcal{D}$ and $\mathcal{J}$ is a finite or countably infinite index set. By Lifting theorem [RS77], every such operator that satisfies the Hörmander rank condition can be approximated by a sub-Laplacian on a stratified Lie algebra. We provide the noncommutative representations of numerous Lie algebras, especially nilpotent Lie algebras. Additionally, we analyse some models constructed using the so called Serre-Chevalley relations and creation/annihilation operators. We provide examples where an interesting quantum stochastic analysis could be developed. These representations can also be utilised to study the noncommutative analogues of hypoellipticity and hypocoercivity.

Since the development of quantum stochastic calculus [Par15], there has been extensive study on quantum Lévy processes [Fra04]. In the final chapter of the thesis, we focus on a more applied statistical model in the classical setup of a generalised Lévy driven OU process known as superposition of OU process introduced in [Bar01] on a graph structure. The model presented in this chapter has a potential to be adapted for the quantum setup by employing the construction of the quantum Ornstein-Uhlenbeck process and drawing on the literature of quantum Lévy processes. The OU-type relationship between the nodes of a continuously-observed graph was studied in [CV22]. In our work, we extend this model to accommodate long memory. Long memory is a desirable property for a process since it may lead to better forecasting accuracy. We show the consistency and asymptotic normality for this model. Additionally, we validate the model by performing Monte Carlo simulations and parameter estimation using Generalised Method of Moments.

Chapter 5 can be thought of as a stand-alone chapter, and hence we include very little relevant background material in Chapter 2. Instead, we include an expanded introduction along with the basic definitions at the beginning of Chapter 5.

The outline of the thesis is as follows. In Chapter 2, we review the literature surrounding the subject area dealt with in Chapters 3 and 4 along with the necessary definitions and basic results.

In Chapter 3, we discuss the dissipative dynamics of numerous infinite dimensional models with multiparticle interaction on CCR algebras. We prove that these models do not have a spectral gap and has polynomial rate of convergence to equilibrium.

Chapter 4 deals with the discussion on the representations of Lie algebras in terms of creation and annihilation operators. The corresponding generators of these representations can be used to construct the dissipative dynamics and to investigate the noncommutative analogues of problems in hypoellipticity and hypocoercivity.

Finally, Chapter 5 is concerned with the extension of Graph OU process to accommodate for long(er) memory. We present a novel, more flexible model along with the simulation study and moment based estimation of parameters. We also provide the asymptotic theory for the moment estimator.

Lastly, Chapter 6 provides a summary of the thesis and suggests possible future research directions.

# Chapter 2

# Background and Definitions

In this chapter, we discuss the basic definitions and background required for Chapters 3 and 4. We begin by discussing important properties of unbounded operators in Section 2.1. In Section 2.2, we provide a necessary introduction to quantum statistical mechanics. Furthermore, we discuss the CCR algebra, which forms the basis of our analysis in Chapter 3, in Section 2.3. In Section 2.4, we give a brief background on quantum Markov semigroups and coercive inequalities. We then discuss noncommutative Dirichlet forms in Section 2.5, followed by a literature review on quantum coercive inequalities in Section 2.6. Finally, we provide brief descriptions of the connections to quantum computing and quantum stochastic calculus in Sections 2.7 and 2.8, respectively.

## 2.1   Unbounded Operators

In the field of mathematical physics and quantum mechanics, the majority of operators are unbounded. Notably, both the position and momentum operators in quantum mechanics are examples of unbounded operators. In this section, we will briefly explore the characteristics of unbounded operators defined on the Hilbert space $\mathcal{H}$, which will be beneficial for subsequent chapters. For more comprehensive information, one can refer [RS72].
The unbounded operators are not defined across the entire Hilbert space. Instead, an unbounded operator is restricted to be defined on a dense linear subset of the Hilbert space $\mathcal{H}$. Thus, to characterize an unbounded operator on a Hilbert space, it is necessary to specify its domain of action.

The definition of the adjoint operator $T^*$ on the domain $D(T^*)$ for unbounded operator $T$ on a Hilbert space $\mathcal{H}$ is given by

$$\langle Tx, y \rangle = \langle x, z \rangle \tag{2.1}$$

where for each $y \in D(T^*)$, $T^*y = z$. To ensure the unique determination of $z$, it is essential that the domain of $T$, denoted as $D(T)$, is dense. In contrast to bounded operators, the domain of the adjoint operator $T^*$ may not necessarily be dense.

A significant number of inquiries regarding the domains and closures of unbounded operators extend beyond mere technical inconveniences. It's not simply a matter of selecting any dense domain that is sufficiently small to render the unbounded operator meaningful. Instead, the choice of an appropriate domain is often intricately linked to the underlying physics of the system being described. Many crucial properties of operators, such as the spectrum, are highly sensitive to the chosen domain.

We recall that a densely defined operator $T$ is called symmetric if and only if

$$\langle Tx, y \rangle = \langle x, Ty \rangle \tag{2.2}$$

for all $x, y \in D(T)$. Additionally, $T$ is self adjoint if $T = T^*$, which holds if and only if $T$ is symmetric and $D(T) = D(T^*)$.

Some operators analysed in this thesis are symmetric operators. The symmetric operators always have closed extensions, that is, they are closable. If $T$ is symmetric, then $T^*$ is the closed extension of $T$. It is important to note that a closed symmetric operator $T$ is self-adjoint if and only if $T^*$ is symmetric. The spectral theorem applies exclusively to self-adjoint operators, and only they can be exponentiated to generate one-parameter unitary groups, which play a crucial role in defining dynamics in quantum mechanics, as discussed in further chapters.

In quantum mechanics, the Spectral theorem for unbounded operators holds particular significance, as it offers insights into probability distributions associated with measuring observables characterized by continuous spectra, such as position and momentum. This is crucial for understanding the probabilistic nature of quantum mechanics and for making predictions about the behavior of quantum systems. For completeness, we give the statement of the spectral theorem. It asserts a direct relationship be-tween self-adjoint operators, denoted as $T$, and projection-valued measures $\{E_\Omega\}_{\Omega \subset \mathbb{R}}$ defined on the Hilbert space $\mathcal{H}$, for definition we refer to Chapter VIII, [RS72]. This relationship is represented by the equation:

$$T = \int_{-\infty}^{\infty} \lambda \, dE_\lambda, \tag{2.3}$$

where $T$ corresponds to the integral of the spectral parameter $\lambda$ with respect to the projection $E_\lambda$. Furthermore, for a real-valued Borel function $g$, its action on $T$ is described by

$$g(T) = \int_{-\infty}^{\infty} g(\lambda) \, dE_\lambda. \tag{2.4}$$

Another important result in mathematical physics and quantum mechanics is the Stone's theorem for unbounded operators. It provides a powerful framework for understanding the time evolution of quantum systems described by unbounded operators.

For bounded operators, such as $S$, the exponential of $S$ is easily defined using the series

$$e^{itS} = \sum_{n=0}^{\infty} \frac{(it)^n S^n}{n!}.$$

This series converges in norm, ensuring $e^{itS}$ is well defined. However, for unbounded and self-adjoint operators $T$, direct use of this power series is not viable. Instead, functional calculus is employed to define $e^{itT}$. The Stone's theorem states that given a strongly continuous (that is, continuous in norm topology) one parameter unitary group $U(t)$ on a Hilbert space $\mathcal{H}$, there is a self adjoint operator $T_a$ on $\mathcal{H}$ such that $U(t) = e^{itT_a}$. Furthermore, this self adjoint operator $T_a$ is the infinitesimal generator of $U(t)$.

It is important to note that the unbounded operators are more complicated than just defining the domain carefully. For instance, the concept of commuting operators say $T$ and $V$ is not straightforward since $TV - VT$ does not always makes sense on the domain. If $T$ and $V$ are self adjoint operators, then they commute if their unitary groups $e^{itT}$ and $e^{itV}$ commute. Although, this definition in not useful in practice since the formal series expansion of $e^{itT}$ and $e^{itV}$ may have no meaning for unbounded $T, V$.

Subsequently, the concept of canonical commutation relations is introduced. Two self adjoint opera-tors $P, Q$ satisfy canonical commutation relations if

$$PQ - QP = -iI. \tag{2.5}$$

For this relation to hold, either $P$ or $Q$ or both has to be unbounded. If both were bounded, then since $PQ^n - Q^n P = -inQ^{n-1}$, we obtain

$$n\|Q\|^{n-1} = \|PQ^n - Q^n P\| \leq 2\|P\|\|Q\|^n \tag{2.6}$$

This implies that for all $n$, $2\|P\|\|Q\| \geq n$ which cannot hold if $P$ and $Q$ are bounded. The relation (2.5) can also be written in terms of the unitary groups using power series in the following way

$$U(t)V(s) = e^{its}V(s)U(t). \tag{2.7}$$

These are called Weyl relations.

As discussed before, for an unbounded operator $T$, the formal expansion of $e^{itT}$ is not well defined. Although, there is a possible way to demonstrate the series expansion for unbounded operators 'on some special set of vectors' which will be useful in further chapters. Let $E_\Omega$ be the projection valued measure of an unbounded essentially self adjoint operator (that is, $\bar{T}$ is self adjoint) $T$ and define a dense set

$$D_c = \{E_{[-M,M]}\phi, \phi \in \mathcal{H}, M < \infty\},$$

contained in $D(T^n)$ for all $n$. Let $\psi = E_{[-M,M]}\phi \in D_c$, then $\|T^n\psi\| \leq M^n\|\psi\|$. Hence,

$$\sum_{n=0}^{\infty} \frac{t^n \|T^n\psi\|}{n!} < \infty$$

converges for all $t$. Such vectors $\psi \in D_c$ are known as analytic vectors.

In this thesis, we will explore various quadratic forms associated with unbounded operators. While the correlation between bounded operators and bounded quadratic forms is established through Riesz's lemma, extending this relation to unbounded operators requires some modifications. A quadratic form is a mapping $q : Q \times Q \to \mathbb{C}$, where $Q$ is a dense linear subset of the Hilbert space $\mathcal{H}$ which is linear in the first variable and conjugate linear in the second variable.

Furthermore, a quadratic form is termed symmetric if $q(\phi, \psi) = \overline{q(\psi, \phi)}$ and it is classified as semibounded if $q(\psi, \psi) \geq -M\|\psi\|^2$ for some $M$.

Moreover, a semibounded quadratic form $q$ is called closed if the space $Q$ equipped with the norm

$$\|\psi\|_{+1} = \sqrt{q(\psi, \psi) + (M+1)\|\psi\|^2}$$

is complete. If $q$ represents a closed semibounded quadratic form, it corresponds uniquely to a self-adjoint operator. In $\mathbb{L}_2$ space with an $n$-dimensional Gaussian measure $d\gamma = \frac{1}{(2\pi)^{n/2}} e^{-\frac{1}{2}x^2} d_n x$, the quadratic form of an operator $-\mathcal{L}$ is represented by the following Dirichlet form

$$\mathcal{E}(f) = \langle \nabla f, \nabla f \rangle_\gamma = \int |\nabla f|^2 \, d\gamma.$$

In Chapter 3, given the infinite dimensional nature of the setting, it becomes necessary to establish the concept of convergence for unbounded operators. Notably, when dealing with unbounded operators $A_n$, their domains may lack common elements due to their definition on dense domains.

To define the convergence of self-adjoint operators $T_n$ towards $T$, the notion of norm resolvent convergence is employed. This entails the convergence of the norm resolvent $R_\lambda(T_n)$ to $R_\lambda(T)$ in norm for all $\lambda$ with nonzero imaginary parts. Here, the resolvent is denoted as $R_\lambda(T) = (\lambda I - T)^{-1}$.

A key result established in [RS72] about the convergence of sequence of self adjoint unbounded operators is as follows. Let $\{T_n\}_{n=1}^\infty$ and $T$ be self-adjoint operators sharing a common domain $D$, where $\|\phi\|_T = \|T\phi\| + \|\phi\|$ defines a norm on this domain. The convergence $T_n \to T$ in the norm resolvent sense is understood to occur when

$$\sup_{\|\phi\|_T = 1} \|R_\lambda(T_n) - R_\lambda(T)\phi\| \to 0.$$

In essence, this convergence criterion ensures that the difference between the operators $A_n$ and $A$, when applied to vectors normalized under the norm $\|\cdot\|_T$, tends to zero as $n$ approaches infinity. The operators $e^{itT}$ and $e^{itV}$ for self adjoint unbounded operators $T$ and $T$ on $\mathcal{H}$ can be approximated using the Trotter product formula. It states that

$$\lim_{n\to\infty}[e^{itT/n}e^{itV/n}]^n = e^{it(T+V)}, \tag{2.8}$$

where $T + V$ is self adjoint on $D = D(T) \cap D(V)$ and the limit is defined in the sense of strong operator convergence.

In quantum mechanics, operators represent various physical observables. Polar decomposition is pivotal for comprehending quantum states, unitary transformations, and measurements. For example, it aids in the Trotter product formula (2.8), enabling the approximation of time-evolution in quantum systems. Moreover, given a function $f$ defined on the spectrum of an operator, one can define $f(T)$ for an operator $T$ using its polar decomposition. This is particularly important in spectral theory and functional analysis.

For bounded operators, say B, the polar decomposition is given by $B = U|B|$ where $|B| = \sqrt{B^*B}$ and $U$ is a partial isometry. In the case of unbounded operators, since it is not clear that $\{x | x \in D(T) \text{ and } Tx \in D(T^*)\}$ is different from $\{0\}$, the polar decomposition is instead constructed by applying the theory of semi bounded quadratic forms. Hence for any closed operator $B$, there is a positive self adjoint operator $|B|$ and a partial isometry $U$ such that $B = U|B|$.

In further sections, in order to discuss quantum spin systems, it is necessary to establish the notion of tensor products for operators. Consider operators $T$ and $V$ acting on Hilbert spaces $\mathcal{H}_1$ and $\mathcal{H}_2$ respectively. Let $D(T) \otimes D(V)$ denote a dense subset of $\mathcal{H}_1 \otimes \mathcal{H}_2$, comprising linear combinations of vectors in the form $\phi \otimes \psi$ where $\phi \in D(T)$ and $\psi \in D(V)$. The tensor product of $T$ and $V$, denoted as $T \otimes V$, is defined as follows

$$(T \otimes V)(\phi \otimes \psi) = T\phi \otimes V\psi.$$

Furthermore, if $T$ and $V$ are closable operators, then $T \otimes V$ is also closable. This can be generalised to an arbitrary finite tensor products of operators. Moreover if we consider a family of self adjoint operators $\{T_k\}_{k=1}^N$ on the Hilbert space, then a monomial of these operators of degree $n_k$ is defined on $\otimes_k D(T^{n_k})$ and is essentially self adjoint. Additionally, the spectrum of the closure of this monomial is closure of the monomial of the spectra of each $T_k$.

## 2.2 Basics of Quantum Statistical Mechanics

We now give brief but essential description of the mathematical formulation of quantum mechanics and some related notions, for more details one can refer [BR87]. We discuss the theory of operator algebras on Hilbert spaces including $C^*$ algebras and von Neumann algebras which are essential for defining these formulations. The two formalisations of quantum mechanics were given by Heisenberg and Schrödinger.

In the formalism established by Heisenberg, the coordinates representing the position and momentum of a particle are denoted by the operators $p_i$ and $q_i$, respectively. These operators adhere to the canonical commutation relations similar to (2.5), as described by the equations:

$$p_i q_j - q_j p_i = -i\hbar \delta_{ij}$$

$$p_i p_j - p_j p_i = 0 = q_i q_j - q_j q_i$$

Here, $\hbar$ represents the reduced Planck constant. The dynamics of an operator $B_t$ within this framework are governed by the following equation:

$$\frac{\partial B_t}{\partial t} = \frac{i}{\hbar}(HB_t - B_t H)$$

where $H$ denotes the Hamiltonian operator. The Hamiltonian is an operator that depends on the particle's position and momentum given by

$$H = \sum_{i=1}^{n} \frac{p_i^2}{2m} + V(q_1, q_2, \ldots, q_n)$$

where the potential energy $V$ proportional to the particle's position operator. One fundamental example of the Hamiltonian is the quantum harmonic oscillator. In one dimensional case, the quantum harmonic oscillator is given by

$$H = \frac{p^2}{2m} + \frac{1}{2}kx^2$$

where $k > 0$ is the force constant, $p$ and $x$ are momentum and position operators.

As discussed in (2.6), at least one among $p_i$ or $q_i$ cannot be a bounded operator. Therefore, an infinite dimensional Hilbert space $\mathfrak{h}$ is considered, upon which these operators act. Each vector $\psi \in \mathfrak{h}$ corresponds to a pure state of the system, and $\langle \psi, B\psi \rangle$ represents the value of the observable or the expectation of $B$ at time $t$.

On the other hand, Schrödinger's formalisation utilises a function $\psi$ of $n$ variables which are the particle coordinates and $\psi$ represents the state of the system and the dynamics are determined by Schrödinger equation

$$i\hbar \frac{\partial \psi_t}{\partial t}(x_1, \ldots, x_n) = -H\psi_t(x_1, \ldots, x_n). \tag{2.9}$$

Then vector $\psi_t$ is a normalised vector of the Hilbert space $L^2(\mathbb{R}^n)$.

These two formalisations are essentially unique and the equivalences can be seen from

$$\mathfrak{h} = L^2(\mathbb{R}^n)$$

$$p_i \psi(x_1, \ldots, x_n) = -i\hbar \frac{\partial \psi}{\partial x_i}(x_1, \ldots, x_n)$$

$$q_i \psi(x_1, \ldots, x_n) = x_i \psi(x_1, \ldots, x_n)$$

$$\langle \psi, B_t \psi \rangle = \langle \psi_t, B\psi_t \rangle$$

where $B$ and $\psi$ are $B_t$ and $\psi_t$ at $t = 0$. We note that if $\psi_t$ is normalised in $L^2(\mathbb{R}^n)$, then $\|\psi_t\|^2$ and $\|\psi\|^2$ are probability densities.

According to Stone's theorem, also discussed in the preceding section, the Schrödinger equation possesses a unique solution $\psi_t$, which satisfies the relations $\|\psi_t\| = \|\psi\|$ if and only if the Hamiltonian $H$ is self-adjoint. If $H$ is self adjoint, then the equation

$$\frac{dU_t}{dt} = iU_t H.$$

determines a unique continuous unitary representation $U_t$ of the real line. Instead of working directly with the unbounded position and momentum operators $p_i$ and $q_i$, it is convenient to write them in terms of unitary operators $U_k(t) = e^{ip_k t}$, $V_j(t) = e^{iq_j t}$. These groups satisfy Weyl form of commutation relations

$$U_k(s)V_j(t) = V_j(t)U_k(s)e^{ist\delta_{kj}},$$

$$U_k(s)U_j(t) - U_j(t)U_k(s) = 0 = V_k(s)V_j(t) - V_j(t)V_k(s).$$

Quantum statistical mechanics introduces the concept of a mixed state denoted by $\omega$ which is defined as a functional over bounded observables, characterized by the expression

$$\omega(B) = \sum_j \lambda_j \langle \psi_j, B\psi_j \rangle$$

where $\lambda_j \geq 0$, $\sum_j \lambda_j = 1$, and $\|\psi_j\| = 1$. If all bounded self-adjoint operators on $\mathfrak{h}$ represent observables, then these mixed states naturally take the form

$$\omega(B) = \mathrm{Tr}(\rho B)$$

where $\rho$ is a positive trace-class operator with a trace equal to one and is called a density matrix.

We recall that $C^*$- algebra is a Banach space $\mathfrak{U}$ with involution $*$ such that $\|B^*B\| = \|B\|^2$ for all $B \in \mathfrak{U}$. A pivotal reformulation of quantum mechanics emerged through von Neumann algebras $\mathfrak{M}$ defined on a Hilbert space $\mathfrak{h}$ with mixed states where the quantum observables contain the self-adjoint elements of these weakly closed *-algebras of operators. A von Neumann algebra is a specialised type of $C^*$-algebra. Within these algebras, mixed states are positive, normalized, linear functionals.

Subsequently, Segal [Seg47] argued that the physical significance of observables lies in their uniform convergence, whereas weak convergence primarily holds analytical significance. As a result, it was proposed that observables could be identified with the self adjoint component of a $C^*$-algebra $\mathfrak{U}$ equipped with an identity, with states forming a subset of the states defined over $\mathfrak{U}$.

If $\mathfrak{U}$ and $\mathfrak{M}$ are the $C^*$- algebra and the von Neumann algebra respectively, generated by the Weyl operators $\{U_k(s), V_j(t); s, t \in \mathbb{R}, k, j = 1, 2, \ldots, n\}$. Then by the uniqueness of operators satisfying Heinsenberg commutation relations and the Schrödinger representation, see [BR87], the choice between $C^*$- algebra and the von Neumann algebra for a finite number of particles is a matter of technical convenience.

In addressing systems involving infinite number of particles, the uniqueness theorem becomes invalid. Consequently, the formalism introduced by Fock is commonly employed. This approach entails the construction of an infinite series of unitary Weyl operators alongside a $C^*$-algebra and the von Neu-mann algebra derived from these operators.

In order to characterize the infinite particle systems systematically, one typically begins by considering a finite subsystem confined within a compact subset $\Lambda$. The associated observables are then constructed as self adjoint elements within a $C^*$-algebra denoted as $\mathfrak{U}_\Lambda$. We assume $\Lambda_1 \subset \Lambda_2$, then $\mathfrak{U}_{\Lambda_1} \subset \mathfrak{U}_{\Lambda_2}$. The observables of a large system would be in the union of $\mathfrak{U}_\Lambda$ as a dense set. The closure of the union of a family of subalgebras $\mathfrak{U}_\Lambda$ is also a $C^*$- algebra which is called as quasi local algebra and $\mathfrak{U}_\Lambda$ are local algebras.

The quasi-local algebras are a type of inductive limit algebra that preserves locality. We recall that an inductive limit algebra of $C^*$ algebras is a $C^*$-algebra that can be written as the closure of the union of a sequence of sub-$C^*$-algebras. The Uniformly Hyperfinite (UHF) Algebras are the inductive limits of sequences of finite dimensional matrix algebras.

The quasi local structures of $C^*$- algebra and von Neumann algebras in the field theoretic models are useful for the analysis of equilibrium state. The equilibrium state over $\mathfrak{U}_\Lambda$ is constructed in the system $\Lambda$ and its thermodynamic limit is studied for each $B \in \mathfrak{U}_\Lambda$ and each $\Lambda$ and is given by

$$\omega(B) = \lim_{|\Lambda'| \to \infty} \omega_{\Lambda'}(B) \tag{2.10}$$

where $\Lambda'$ invades all the space. There are two ways to analyse the properties of the set of these equilibrium states. First, one can start with the Hamiltonian operator $H_\Lambda$. If $H_\Lambda$ is self adjoint, $\beta > 0$ is the inverse temperature and $e^{-\beta H_\Lambda}$ is of trace class, then we construct the Gibbs equilibrium state of the form

$$\omega_\Lambda(B) = \frac{Tr(e^{-\beta H_\Lambda} B)}{Tr(e^{-\beta H_\Lambda})}. \tag{2.11}$$

Then the limit of $\omega_\Lambda(B)$ when $\Lambda \to \infty$ is analysed. Another method of analysing the equilibrium states is that one starts with the assumption for the dynamics of the infinite system to be given by a continuous one parameter group $\alpha_t$ of $*-$ automorphisms of the $C^*$- algebra $\mathfrak{U}$ of all observables. We want to construct the equilibrium states which are invariant with respect to time

$$\omega(\alpha_t(B)) = \omega(B). \tag{2.12}$$

For the state (2.11), the corresponding automorphism $\alpha_t(B)$ is defined as follows

$$\alpha_{t,\Lambda}(B) = \lim_{\Lambda \to \infty} e^{-it\beta H_\Lambda} B e^{it\beta H_\Lambda}.$$

Under some technical conditions, the limit of $\omega_\Lambda$ (2.10) would satisfy the condition

$$\omega(\alpha_t(C)B) = \omega(B\alpha_{t+i\beta}(C))$$

for all $B, C \in \mathfrak{U}$ and $t \in \mathbb{R}$. Abstractly, in the literature, KMS (Kubo-Martin-Schwinger) condition serves as a criterion for equilibrium, indicating a complicated commutation of observables under a given state $\omega$. When analyzing equilibrium states, it is essential to ensure that these states adhere to this condition.

A similar relationship is considered in the Tomita-Takesaki theory [Tak70] of von Neumann algebras. In this theory, authors associate a canonical one-parameter group of $*$-automorphisms, denoted as $\alpha_t^\omega$, to each normal faithful state $\omega$ defined over a von Neumann algebra $\mathfrak{M}$. While the state $\omega$ satisfies the KMS condition, $\alpha_t^\omega(B)$ may not necessarily be continuous in norm.

Both these approaches prove valuable in analysing equilibrium states, often considering system properties like homogeneity, which manifest through the model's symmetry properties. Specifically, homogeneity is often expressed through the invariance of equilibrium states under the action of the group of space translations which can be expressed as $*$-automorphisms of the $C^*$-algebra $\mathfrak{U}$ of all observables.

Subsequently, one can study the noncommutative counterpart of Ergodic theory, which involves analysing the dynamical system $(\mathfrak{U}, \omega, \alpha_t)$, where $\mathfrak{U}$ represents the $C^*$-algebra, $\omega$ denotes the invariant state signifying homogeneity property, and $\alpha_t$ denotes the group of $*$-automorphisms.

Since statistical mechanics primarily concerns the macroscopic examination of systems composed of a large number of particles, there is a particular emphasis on analysing the properties of equilibrium states within infinite particle quantum systems. A common strategy in approaching this analysis involves initially describing finite systems and their respective equilibrium states. This description is then reformulated using an algebraic framework, wherein equilibrium states are identified as states over a quasi-local $C^*$-algebra generated by subalgebras corresponding to observables of subsystems.

Then these states are approximated by taking their limit as the volume of the system increases to infinity. This is known as taking the thermodynamic limit (2.10).

In order to construct the algebraic structure for particle systems, certain structural features of a $C^*$ algebra of observables are utilised. In the next section, we give a brief description of algebras generated by operators that satisfy Canonical Commutation Relations (CCR) and quantum spin systems.

## 2.3 CCR algebra and Quantum Spin system

Consider a complex separable Hilbert space $\mathfrak{h}$ and $\{e_n\}_{n \in \mathbb{Z}^+}$ be an orthonormal basis, so that a vector $\alpha \in \mathfrak{h}$ is represented in this basis as $\alpha = (\alpha_n)$. We define the number operator N with domain

$$D(N) = \left\{ \alpha = (\alpha_n) \in \mathfrak{h} : \sum_{n \geq 0} |n\alpha_n|^2 < \infty \right\}$$

by

$$D(N) \ni \alpha = (\alpha_n) \mapsto N\alpha = (n\alpha_n).$$

This number operator N also known as the Beltrami-Laplacian can be understood as infinite dimensional analog of a finite dimensional Laplacian. In classical theory, the solution of the heat equation associated to $N$ is known the Ornstein-Uhlenbeck (OU) semigroup. We discuss the quantum Ornstein-Uhlenbeck semigroup introduced by [CFL00] later in this section. It is evident that $N$ is a self adjoint operator. For self adjoint operators, there is a method known as second quantisation. In $\mathfrak{h}$, we define the annihilation and creation operators $A$ and $A^*$ respectively, on $Dom(A) = Dom(A^*) = Dom(\sqrt{N})$ by the action on the basis as follows

$$Ae_n = \sqrt{n} e_{n-1}, \quad A^* e_n = \sqrt{n+1} e_{n+1}.$$

The number operator $N$ can be written in terms of $A$ and $A^*$ as $N = A^*A$. These operators satisfy the canonical commutation relations discussed in the previous sections (2.5) given as

$$[A, A] = 0 = [A^*, A^*]$$

$$[A, A^*] = id.$$

We reiterate that at least one of these operators are necessarily unbounded.

**Remark 2.1.** *The following relations will be useful in proving many claims involving creators, annihilators and particle number operators. For analytic function h, (using a basis of eigenvectors for N), one can see that*

$$\begin{aligned} A^*h(N) &= h(N-1)A^* \quad \text{and} \quad h(N)A^* = A^*h(N+1) \\ Ah(N-1) &= h(N)A \quad \text{and} \quad h(N+1)A = Ah(N). \end{aligned} \tag{2.13}$$

The algebra generated by $A$ and $A^*$, that is, the linear combinations of the monomials of $A$ and $A^*$ which can be defined on a dense domain containing finite linear combinations of $(e_n)$ is called a CCR algebra. In the next two chapters of the thesis, the dynamics are analysed on the CCR algebra. One can also perform analysis on a $C^*$-algebra generated by Weyl form of creation and annihilation operators, see [OZ08], [Par00].

Now we consider the quantum spin systems with lattice $L = \mathbb{Z}^d$, and $\Lambda$ be the finite subset of this lattice. Before we move to our results and setup for analysing quantum systems associated to CCR algebra on infinite lattices, we discuss a simpler model like quantum spin system involving bounded operators for which a well-developed analysis exists, refer [BR87].

Consider a $d$-dimensional lattice $\mathbb{Z}^d$ and for each point $x \in \mathbb{Z}^d$, we associate a Hilbert space $\mathfrak{h}_{\{x\}}$ of dimension $n \in \mathbb{N}$, and for each finite $\Lambda \subset \mathbb{Z}^d$, we define the tensor product space

$$\mathfrak{h}_\Lambda = \bigotimes_{x \in \Lambda} \mathfrak{h}_{\{x\}}$$

The associated $C^*$-algebra of bounded operators on $\mathfrak{h}_\Lambda$ is denoted by $\mathfrak{U}_\Lambda$. If $\Lambda_1 \subset \Lambda_2 \subset\subset \mathbb{Z}^d$, then we can identify $\mathfrak{U}_{\Lambda_1}$ as a subalgebra of $\mathfrak{U}_{\Lambda_2}$ in a natural way, by tensorising the elements of $\mathfrak{U}_{\Lambda_1}$ with unit operator of $\mathfrak{U}_{\Lambda_2 \setminus \Lambda_1}$, see [BR87]. For describing the dynamics of the quantum spin model, we need to define interaction of finite range $R \in (0, \infty)$ which is a family $\Phi = (\Phi(X) : X \subset\subset \mathbb{Z}^d)$ of self adjoint operators $\Phi(X) \in \mathfrak{U}_X$ for any $X \subset\subset \mathbb{Z}^d$. Now for a finite set $\Lambda \subset\subset \mathbb{Z}^d$ the Hamiltonian is defined by

$$U_\Lambda = \sum_{X \subset \Lambda} \Phi(X). \qquad (2.14)$$

The associated modular dynamics is defined by

$$\alpha_{t,\Lambda}(B) = e^{-it\beta U_\Lambda} B e^{it\beta U_\Lambda}. \qquad (2.15)$$

where $\beta > 0$ is the inverse temperature.

It satisfies the following condition

$$\frac{d\alpha_{t,\Lambda}(B)}{dt} = \delta_\Lambda(\alpha_{t,\Lambda}(B))$$

where $\delta_\Lambda$ is the bounded derivation defined by

$$\delta_\Lambda(B) = i[U_\Lambda, B]$$

and is known as inner derivation.

When the operator $B$ is localised and $U_\Lambda$ takes the form (2.14), the following limit provides the non-inner derivations

$$\delta(B) = \lim_{\Lambda \to \mathbb{Z}^d} i[U_\Lambda, B].$$

The limit

$$\alpha_t(f) \equiv \lim_{\Lambda \to \mathbb{Z}^d} \alpha_{t,\Lambda}(f)$$

exists and is generated by $\delta$ under the the conditions given in the following theorem, Prop 6.2.9, [BR87].

In the following theorem, we use the following notation. For the interaction $\Phi = (\Phi(X) : X \subset\subset \mathbb{Z}^d)$,

we define a norm
$$\|\Phi\|_\lambda = \sup_{x\in\mathbb{Z}^d} \sum_{X \ni x} |X|(n+1)^{2|X|} e^{\lambda D(X)} \|\Phi(X)\| < \infty$$

where $\lambda > 0$, $n$ is the dimension of $\mathfrak{h}_{\{x\}}$, $|X|$ and $D(X)$ denotes the cardinality and diameter of $X$, respectively. Let $\alpha_{t,\Lambda}$ be associated to potential $\Phi$.

**Proposition 2.1.** *(Finite speed of propagation) Let $\Phi$ be any interaction which satisfies the condition*

$$\|\Phi\|_\lambda < \infty$$

*for some $\lambda > 0$. Let $\alpha_{t,\Lambda}$ denote the evolution associated with $\Phi$ and $\Lambda \subset\subset \mathbb{Z}^d$ and $x \in \mathbb{Z}^d \to T_x$ the action of space translations.*

*It follows that*
$$\|[\alpha_{t,\Lambda}(A), B]\| \leq \|A\| \sum_{x\in\mathbb{Z}^d} \sup_{C\in\mathfrak{U}_{\{0\}}} \left(\frac{\|[T_x(C), B]\|}{\|C\|}\right) e^{-|x|\lambda + 2|t|\|\Phi\|_\lambda}$$

*for all $A \in \mathfrak{U}_{\{0\}}$ and $B \in \mathfrak{U}_\Lambda$, uniformly in $\Lambda$.*

Since for $t$, as compared to $|x|$ the exponential of the right hand side will be small. However, for large $t$ this is not the case. In the literature, this property is called finite speed of propagation.

In the next two chapters, we deal with more general cases of dynamics which includes extension of quantum harmonic oscillators and involving unbounded operators.

We now delve into exploring the dynamic evolution of various quantum systems. Central to this exploration are Markov semigroups and related coercive inequalities, which serve as fundamental tools for analysing the nuanced behaviours intrinsic to these systems. The next section provides a concise exposition on the background theory.

## 2.4 Quantum Markov Semigroups and Coercive Inequalities

The theory of operator semigroups was developed to describe physical dissipative evolutions which satisfy 'Markov property'. The quantum generalisation of the Markov semigroup is defined on von

Neumann algebra say $\mathfrak{M}$ which is a subalgebra of bounded operators $B(\mathfrak{h})$ on the Hilbert space $\mathfrak{h}$.

A map $\varphi : \mathfrak{M}_+ \to [0, \infty]$ is a normal semifinite faithful weight on $\mathfrak{M}$ where $\mathfrak{M}_+$ is the positive cone if it satisfies the following conditions:

1. $\{x \in \mathfrak{M} : \varphi(x^*x) < \infty\}$ is weak $*$-dense in $\mathfrak{M}$(semifiniteness),

2. if $x_i \to x$, then $\varphi(x) \leq \limsup \varphi(x_i)$ (normality),

3. $\varphi(x^*x) = 0$ implies $x = 0$ (faithfulness).

Note that a weight is a state described in the previous section if $\varphi(\mathbf{1}) = \mathbf{1}$. Moreover, a weight $\varphi$ is tracial if $\varphi(xy) = \varphi(yx)$ for all $x, y$ in $\mathfrak{M}$ where it is semifinite.

For the purpose of defining Quantum Markov semigroups, we also need to discuss the concept of completely positive maps.

A map $\phi : \mathfrak{M} \to \mathfrak{M}$ is called positive if $\phi(\mathfrak{M}_+) \subset \mathfrak{M}_+$ and is completely positive if for each $n \in \mathbb{N}$, $\phi \otimes id_{M_n} : \mathfrak{M} \otimes M_n \to \mathfrak{M} \otimes M_n$ is positive. Here, $\mathfrak{M} \otimes M_n$ is von Neumann algebra of $n$ by $n$ matrices with entries in $\mathfrak{M}$.

**Definition 2.1.** *Given a von Neumann algebra with a weight* $(\mathfrak{M}, \varphi)$. *A quantum Markov semigroup is a continuous one parameter family of linear transformations* $(P_t)_{t \geq 0}$ *on* $\mathfrak{M}$ *such that*

*i)* $P_{t+s} = P_t P_s$ *for all* $s, t \geq 0$;

*ii) For each* $t \geq 0$, $P_t$ *is completely positive;*

*iii)* $P_t 1 \leq 1$.

The quantum Markov semigroup preserves the normal semifinite faithful weight if $\varphi(f) = \varphi(P_t f)$, $f \in \mathfrak{M}_+, t \geq 0$. There are several possibilities for the notion of continuity in the definition of Markov semigroups. For Markov semigroups on a Banach space, the standard notion of continuity considered is strong continuity. Conversely, for semigroups on a von Neumann algebra, continuity is typically defined with respect to the weak topology. Additionally, continuity can be characterized in terms of a normal semifinite faithful weight $\varphi$. One such definition asserts that a Markov semigroup $(P_t)_{t \geq 0}$ is continuous if $\varphi(P_t f)$ converges weakly to $\varphi(f)$ as $t \to 0$ for any normal semifinite faithful weight $\varphi$.

We note that this type of continuity is always satisfied if the semigroup is invariant with respect to the weight $\varphi$. Another approach to defining continuity in this context is to require that $\varphi(|P_t f - f|)$ converges to 0 as $t \to 0$.

For a tracial weight $\varphi$, a noncommutative $L_p$ space $(\mathfrak{M}, \varphi)$ is defined as follows

$$\{x \in \mathfrak{M} : \varphi(|x|^p) < \infty\}$$

for $p \in [1, \infty)$, and the corresponding norm given by $\|x\|_p = \varphi(|x|^p)^{\frac{1}{p}}$. For the non-tracial case, this is not the norm. Instead, the construction for the non-tracial case by Haagerup, which is based on the Tomita-Takesaki theory, is utilised. The Tomita-Takesaki theory and KMS symmetry is discussed in the previous section to tackle nontracial weights, for more details we refer [BR87]. For a state $\omega = \text{Tr}(\rho \cdot)$, the $L_p(\omega)$ norm (see [OZ99] and references therein) is defined as

$$\|x\|^p_{L_p(\omega)} = \text{Tr} \left| \rho^{\frac{1-s}{p}} x \rho^{\frac{s}{p}} \right|^p, \quad p \in [1, \infty), \quad s \in [0, 1].$$

In particular, for $p = 2$ the norm is given by the scalar product

$$\langle x, y \rangle_{\omega, s} = \text{Tr}\left( \rho^{\frac{1-s}{2}} x^* \rho^{\frac{s}{2}} y \right).$$

We define an operator $\mathfrak{L}$ for $x \in \mathbb{L}_p(\omega)$ for which the following limit exists in the norm

$$\mathfrak{L}x \equiv \lim_{t \to 0^+} \frac{P_t x - x}{t}.$$

Then the operator $\mathfrak{L}$ is called the generator of the semigroup $(P_t)_{t \geq 0}$ and we use the notation $P_t = e^{t\mathfrak{L}}$. In the classical theory, a result by Hille and Yoshida provides necessary and sufficient condition for an operator to be the generator of a strongly continuous semigroup. Additionally, Beurling-Deny theorem [BD59] states that there is a one to one correspondence between the generator $\mathfrak{L}$ and a quadratic form called the Dirichlet form. There is a similar generalisation for the quantum setting explored in [Cip97],[Cip08],[Par00], [AHK77] which states that there is a one to one correspondence between quantum KMS symmetric Markov semigroups on $(\mathfrak{M}, \varphi)$ and Dirichlet forms say $\mathcal{E}(f)$ on

$L^2(\mathfrak{M}, \varphi)$ such that

$$\mathcal{E}(f) = \langle f, -\mathfrak{L}f \rangle$$

where scalar product corresponds to the Hilbert space $L^2(\mathfrak{M}, \varphi)$. The Markovian form corresponding to the operator $\mathfrak{L}$ is defined as

$$\Gamma(f) = \frac{1}{2}\left(\mathfrak{L}(f^*f) - f^*\mathfrak{L}(f) - \mathfrak{L}(f^*)f\right).$$

In exploring the evolution from classical to quantum Dirichlet forms, some key insights can be drawn from [Ska19]. We will discuss the noncommutative/quantum Dirichlet forms in detail in the next section. Analysing various quantum systems include determining the convergence of the corresponding Quantum Markov semigroup. Quantum dissipative systems and the convergence of quantum Markov semigroups are interconnected concepts that describe the behaviour of open quantum systems under the influence of dissipative processes. The study of their relationship provides insights into the long-term dynamics and equilibrium properties of quantum systems in contact with their environment. Similar to classical theory, the convergence of the quantum Markov semigroups are studied using a tool known as quantum coercive inequalities. The two extensively studied coercive inequalities are Poincaré inequality(also known as Spectral gap inequality) and Logarithmic Sobolev inequality [OZ99]. We first describe these inequalities in the classical setup. In the next section, we will discuss the literature review for quantum coercive inequalities. We recall that a generator $\mathfrak{L}$ satisfies Poincaré inequality if there exists a constant $0 < c_p < \infty$ such that

$$\text{var}(f) \leq c_p \mathcal{E}(f) \tag{2.16}$$

for any $f$ for which the right hand side is well defined and where $\text{var}(f)$ is the variance of $f$.
In the classical theory of Riemannian manifolds [Bak04], under condition that the Ricci curvature Ric $> 0$, the Markov semigroup satisfies the following gradient bound for a constant $\rho$

$$\Gamma(P_t f) \leq e^{-2\rho t} P_t \Gamma(f) \tag{2.17}$$

Such inequalities are applied in the theory of hypercontractivity. The semigroup $P_t$ is said to be hypercontractive if for all $1 \leq p < q < \infty$, there exists a constant $M$ such that

$$\|P_t f\|_q \leq \|f\|_p$$

for all $t > M$. Hypercontractivity provides analysis of the smoothing properties of a semigroup. Additionally, in order to prove the hypercontractivity for the Ornstein-Uhlenbeck semigroup the inequality known as Logarithmic Sobolev inequality introduced by Gross[Gro75] is utilised. For all positive functions $f$ such that $\mathcal{E}(f)$ is well-defined, the Logarithmic Sobolev inequality holds if there exists a constant $0 < c_{LS} < \infty$ independent of $f$ such that

$$\text{Ent}(f^2) \leq c_{LS} \mathcal{E}(f),$$

where Ent denotes the relative entropy of the positive function $f$. For a more detailed discussion on coercive inequalities in the classical setup, we refer the reader to [GZ02].

Before delving into the literature review on quantum coercive inequalities, we will first discuss the theory of noncommutative Dirichlet forms in the following section.

## 2.5 Noncommutative Dirichlet Forms

General theory of noncommutative Dirichlet forms was developed in [Cip08], following earlier contributions [MZ96, SQV84]. In the thesis we study a special class of noncommutative Dirichlet forms as developed in [Par00], see also [CZ24]. Formally, the Dirichlet form which is of interest to us is given by the following expression

$$\mathcal{E}_X(f) \equiv \int_{\mathbb{R}} \left( \langle \delta_{\alpha_t(X)}(f), \delta_{\alpha_t(X)}(f) \rangle_\omega + \langle \delta_{\alpha_t(X^*)}(f), \delta_{\alpha_t(X^*)}(f) \rangle_\omega \right) \eta(t) dt$$

where $X$ is an element of a noncommutative space, $\alpha_t$ is the modular dynamics associated to a state $\omega$, $\eta$ belongs to a special class of functions specified below and operators $f$ are elements of $\mathbb{L}_2(\omega)$ for which the expression given above is well defined. We are interested in studying specific models in which we can prove that such quadratic form satisfies all necessary conditions of Dirichlet form, that is densely defined, positive and satisfies the so called contraction property, see [Cip08, Par00]. According to general theory of noncommutative Dirichlet forms, via Beurling-Deny theorem, we can associate a Markov generator denoted later by $\mathfrak{L}$. Before proceeding with further study, we define the admissible function following [Par00].

**Definition 2.2.** *An analytic function $\eta : D \to \mathbb{C}$ on a domain $D$ containing the strip $\text{Im} z \in [-1/4, 1/4]$ is said to be admissible function if the following holds:*

1. $\eta(t) \geq 0$ for $t \in \mathbb{R}$,

2. $\eta(t + i/4) + \eta(t - i/4) \geq 0$ for $t \in \mathbb{R}$,

3. *there exists $M > 0$ and $p > 1$ such that the bound*

$$|\eta(t + is)| \leq M(1 + |t|)^{-p}$$

*holds uniformly in $s \in [-1/4, 1/4]$.*

Condition 1 ensures the positivity of the Dirichlet form $\mathcal{E}$. Furthermore, condition 2 implies the dissipativity of the corresponding generator $\mathfrak{L}$ (see [Par00]), given by

$$\mathfrak{L}(f^*f) - f^*\mathfrak{L}(f) - \mathfrak{L}(f^*)f = \int \left|\delta_{\sigma_{t-\frac{i}{4}}(X)}(f)\right|^2 \left(\eta^2_{s-i/4} + \eta^2_{s+i/4}\right) dt \geq 0.$$

Lastly, condition 3 is imposed as a technical assumption to ensure the well-definedness of the Dirichlet form and the operator $\mathfrak{L}$.

Examples of $\eta$ include Gaussian smoothing of the following function

$$\eta(t) \equiv \frac{e^{i\kappa t}}{\cosh(2n\pi t)}.$$

For a variety of other examples, refer to [CZ24] and [Par00].

In special cases, the Dirichlet form has a simpler form (see quantum OU, [CFL00], Alicki's operator

[CM17]) as follows

$$\mathcal{E}'(f) \equiv \sum_{j \in \mathcal{J}} \left( \langle \delta_{E_j}(f), \delta_{E_j}(f) \rangle_\omega + \langle \delta_{E_j^*}(f), \delta_{E_j^*}(f) \rangle_\omega \right)$$

for all operators $f$ for which the right hand side is well defined. Here, $E_j$ are the eigenvectors of the modular operator such that $\alpha_{\omega, \pm \frac{i}{2}}(E_j) = e^{\pm \xi} E_j$, for some $\xi \in \mathbb{R}$. It is important to note that here unlike in the classical case, we need to consider the derivations which depends on the choice of the state so that the corresponding operator defines a semigroup which is positivity preserving and symmetric. One can also consider general quantum OU by considering for $E_j$ more general Wick monomials in annihilation and creation operators, see also Chapter 3 below.

## 2.6 Quantum Coercive Inequalities

In the classical setup, some coercive inequalities can be obtained when the Ricci curvature has a lower bound. For the quantum markov semigroups on a finite dimensional $C^*$-algebra of full matrix algebras, known as the Lindbladian semigroups, the quantum generalisation of the Ricci curvature was introduced in [CM17]. We reiterate that in an infinite dimensional setting, it would be necessary to consider a von Neumann algebra in addition to a $C^*$-algebra. The generators of quantum markov semigroups in the finite dimensional setting described above is given by the generalisation of Alicki's theorem. Given $\{V_j\}_{j \in \mathcal{J}}$ consists of eigenvectors of the modular operator, with certain properties (refer Theorem 3.1,[CM17]), the generator of the corresponding semigroups for any operator $f$ in $C^*$-algebra has the form

$$\mathfrak{L}f = \sum_j \left( -e^{-\frac{\beta}{2}} [V_j, f] V_j^* + e^{\frac{\beta}{2}} V_j^* [V_j, f] \right). \tag{2.18}$$

The corresponding quantum coercive inequalities and interlinks are studied in [DR20](see Figure 2). The corresponding dimension dependent gradient bound estimates of the form (2.17) on finite dimensional Hilbert space were proved in [WZ23]. The hypercontractivity for noncommutative semigroups acting on operator algebras was studied in [OZ99] and later for finite dimensional quantum systems in [TPK14, Kin14] and references therein. We give a brief description of their applications to the study of quantum information theory and quantum computing in the next section.

The analysis of these inequalities have been extended in the case of Bosonic Ornstein-Uhlenbeck semigroups which are always infinite dimensional. The quantum generalisation of the Ornstein-Uhlenbeck semigroup was introduced in [Rgu15] and was later analysed in [CFL00],[CS07]. The quantum OU semigroup describes the dissipative dynamics of the quantum harmonic oscillator which is described in Section 2.2. For the interaction $U = N$, where the number operator is defined as $N = A^*A$ and state $\omega = Tr(e^{\beta N} \cdot)$, the corresponding Markov generator can be given as follows

$$\mathfrak{L}f = -\frac{\mu^2}{2}(A^*Af - 2A^*fA + fA^*A) - \frac{\lambda^2}{2}(AA^*f - 2AfA^* + fAA^*) \tag{2.19}$$

for constants $\lambda, \mu$ such that $\mu > \lambda > 0$. This generator describes the quantum OU semigroup. In [CFL00], the spectral properties of quantum OU semigroup are discussed where these semigroups are constructed by the means of noncommutative Dirichlet forms defined in the previous section. In the next chapter, we would be using similar construction for more general models. In Section 8 of [CFL00], the authors discuss a model for the limiting case $\lambda = \mu$, referring to it as the quantum Brownian motion semigroup. In this semigroup, the dynamics lack an invariant state. The associated Dirichlet form generates a symmetric Markov semigroup on $L^2$ and determines a semigroup on the algebra. The following result, demonstrating the absence of a spectral gap, was established.

**Theorem 2.1.** *The $L^2$ generator of the quantum Brownian motion has spectrum $[0, \infty[$.*

In this thesis, we present several more general models that exhibit a similar lack of a spectral gap but also possess an equilibrium state, which is not present in the quantum Brownian motion model.
We analyse the dynamics of such models on infinite lattice systems. In classical setup, the polynomial decay of dissipative dynamics on infinite lattice systems was shown in [INZ12]. In our project, we provide the quantum generalisation of Corollary 6.2, [INZ12]. The no spectral gap property implies that the Poincaré inequality does not hold. The hypercontractivity and hence a logarithmic Sobolev inequality for quantum OU semigroup via spectral theory was established in [CS07, CM15]. The authors also proved that the logarithmic Sobolev inequalities implies spectral gap inequalities for Markov evolutions on von Neumann algebras. It is an open problem to prove hypercontractivity and logarithmic Sobolev inequalities for more general infinite dimensional models.

# 2.7 Connections to Quantum Information Theory and Quantum Computing

The development of quantum computers is the focus of many mathematicians, physicists and computer scientists nowadays. One of the major challenges that occur in their development is the external noise which needs to be suppressed [DR20]. The quantum Markov semigroups defined in the previous section is used to model some of the standard forms of noise in quantum computers. Understanding the convergence of quantum Markov semigroups which is done using quantum coercive inequalities is useful in determining the possible well behaving systems.

It was proposed in [VWC09] that analysing these evolution systems can identify relevant quantum states and develop quantum computing algorithms. In the cases where the evolutions converge fast enough, often defined as 'rapid mixing' in the quantum literature, the systems are stable against local perturbations. In the context of designing lifetime quantum memories, rapid mixing in a quantum Markov semigroup indicates quick quantum decoherence. Quantum decoherence is the inability of a quantum state to maintain superposition due to interactions with the environment, thus making the system ineffective for preserving quantum information.

The quantum decoherence creates the barrier for development of quantum information processing due to the decay of quantum correlations. Hence, it becomes essential to study the speed of decoherence. This has been formalised using the quantum Markov semigroups (QMS) which have unique equilibrium state known as primitive QMS in [OZ99] and non primitive case in [BR22]. The functional inequalities like Poincaré inequality, modified Log Sobolev inequality and its equivalent to hypercontractivity have been explored for finite open quantum systems in [DR20, CM17]. The hypercontractivity for specific classes of quantum semigroups have been discussed in [TPK14]. In quantum information theory, the hypercontractivity is analysed primarily for quantum channel semigroups on full matrix algebras. The quantum channels can be defined as a completely positive trace-preserving elements of the space of linear maps on the algebra 2.4 of complex-valued matrices. The contractive properties on such quantum channels are discussed in [Kin14].

The analysis of rapid mixing for various finite systems with finite range commuting interactions has been studied in [BCP22, Cap+23] . In order to explore the general case with non commuting long range interactions, one needs to establish certain Lieb-Robinson bounds. Once these bounds are established, it can be worth extending the results of [Cap+23] for infinite lattice systems.

## 2.8 Quantum Stochastic Calculus

The creation and annihilation operators were also used to generalise stochastic calculus for the quantum setup. In [GM91], a dynamical quantum system is constructed as an extension of the classical system where they provide relations between ground state expectations for the Hamiltonian and the expected value of functions of the configurations on which the appropriate classical stochastic process exists. In [Par15, Par18, Par86], the basic operator processes of quantum stochastic calculus are creation, annihilation and guage processes in the Hilbert space of square integrable functionals. The creation and annihilation operators replace the Brownian motion and the guage process replace Poisson process [Par15]. Using canonical commutation relations, the noncommutative stochastic differential equations and quantum Itô's formula are constructed. Further generalisations of important concepts in stochastic calculus including quantum martingales, Itô formula on some specific interesting examples like Heinsenberg algebra are explored in [Bia10, HS81]. This discussion on quantum stochastic calculus would be utilised to discuss possible open problems including interesting extensions of the work in this thesis.

# Chapter 3

# Dirichlet Forms and Poincaré Inequalities for Infinite Dimensional Models

This chapter is devoted to the study of dissipative dynamics of various infinite lattice systems and is based on the paper [MZ24a]. We provide numerous explicit examples of such models of large interacting systems and discuss their Dirichlet forms and corresponding Markov generators.

For some examples of models with locally conserved quantities associated to an infinite lattice, we discuss the existence of the Poincaré inequality and the corresponding dynamics.

As we discussed in the background of quantum spin systems in Section 2.3, in order to define the Dirichlet forms and Markov generators on dense domain for infinite systems we need to establish the finite speed of propagation of information (also known as Lieb-Robinson bounds) [BR87]. In this chapter, we establish this propagation bound for the systems with multiparticle interactions and unbounded potentials.

This chapter is organised as follows. In Section 3.1, we set up the framework for analyzing dissipative dynamics on infinite lattice systems. Section 3.2 explores potential domain issues that may arise within this setup. Then, in Section 3.3, we derive explicit expressions for the adjoint of the derivations and establish a modified Leibniz rule. Section 3.4 presents the finite speed of information propagation, followed by Section 3.5, where we demonstrate convergence in $\mathbb{L}_p$ norms. Section 3.6 delves into the corresponding Dirichlet and Markovian forms. In Section 3.7, we introduce a variety of explicit models, with Section 3.8 examining their spectral properties.

In Section 3.9, we discuss the algebra of invariant derivations. Finally, in Section 3.10, we show that these models exhibit algebraic decay to equilibrium.

## 3.1 Infinite Quantum Systems

Let $\mathbb{Z}^d$ be the $d$-dimensional square lattice for some fixed $d \in \mathbb{N}$, equipped with the $l_1$ lattice metric $dist(\cdot, \cdot)$ defined by
$$dist(\mathbf{i}, \mathbf{j}) := |i - j|_1 \equiv \sum_{l=1}^{d} |i_l - j_l|$$
for $\mathbf{i} = (i_1, \ldots, i_d), \mathbf{j} = (j_1, \ldots, j_d) \in \mathbb{Z}^d$. For $\mathbf{i}, \mathbf{j} \in \mathbb{Z}^d$, we say that $\mathbf{i}$ and $\mathbf{j}$ are neighbours in the lattice whenever $0 \le dist(\mathbf{i}, \mathbf{j}) \le R$ given $R \in (0, \infty)$ and write $\mathbf{i} \sim \mathbf{j}$. If $O$ is a finite subset of $\mathbb{Z}^d$ we will write $O \subset\subset \mathbb{Z}^d$, that is, $O$ is relatively compact in $\mathbb{Z}^d$.

For each $\Lambda \subseteq \mathbb{Z}^d$ which is not necessarily a bounded set, we associate a separable Hilbert space $\mathcal{H}_\Lambda$. This space satisfies a property that for bounded sets $\Lambda_1, \Lambda_2 \subset\subset \mathbb{Z}^d$, $\Lambda_1 \subset \Lambda_2$, we have $\mathcal{H}_{\Lambda_1} \subset \mathcal{H}_{\Lambda_2}$ and
$$\mathcal{H}_{\mathbb{Z}^d} = \overline{\cup_{\Lambda \subset\subset \mathbb{Z}^d} \mathcal{H}_\Lambda}.$$

For a finite set $O \subset\subset \Lambda$, let $\mathcal{A}_O$ denote the algebra of bounded operators. Let $\mathcal{A}_\Lambda = \cup_{O \subset\subset \Lambda} \mathcal{A}_O$ be an inductive limit algebra [BR87] generated by all local algebras.

Assume that we have a family of mutually commuting copies of creation and annihilation operators $\{A_j, A_j^*\}_{j \in \mathbb{Z}^d}$ satisfying the canonical commutation relation(CCR) given by

$$[A_j, A_j^*] = id, \quad [A_j, A_k^\sharp] = 0, \ j \ne k. \tag{3.1}$$

where $A_k^\sharp \in \{A_k, A_k^*\}$.

Since $A_k^\sharp$ are unbounded operators, for $O \subset\subset \mathbb{Z}^d$, we define $\mathcal{D}_O$ which denotes the family of finite polynomials in the creation and annihilation operators $A_k^\sharp$, $k \in O$ and let $\mathcal{D} \equiv \cup_{O \subset\subset \mathbb{Z}^d} \mathcal{D}_O$. A particle number operator at $j \in \mathbb{Z}^d$ is defined by
$$N_j \equiv A_j^* A_j$$

Since the eigenvalues of $N_j$ are $n_j \in \mathbb{N}$, it is a positive operator. Since these $N_j$'s are copies, we define for $\beta \in (0, \infty)$, a finite and positive operator

$$Z_o \equiv Tr_j\left(e^{-\beta N_j}\right)$$

which is independent of $j \in \mathbb{Z}^d$. The partial trace $Tr_j$ is defined as follows. We can choose an orthonormal basis $\{k_j\}$ of the considered Hilbert space $\mathcal{H}_j$ consisting of eigenvectors of $N_j$. Then we can write

$$Z_o \equiv Tr_j\left(e^{-\beta N_j}\right) \equiv \sum_{k_j} \langle e^{-\beta N_j} k_j, k_j \rangle.$$

Since by choice of the basis we have, $e^{-\beta N_j} k_j = e^{-\beta n_j} k_j$, where $n_j$ is the corresponding eigenvalue, so

$$Z_o \equiv \sum_{k_j} \langle e^{-\beta n_j} k_j, k_j \rangle = \sum_{k_j} e^{-\beta n_j} \langle k_j, k_j \rangle = \sum_j e^{-\beta n_j}$$

which is a geometric series.

Further, for $\Lambda \subset\subset \mathbb{Z}^d$, define a density matrix

$$\rho_{o,\Lambda} \equiv \bigotimes_{j \in \Lambda} \frac{1}{Z_o} e^{-\beta N_j}. \tag{3.2}$$

With $Tr_\Lambda = \otimes_{j \in \Lambda} Tr_j$, define on $\mathcal{D} \cup \mathcal{A}$ the following maps

$$\mathbb{E}_{o,\Lambda,s,p}(f) \equiv Tr_\Lambda \left|\rho_{o,\Lambda}^{(1-s)/p} f \rho_{o,\Lambda}^{s/p}\right|^p$$

with $0 \leq s \leq 1$ and $p \in [1, \infty)$. We consider an infinite product state

$$\omega_o(f) \equiv \lim_{\Lambda \to \mathbb{Z}^d} \mathbb{E}_{o,\Lambda,s,p=1}(f)$$

which is well defined for nonnegative operators on $\mathcal{D} \cup \mathcal{A}$ by definition of these spaces and can be extended to more general operators using decomposition of operators in positive and negative part for which the above formula is finite.

The corresponding $\mathbb{L}_p(\omega_o)$, $p \in [1, \infty)$, norm is well defined on $\mathcal{D}$ (and $\cup \mathcal{A}$) by

$$\|f\|_{\omega_o,p,s}^p = \lim_{\Lambda \to \mathbb{Z}^d} Tr_\Lambda \left|\rho_{o,\Lambda}^{(1-s)/p} f \rho_{o,\Lambda}^{s/p}\right|^p$$

where $|g| \equiv (g^*g)^{\frac{1}{2}}$. For any unbounded $O \subset \mathbb{Z}^d$, if we replace the limits by $\Lambda \to O$, we can introduce the corresponding state $\omega_{o,O}$ and the corresponding $\mathbb{L}_p(\omega_{o,O})$, $p \in [1, \infty)$, norms in a similar fashion. Next we define the finite range interaction for our setup. Consider a family of self adjoint densely defined operators $\Phi_O \in \mathcal{A}_O$ (or $\mathcal{D}_O$), such that

$$\Phi_O = 0 \text{ if } diam(O) \geq R.$$

This family of operators $\{(\Phi_O)_{O \subset \subset \mathbb{Z}^d}\} = \Phi$ is called a potential of finite range $R \in (0, \infty)$. The potential energy for a finite set $\Lambda \subset \subset \mathbb{Z}^d$ is defined by

$$U_\Lambda = \sum_{O \subset \Lambda} \Phi_O.$$

Following [BR87], we assume that the potential has the following thermodynamic stability property

$$0 < Z_\Lambda \equiv Tr_\Lambda e^{-U_\Lambda} < \infty,$$

$$0 < \left|\limsup_{\Lambda \to \mathbb{Z}^d} \frac{1}{|\Lambda|} \log Z_\Lambda\right| < \infty.$$

In our definition of density matrix (3.2), if the inverse temperature $\beta$ varies across each copy of creation and annihilation operators, denoted as $\beta_j$, with $j$ indexing each copy, a notable circumstance arises when $\beta_j$ tends towards infinity as the index $j$ increases indefinitely. Under such conditions, this conventional stability property associated with statistical mechanics and thermodynamics may no longer hold.

For this chapter, we consider the bounded multiparticle interaction in the form of polynomials in bounded operators $a_k^\sharp \equiv (1 + \varepsilon N_k^{\frac{1}{2}})^{-1} A_k^\sharp$, $k \in \mathbb{Z}^d$, $\varepsilon \in (0, \infty)$,

$$\Phi_O \equiv \Phi_O(a_k^\sharp, k \in O)$$

where the right hand side is a function of bounded operators indicated, for example, a polynomial function.

One can see that $a_k^\sharp$ is bounded by considering $|a_k^\sharp|^2$ and writing it as a function of $N$ using the relations (2.13).

In this context, $\Phi_O$ is given by bounded operators. The set $O$ considered here is bigger than one point set. For the $Tr_\Lambda$ to be finite, we need atleast one point of lattice interaction to be unbounded since we have infinite dimensional Hilbert space. This behavior enables a method for estimating $Z_\Lambda$, constrained by the trace, wherein the partition function involves the exponential of the negative sum of one-point interactions. With this observation, it becomes feasible to evaluate the bound of the operator $\Phi_O$ where $O$ denotes a region or set encompassing more than one point. By employing similar principles as applied to the one-point case, an estimation of $\Phi_O$ can be derived, facilitating the analysis of systems characterized by interactions across larger spatial regions.

We can then define a state on $\mathcal{D} \cup \mathcal{A}$ with respect to the interaction $U_\Lambda$

$$\omega_\Lambda(f) = \omega_o Tr_\Lambda(\rho_\Lambda f)$$

where the density matrix is given by $\rho_\Lambda \equiv \frac{1}{Z_\Lambda} e^{-\beta U_\Lambda}$ and the associated modular dynamics is defined by

$$\alpha_{t,\Lambda}(B) = \lim_{\tilde{\Lambda} \to \mathbb{Z}^d} \rho_{o,\tilde{\Lambda} \cap \Lambda^c}^{it} \rho_\Lambda^{it} B \rho_\Lambda^{-it} \rho_{o,\tilde{\Lambda} \cap \Lambda^c}^{-it}. \tag{3.3}$$

In this definition, for localised operators $B$ in $\Lambda$, the $o$ in the indices of the density matrix can be neglected and we have

$$\alpha_{t,\Lambda}(B) = \lim_{\Lambda \to \mathbb{Z}^d} \rho_\Lambda^{it} B \rho_\Lambda^{-it}. \tag{3.4}$$

Notice that the limit 3.3 is considered over $\tilde{\Lambda}$ such that $\Lambda \subset \tilde{\Lambda}$. This deliberate selection is made in anticipation of employing interpolation techniques later, particularly when establishing the finite speed of information propagation.

The scalar product associated to this state is given as follows

$$\langle f, g \rangle_{\omega^{(\Lambda)}} = \omega^{(\Lambda)}((\alpha_{-i/2,\Lambda}(f))^* g) = \omega_\Lambda(f^*(\alpha_{-i/2,\Lambda}(g))).$$

To be able to define these quantities for the whole lattice, we need to establish the following limits where $\Lambda$ evades the whole lattice

$$\omega(f) \equiv \lim_{\Lambda \to \mathbb{Z}^d} \omega_\Lambda(f) \tag{3.5}$$

$$\alpha_t(f) \equiv \lim_{\Lambda \to \mathbb{Z}^d} \alpha_{t,\Lambda}(f) \tag{3.6}$$

These limits exist in the case of bounded quantum spin systems, as discussed in Chapter 6 of [BR87]. We provide suitable conditions under which these limit exists in the next few sections for finite range uniformly bounded weak multiparticle potential or quadratic ones.

Consequently, the scalar product given by

$$\langle f, g \rangle_\omega = \lim_{\Lambda \to \mathbb{Z}^d} \langle f, g \rangle_{\omega_\Lambda} = \omega(f^*(\alpha_{-i/2}(g))) = \omega((\alpha_{-i/4}(f))^*(\alpha_{-i/4}(g)))$$

would be well defined. Again, the corresponding $\mathbb{L}_p(\omega)$, for any $p \in [1, \infty)$, functional is well defined on $\mathcal{D} \cup \mathcal{A}$ by

$$\|f\|_{\omega,p,s}^p = \lim_{\Lambda \to \mathbb{Z}^d} Tr_\Lambda \left| \rho_\Lambda^{(1-s)/p} f \rho_\Lambda^{s/p} \right|^p$$

where $|g| \equiv (g^*g)^{\frac{1}{2}}$. Note that in particular case $p \in \mathbb{N}$, we have the following expression, see e.g. [MZ96],

$$\|f\|_{\omega,p,1/2}^p = \omega\left(\alpha(i/2p)(f)\alpha(3i/2p)(f)...\alpha((2p-1)i/2p)(f)\right)$$

where $\alpha(qi/2p)(f) \equiv \alpha_{qi/2p}(f)$, for a suitable class of operators $f$.

The derivations of the bounded operators are well defined on assumed Hilbert space (see [BR87]). Although to define them for unbounded operators, we need some extra conditions to be able to define them on some domain in the Hilbert space.

A derivation $\delta_X$ in direction $X \in \mathcal{A}$ (resp. $\mathcal{D}$) on $D(\delta_X) = \mathcal{A}$ (resp. $\mathcal{D}$) is defined by

$$\delta_X(B) = i[X, B].$$

The following proposition defines the domain on which the derivation is well defined.

**Proposition 3.1.** *If $X^*X, XX^* \in \mathbb{L}_2(\omega)$, then the derivation is well defined on a domain*

$$D(\delta_X) \supset \{B \in \mathbb{L}_2 : B^*B, BB^* \in \mathbb{L}_2(\omega)\}.$$

*In particular for any $X \in \mathcal{D} \cup \mathcal{A}$*

$$\mathcal{D} \cup \mathcal{A} \subset D(\delta_X).$$

*The above condition holds if $X, \alpha_{\pm i/4}(X) \in \mathbb{L}_4(\omega)$ and then we have $D(\delta_X) \supset \{B, \alpha_{\pm i/4}(B) \in \mathbb{L}_4(\omega)\}$.*

*Proof.* We note if $XB, BX \in \mathbb{L}_2(\omega)$, then

$$\langle \delta_X(B), \delta_X(B) \rangle_\omega = \langle XB - BX, XB - BX \rangle_\omega \leq 2\|XB\|_\omega^2 + 2\|BX\|_\omega^2$$

If $\tilde{\omega}(\cdot) = Tr(\tilde{\rho} \cdot)$, with a density matrix $\tilde{\rho}$, then using the definition of the norm and Cauchy-Schwartz inequality(in $\mathbb{L}_2(\omega)$) for trace we get

$$\begin{aligned}
\|XB\|_{\tilde{\omega}}^2 &= Tr\left(\tilde{\rho}^{\frac{1}{2}} B^* X^* \tilde{\rho}^{\frac{1}{2}} XB\right) = Tr\left(B\tilde{\rho}^{\frac{1}{2}} B^* \cdot X^* \tilde{\rho}^{\frac{1}{2}} X\right) \\
&\leq \left(Tr\left(B\tilde{\rho}^{\frac{1}{2}} B^* B \tilde{\rho}^{\frac{1}{2}} B^*\right)\right)^{1/2} \left(Tr\left(X\tilde{\rho}^{\frac{1}{2}} X^* X \tilde{\rho}^{\frac{1}{2}} X^*\right)\right)^{1/2} \\
&= \left(Tr\left(\tilde{\rho}^{\frac{1}{2}} B^* B \tilde{\rho}^{\frac{1}{2}} B^* B\right)\right)^{1/2} \left(Tr\left(\tilde{\rho}^{\frac{1}{2}} X^* X \tilde{\rho}^{\frac{1}{2}} X^* X\right)\right)^{1/2} = \|B^*B\|_{\tilde{\omega}} \|XX^*\|_{\tilde{\omega}}
\end{aligned}$$

Similarly we have

$$\|BX\|_{\tilde{\omega}}^2 = Tr\left(\tilde{\rho}^{\frac{1}{2}} X^* B^* \tilde{\rho}^{\frac{1}{2}} BX\right) = Tr\left(B^* \tilde{\rho}^{\frac{1}{2}} B \cdot X \tilde{\rho}^{\frac{1}{2}} X^*\right)$$

$$\leq \|BB^*\|_{\tilde{\omega}} \|X^*X\|_{\tilde{\omega}}$$

Next we remark that, with $\tilde{\alpha}_{\frac{i}{4}}(X) \equiv \tilde{\rho}^{-\frac{1}{4}} X \tilde{\rho}^{\frac{1}{4}}$, we have

$$\begin{aligned}
\|X^*X\|_{\tilde{\omega}}^2 &= Tr\left(\left(\tilde{\rho}^{\frac{1}{8}} X^* \tilde{\rho}^{\frac{1}{8}}\right)\left(\tilde{\rho}^{\frac{1}{8}} \tilde{\rho}^{-\frac{1}{4}} X \tilde{\rho}^{\frac{1}{4}} \tilde{\rho}^{\frac{1}{8}}\right)\left(\tilde{\rho}^{\frac{1}{8}} X^* \tilde{\rho}^{\frac{1}{8}}\right)\left(\tilde{\rho}^{\frac{1}{8}} \tilde{\rho}^{-\frac{1}{4}} X \tilde{\rho}^{\frac{1}{4}} \tilde{\rho}^{\frac{1}{8}}\right)\right) \\
&= Tr\left(\left(\tilde{\rho}^{\frac{1}{8}} X^* \tilde{\rho}^{\frac{1}{8}}\right)\left(\tilde{\rho}^{\frac{1}{8}} \tilde{\alpha}_{\frac{i}{4}}(X) \tilde{\rho}^{\frac{1}{8}}\right)\left(\tilde{\rho}^{\frac{1}{8}} X^* \tilde{\rho}^{\frac{1}{8}}\right)\left(\tilde{\rho}^{\frac{1}{8}} \tilde{\alpha}_{\frac{i}{4}}(X) \tilde{\rho}^{\frac{1}{8}}\right)\right) \\
&\leq \|X^*\|_{\tilde{\omega},4}^2 \|\tilde{\alpha}_{\frac{i}{4}}(X)\|_{\tilde{\omega},4}^2
\end{aligned}$$

and similarly

$$\|XX^*\|_{\tilde{\omega}}^2 \leq \|X\|_{\tilde{\omega},4}^2 \|\tilde{\alpha}_{\frac{i}{4}}(X^*)\|_{\tilde{\omega},4}^2$$

The rest follows by approximation of norms associated to a state by the corresponding norms associated to normal states, see [Zeg02] and references therein. □

## 3.2  Discussion of Domain Issues

Consider single point potential given by $\Phi_j \equiv V(N_j)$ with unbounded real function $V$ (and the rest of the potential bounded of finite range). Consider the one point case with the state $\omega_V$ given by density matrix $\exp\{-\beta V(N)\}/Z$. Then we notice that, using (2.13), we have

$$\alpha_{is}(A^*) \equiv e^{sV(N)} A^* e^{-sV(N)} = e^{s(V(N+1)-V(N))} A^*$$
$$\alpha_{-is}(A) \equiv e^{-sV(N)} A\, e^{sV(N)} = e^{s(V(N+1)-V(N))} A$$

If $V$ is linear or sublinear the exponential multiplier is bounded. However for $V$ growing faster than linearly, the exponential multiplier is an unbounded operator and $\alpha_{\pm is}(A^\sharp)$ may not be in $L_p(\omega_V)$ for some $s \neq 0$ and all sufficiently large $p \in [1, \infty)$. In this case it is necessary to consider a suitable replacement for the space $\mathcal{D}$ by considering mollification such that it kills the exponential factor. We propose to consider for $\varepsilon > 0$ the following set

$$\mathcal{D}_{\varepsilon,\Lambda} \equiv \{F e^{-\sum_{j \in \Lambda} \varepsilon V(N_j) N_j^\delta} : F \in \mathcal{D}_\Lambda,\ \delta \in (0, \infty)\}$$

and for infinite system

$$\mathcal{D}_\varepsilon \equiv \cup_{\Lambda \subset\subset \mathbb{Z}^d} \mathcal{D}_{\varepsilon,\Lambda}.$$

Such set is dense in the closure of $\mathcal{D}$ with respect to $\mathbb{L}_p$ norms. This dense set approximates every $F \in \mathcal{D}_\Lambda$ and leads to well defined quantities.

## 3.3 Adjoint Operators

In order to define the adjoint of the derivations, we first define the left and right multiplication operator(similar to [CFL00]) as follows

$$L_X f \equiv Xf, \qquad R_X f \equiv fX$$

provided $Ran(f) \subset D(X)$ and $Ran(X) \subset D(f)$, respectively.

We obtain the following formulas for the corresponding adjoints with respect to the scalar product.

$$\langle L_X(g), f \rangle = Tr\left(\rho^{\frac{1}{2}}(Xg)^* \rho^{\frac{1}{2}} f\right) = Tr\left(\rho^{\frac{1}{2}} g^* \rho^{\frac{1}{2}} \left(\rho^{-\frac{1}{2}} X^* \rho^{\frac{1}{2}}\right) f\right)$$
$$= \langle g, L_{\alpha_{i/2}(X^*)} f \rangle$$

where for the first equality, we use the definition of scalar product associated to a normal state $\omega(\cdot) \equiv Tr(\rho \cdot)$, with density operator $\rho$ with respect to a trace $Tr$. In the second equality, the property of operator adjoint $(Xg)^* = g^* X^*$ is used and inserted $\rho^{\frac{1}{2}} \rho^{-\frac{1}{2}}$ between $g^*$ and $X^*$ in order to represent it as a scalar product.

Similarly we get

$$\langle R_X(g), f \rangle = \langle g, R_{\alpha_{-i/2}(X^*)} f \rangle.$$

Denoting the adjoint operation associated to the scalar product by $\star$, we summarise this as follows

$$L_X^\star = L_{\alpha_{i/2}(X^*)}, \qquad R_X^\star = R_{\alpha_{-i/2}(X^*)}.$$

Consequently, the adjoint of derivation with respect to a scalar product can be derived as follows

$$\langle \delta_X(f), g \rangle_\omega \equiv Tr(\rho^{1/2}(\delta_X(f))^* \rho^{1/2} g) = Tr(\rho^{1/2}(i[X^*, f^*])\rho^{1/2} g)$$
$$= Tr(\rho^{1/2} f^* \rho^{1/2}(g\rho^{1/2} iX^* \rho^{-1/2} - \rho^{-1/2} iX^* \rho^{1/2} g)))$$
$$\equiv \langle f, \delta_X^\star(g) \rangle_\omega$$

As a result, we have the following proposition.

**Proposition 3.2** *For $X \in \mathcal{D}_\varepsilon$*

$$\delta_X^\star(g) = i\left(R_{\alpha_{-i/2}(X^*)}(g) - L_{\alpha_{i/2}(X^*)}(g)\right)$$

$$= i\left(g\alpha_{-i/2}(X^*) - \alpha_{i/2}(X^*)g\right)$$

$$= -\delta_{\alpha_{-i/2}(X^*)}(g) - i\left(\alpha_{i/2}(X^*) - \alpha_{-i/2}(X^*)\right)g$$

$$= -\delta_{\alpha_{i/2}(X^*)}(g) - gi\left(\alpha_{i/2}(X^*) - \alpha_{-i/2}(X^*)\right)$$

$$= -\delta_{\frac{1}{2}(\alpha_{i/2}(X^*)+\alpha_{-i/2}(X^*))}(g) + \left\{\frac{1}{2}i\left(\alpha_{-i/2}(X^*) - \alpha_{i/2}(X^*)\right), g\right\}$$

*on a dense domain $\mathcal{D}(\delta_X^\star) \supset \mathcal{D}$. Moreover, for $f, g \in \mathcal{D}_\varepsilon$, we have modified Leibnitz rule*

$$\delta_X^\star(fg) = \delta_X^\star(f)g - f\delta_{\alpha_{-i/2}(X^*)}(g)$$

$$= f\delta_X^\star(g) - \delta_{\alpha_{i/2}(X^*)}(f)g$$

$$= \delta_X^\star(f)g + f\delta_X^\star(g) - if\left(\alpha_{-i/2}(X^*) - \alpha_{i/2}(X^*)\right)g.$$

**Corollary 3.1.** *Let $X \in \mathcal{D}_\varepsilon$. If $\xi \in \mathbb{R}$, we have*

$$\alpha_{\pm i/2}(X) = e^{\pm\xi}X$$

*and so*

$$\alpha_{\mp i/2}(X^*) = e^{\pm\xi}X^*,$$

*then*

$$\delta_X^\star(g) = i\left(g(e^\xi X^*) - (e^{-\xi}X^*)g\right)$$

$$= -e^\xi \delta_{X^*}(g) + 2i\sinh(\xi)X^*g$$

$$= -e^{-\xi}\delta_{X^*}(g) + 2i\sinh(\xi)gX^*$$

$$= -\cosh(\xi)\delta_{X^*}(g) + i\sinh(\xi)\{X^*, g\}.$$

*Moreover, one has the following modified Leibnitz rule*

$$\delta_X^\star(fg) = \delta_X^\star(f)g - e^\xi f \delta_{X^*}(g)$$
$$= f\delta_X^\star(g) - e^{-\xi}\delta_{X^*}(f)g.$$

**Remark 3.1.** *In the algebra related to $\xi = 0$, such as the algebra of functions of number operator N as seen in the quantum harmonic oscillator, we rediscover perfect Leibnitz rule.*

## 3.4 Modular Dynamics and Finite Speed of Propagation of Information

In this section, we establish the limit of the modular dynamics (3.6). In order to define the modular dynamics $\alpha_t(f)$ for infinite sets, we need a sequence $\alpha_{t,\Lambda}$ such $\Lambda \to \infty$. The operator estimates exist in the case of bounded operators since there are no domain issues which is not the case the unbounded operators.

The methodology employed to derive the estimates within this section bears resemblance to techniques used in the examination of Poincaré inequalities within Heisenberg groups and Log Sobolev inequalities on lattice structures. Central to our approach is the utilisation of interpolation principles and the strategic addition of points within the set $\Lambda$, thereby ensuring that estimates characterized by a single-point disparity exert significant influence over the required estimations.

Given the domain $\mathcal{D}$, it is adequate to define modular dynamics concerning creation and annihilation operators. Initially, we establish the modular dynamics concerning bounded operators, which are specified by polynomials involving the modified creation and annihilation operators.

$$a_j^\sharp \equiv \frac{1}{1 + \epsilon N_j^{\frac{1}{2}}} A_j^\sharp.$$

For $\Lambda \subset\subset \mathbb{Z}^d$, $\alpha_{t,\Lambda}$ is well defined and for a given $j \in \Lambda$ and $k \notin \Lambda$, we have

$$\alpha_{t,\Lambda\cup\{k\}}(a_j^\sharp) - \alpha_{t,\Lambda}(a_j^\sharp) = -\int_0^t ds \frac{d}{ds}\alpha_{s,\Lambda}\left(\alpha_{t-s,\Lambda\cup\{k\}}(a_j^\sharp)\right)$$

$$= \int_0^t \sum_{\substack{O \ni k \\ O \neq \{k\}}} \alpha_{s,\Lambda}\delta_{\Phi_O}\left(\alpha_{t-s,\Lambda\cup\{k\}}(a_j^\sharp)\right)ds$$

Assuming $\Phi_O$ are bounded if $O$ is not one point set, this implies the following bound involving operator norm

$$\|\alpha_{t,\Lambda\cup\{k\}}(a_j^\sharp) - \alpha_{t,\Lambda}(a_j^\sharp)\| \leq \int_0^t \sum_{\substack{O \ni k \\ O \neq \{k\}}} \|\delta_{\Phi_O}\left(\alpha_{s,\Lambda\cup\{k\}}(a_j^\sharp)\right)\|ds$$

Let

$$c_\Phi \equiv 2\sup_{|diam(O)|\leq 2R} \widetilde{\sum_{O' \subset O_R}} \|\Phi_{O'}\|,$$

where the summation with ~ runs over sets different than one point sets and $O_R$ denotes a set of points with distance from $O$ bounded by $R$ which will be used to estimate the right hand side of the above inequality.

**Theorem 3.1.** *(Finite speed of propagation of information estimate)*
*Assume the potential is of finite range $R \in (0, \infty)$ and $c_\Phi < \infty$. There exist constants $D, C, m \in \mathbb{R}^+$ such that for any $j \in \Lambda \subset\subset \mathbb{Z}^d$ and any $t \in \mathbb{R}^+$ we have*

$$\|\delta_{\Phi_O}\left(\alpha_{t,\Lambda}(a_j^\sharp)\right)\| \leq De^{Ct-md(O,j)}$$

*where $\Phi_O$ is a bounded part of potential localised in a set of size $2R$. The estimate remains valid for $t \in \mathbb{C}$, $|\Im(t)| \leq 1$, with $\Re(t) \geq 0$, provided one point interaction $V(N_k)$ is at most linear or $a_j \in \mathcal{D}_\varepsilon$.*

*Proof.* If the interaction is of range $R \in (0, \infty)$, we note that when $dist(O, j) > 2R$ and $j \in \Lambda \setminus O_R$, we have

$$\delta_{\Phi_O}\left(\alpha_{s,\Lambda\setminus O_R}(a_j^\sharp)\right) = 0,$$

where $O_R \equiv \{l : dist(l, O) \leq 2R\}$. Hence

$$\|\delta_{\Phi_O}\left(\alpha_{t,\Lambda}(a_j^\sharp)\right)\| = \|\delta_{\Phi_O}\left(\alpha_{t,\Lambda}(a_j^\sharp) - \alpha_{t,\Lambda\setminus O_R}(a_j^\sharp)\right)\|$$

We have

$$\alpha_{t,\Lambda}(a_j^\sharp) - \alpha_{t,\Lambda\setminus O_R}(a_j^\sharp) = -\int_0^t ds \frac{d}{ds}\alpha_{t-s,\Lambda\setminus O_R}(\alpha_{s,\Lambda}(a_j^\sharp))$$
$$-\int_0^t ds \, \alpha_{t-s,\Lambda\setminus O_R}\left(\sum_{O'\subset O_R}^{\sim} \delta_{\Phi_{O'}}\left(\alpha_{s,\Lambda}(a_j^\sharp)\right)\right)$$

where $\sim$ over the sum indicates summation over the bounded part of the potential, since $\alpha_{\tau,\Lambda'}$ is defined outside the set $\Lambda'$ by using the one point potential only and this is cancelled out in the derivation with respect to $s$. (Here we may need to consider first bounded approximation of the one point potential.) Hence we get

$$\|\delta_{\Phi_O}\left(\alpha_{t,\Lambda}(a_j^\sharp)\right)\| \leq 2\|\delta_{\Phi_O}\left(\int_0^t ds \, \alpha_{t-s,\Lambda\setminus O_R} \sum_{O'\subset O_R}^{\sim} \delta_{\Phi_{O'}}\left(\alpha_{s,\Lambda}(a_j^\sharp)\right)\right)\|$$
$$\leq 2\|\Phi_O\| \sum_{O'\subset O_R}^{\sim} \int_0^t ds \|\delta_{\Phi_{O'}}\left(\alpha_{s,\Lambda}(a_j^\sharp)\right)\|$$

Since by our assumption

$$c_\Phi \equiv 2 sup_{|diam(O)|\leq 2R} \sum_{O'\subset O_R}^{\sim} \|\Phi_{O'}\| < \infty,$$

we get the following bound

$$\|\delta_{\Phi_O}\left(\alpha_{t,\Lambda}(a_j^\sharp)\right)\| \leq c_\Phi \sum_{O'\subset O_R}^{\sim} \int_0^t ds \|\delta_{\Phi_{O'}}\left(\alpha_{s,\Lambda}(a_j^\sharp)\right)\|$$

We repeat this bound iteratively, each time acquiring a finite constant multiplier $c_\Phi$, a finite summation comprising at most $2^{(2R)^d}$ terms (which represents the maximal count of subsets $O' \subseteq O_R$), and iterated integrals. The iteration halts when one of the sets $O'_R$ encompasses $j$. The minimum number of iterations required to reach $j$ is no less than $n \equiv \left\lfloor \frac{dist(O,j)}{2R} \right\rfloor$ (where $\lfloor \cdot \rfloor$ denotes the integer part). Thus, defining $C \equiv 2^{(2R)^d} c_\Phi$, we derive the ensuing bound.

$$\|\delta_{\Phi_O}\left(\alpha_{t,\Lambda}(a_j^\sharp)\right)\| \leq C^n \frac{t^n}{n!} e^{Ct} \|a_j^\sharp\|$$

At this point one needs to use additional assumptions on $a_j^\sharp$. Using Stirling-de Moivre bound

$$n! \geq n^{n+1/2} e^{-n}$$

we get

$$\|\delta_{\Phi_O}\left(\alpha_{t,\Lambda}(a_j^\sharp)\right)\| \leq \exp\{n(\log(Cet) - \log n)\} e^{Ct} \|a_j^\sharp\|$$

which for any fixed $t \in \mathbb{R}^+$ and $n \geq Cet \exp(m(2R)^d)$, yields

$$\|\delta_{\Phi_O}\left(\alpha_{t,\Lambda}(a_j^\sharp)\right)\| \leq e^{Ct-mn} \|a_j^\sharp\|$$

$\square$

This result generalises the result of [LR72], [Mat93] where a system of bounded spin on a lattice was considered. In our result, we are considering one special case where one may have unbounded single point potential but bounded for multiparticle potential. Within the set $\Lambda$, $\alpha_{t,\Lambda}$ is associated to product state or Gibbs state $\omega_\Lambda$. Conversely, for the points outside the set $\Lambda$ the product of states corresponds to single point potentials.

## 3.5 Convergence of $\mathbb{L}_p$ Norms

In this section, we demonstrate the limit (3.5) of the sequence of states $\omega^{(\Lambda)}$, $\Lambda \subset\subset \mathbb{Z}^d$ and corresponding $\mathbb{L}_p$ norms. Let $\Lambda_0 \subset \Lambda$ and $\partial_{2R}\Lambda_0 \equiv \{k \notin \Lambda_0 : dist(k, \Lambda_0) \leq 2R\}$. Consider the interpolation of the potential

$$\Phi_s \equiv \{\Phi_O, O \subset \Lambda \setminus \partial_{2R}\Lambda_0, s\Phi_{O'}, O' \cap \partial_{2R}\Lambda_0 \neq \emptyset\}_\sim, \quad s \in [0,1]$$

where $\sim$ signifies that we exclude one point potential and $s$ is the interpolation parameter. Denote by $\rho_{\Lambda,s} \equiv \rho_{\Lambda,\Lambda_0,s}$ the corresponding density matrix localised in $\Lambda$.

When $s = 0$, this density matrix is

$$\rho_{\Lambda,s=0} = \rho_{\Lambda\setminus\partial_{2R}\Lambda_0} \rho_{\Lambda_0} \rho_{o,\partial_{2R}\Lambda_0}$$

where the density matrices on the right hand side commute and the corresponding state is a product state. For $p \in \mathbb{N}$ and $f$ localised in $\Lambda_0$, we have

$$Tr_\Lambda \rho_{\Lambda,s=0} f = Tr_{\Lambda_0} \rho_{\Lambda_0} f$$

and

$$\|f\|_{p,\Lambda,s=0} = \|f\|_{p,\Lambda_0}.$$

We begin with product state with respect to the one point interaction. Further, we add points successively to obtain multiparticle interaction since we need to see the effect of multiparticle interaction. Next for bounded operator $f$, using fundamental theorem of calculus we have

$$Tr_\Lambda \left((\rho_\Lambda)^{\frac{1}{4p}} f^* (\rho_\Lambda)^{\frac{2}{4p}} f (\rho_\Lambda)^{\frac{1}{4p}}\right)^{2p} - Tr_\Lambda \left((\rho_{\Lambda,s=0})^{\frac{1}{4p}} f^* (\rho_{\Lambda,s=0})^{\frac{2}{4p}} f (\rho_{\Lambda,s=0})^{\frac{1}{4p}}\right)^{2p}$$
$$= \int_0^1 ds \frac{d}{ds} Tr_\Lambda \left((\rho_{\Lambda,s})^{\frac{1}{4p}} f^* (\rho_{\Lambda,s})^{\frac{2}{4p}} f (\rho_{\Lambda,s})^{\frac{1}{4p}}\right)^{2p}$$

Next we note that

$$\frac{d}{ds} (\rho_{\Lambda,s})^{\frac{1}{4p}} = \left(\frac{d}{ds} \exp\{-\frac{1}{4p} U_\Lambda(\Phi_s)\}\right) \frac{1}{Z_{\Lambda,s}^{\frac{1}{4p}}} - \frac{1}{4p} (\rho_{\Lambda,s})^{\frac{1}{4p}} \cdot \frac{1}{Z_{\Lambda,s}} \frac{d}{ds} Z_{\Lambda,s}$$

Introducing the interpolation between $\exp\{-\frac{1}{4p} U_\Lambda(\Phi_{s+h})\}$ and $\exp\{-\frac{1}{4p} U_\Lambda(\Phi_s)\}$ given by,

$$\exp\{-\frac{\tau}{4p} U_\Lambda(\Phi_{s+h})\} \exp\{-\frac{1-\tau}{4p} U_\Lambda(\Phi_s)\}$$

we can use an analogue of Fundamental theorem of Calculus for the derivative of the first factor on the right hand side as follows

$$\frac{d}{ds}\exp\{-\frac{1}{4p}U_\Lambda(\Phi_s)\} = \lim_{h\to 0}\frac{1}{h}\left(\exp\{-\frac{1}{4p}U_\Lambda(\Phi_{s+h})\} - \exp\{-\frac{1}{4p}U_\Lambda(\Phi_s)\}\right)$$

$$= \lim_{h\to 0}\frac{1}{h}\int_0^1 d\tau \frac{d}{d\tau}\exp\{-\frac{\tau}{4p}U_\Lambda(\Phi_{s+h})\}\exp\{-\frac{1-\tau}{4p}U_\Lambda(\Phi_s)\}$$

$$= -\frac{1}{4p}\int_0^1 d\tau \exp\{-\frac{\tau}{4p}U_\Lambda(\Phi_{s+h})\}\lim_{h\to 0}\frac{1}{h}(U_\Lambda(\Phi_{s+h}) - U_\Lambda(\Phi_s))\exp\{-\frac{1-\tau}{4p}U_\Lambda(\Phi_s)\}$$

$$= -\frac{1}{4p}\sum_{O\subset\partial_{2R}\Lambda}^{\sim}\int_0^1 d\tau\, (\rho_{\Lambda,s})^{\frac{\tau}{4p}}(\Phi_O)(\rho_{\Lambda,s})^{-\frac{\tau}{4p}}\rho_{\Lambda,s}^{\frac{1}{4p}}$$

$$\equiv -\frac{1}{4p}\sum_{O\subset\partial_{2R}\Lambda}^{\sim}\int_0^1 d\tau \left(\alpha_{\Lambda,s}(-i\frac{\tau}{4p})(\Phi_O)\right)\rho_{\Lambda,s}^{\frac{1}{4p}}$$

where $\alpha_{\Lambda,s}(-i\frac{\tau}{4p})$ denotes the automorphism corresponding to $\Phi_s$ at time $-i\frac{\tau}{4p}$, and similarly for the second power of the density(here the derivative of $U_\Lambda$ is trivial). Hence, we also get

$$\left|\frac{1}{Z_{\Lambda,s}}\frac{d}{ds}Z_{\Lambda,s}\right| = \left|-\sum_{O\subset\partial_{2R}\Lambda_0}^{\sim}\int_0^1 d\tau\frac{1}{Z_{\Lambda,s}}Tr_\Lambda(\rho_{\Lambda,s})^\tau(\Phi_O)(\rho_{\Lambda,s})^{-\tau}\rho_{\Lambda,s}\right|$$

$$= \left|-\sum_{O\subset\partial_{2R}\Lambda_0}^{\sim}\int_0^1 d\tau\omega_{\Lambda,s}(\Phi_O)\right| \le \sum_{O\subset\partial_{2R}\Lambda_0}^{\sim}\|\Phi_O\|$$

Using the above, we obtain

$$\frac{d}{ds}Tr_\Lambda\left((\rho_{\Lambda,s})^{\frac{1}{4p}}f^*(\rho_{\Lambda,s})^{\frac{2}{4p}}f(\rho_{\Lambda,s})^{\frac{1}{4p}}\right)^{2p} \le$$

$$\le Tr_\Lambda\left((\rho_{\Lambda,s})^{\frac{1}{4p}}f^*(\rho_{\Lambda,s})^{\frac{2}{4p}}f(\rho_{\Lambda,s})^{\frac{1}{4p}}\right)^{2p}\left(\sum_{O\subset\partial_{2R}\Lambda_0}^{\sim}\left(\|\Phi_O\| + \sup_{\Lambda,s,\tau\in[0,1]}\|\alpha_{\Lambda,s}(-i\frac{\tau}{4p})(\Phi_O)\|\right)\right)$$

and hence we arrive at the following bound

$$\|f\|_{\Lambda,p} \le \|f\|_{\Lambda_0,p}e^{C|\partial_{2R}\Lambda_0|} \tag{3.7}$$

with a constant

$$C \le \sum_{O\subset\partial_{2R}\Lambda_0}^{\sim}\left(\|\Phi_O\| + \sup_{\Lambda,s,\tau\in[0,1]}\|\alpha_{\Lambda,s}(-i\frac{\tau}{4p})(\Phi_O)\|\right)$$

which is finite under suitable assumptions on bounded part of the potential as discussed in previous section. This can be extended for any $f$ such that the right hand side of (3.7) is finite.

The above bounds (3.7) provide compactness of the set of states $\omega_\Lambda$, $\Lambda \subset\subset \mathbb{Z}^d$, and we have the following possibility of defining $\mathbb{L}_p(\omega)$ norms associated to a state $\omega \equiv \lim_{\Lambda_k \to \mathbb{Z}^d} \omega_{\Lambda_k}$ for some subsequence $\Lambda_k \subset \Lambda_{k+1}$,

$$\|f\|_{\omega,p} \equiv \limsup_{\Lambda \to \mathbb{Z}^d} \|f\|_{\omega_\Lambda,p}$$

Under additional assumptions on the interactions it is possible to use the ideas utilised above to prove convergence of the $\|f\|_{\omega_\Lambda,p}$ as $\Lambda \to \infty$.

Note that for positive $f$ the symmetric $\mathbb{L}_1(\omega_\Lambda)$ coincides with $\omega_\Lambda(f)$, so the problem of convergence is the same for both. On the other hand for the symmetric $\mathbb{L}_2(\omega_\Lambda)$, given the convergence of the sequence of state and the modular operator, we get convergence for corresponding norms. Given $\mathbb{L}_1(\omega_\Lambda)$ and $\mathbb{L}_2(\omega_\Lambda)$ one can use interpolation theory [RS72] to get all the intermediate norms and spaces and then by duality one can define the norms and $\mathbb{L}_2(\omega)$ spaces for $p \in (2, \infty)$.

## 3.6 Dirichlet Form and $\Gamma_1$ Function

Now, we define the Dirichlet form for our setup using the definition in Section 2.5.

Consider a family of local elements $X_j \in \mathcal{D}$, $j \in \mathbb{Z}^d$. As discussed in the Section 3.4, we can define the derivations in the directions $\alpha_t(X_j)$.

The Dirichlet form in $\mathbb{L}_{\omega,2}$ can be defined on the dense domain containing $\mathcal{D}_\varepsilon$ in the following way

$$\mathcal{E}_j(f) \equiv \int_\mathbb{R} \left(\nu_j \langle \delta_{\alpha_t(X_j)}(f), \delta_{\alpha_t(X_j)}(f) \rangle_\omega + \mu_j \langle \delta_{\alpha_t(X_j^*)}(f), \delta_{\alpha_t(X_j^*)}(f) \rangle_\omega \right) \eta(t) dt$$

for some constants $\nu_j, \mu_j$. For a finite $\Lambda \subset \mathbb{Z}^d$,

$$\mathcal{E}_\Lambda(f) \equiv \sum_{j \in \Lambda} \mathcal{E}_j(f).$$

The Markov generator corresponding to this Dirichlet form is given by

$$\langle f, -\mathfrak{L}_\Lambda f \rangle = \mathcal{E}_\Lambda(f).$$

The generator can be written explicitly in the following way

$$-\mathfrak{L}_j(f) = \int \left(\nu_j \delta^\star_{\alpha_t(X_j)} \delta_{\alpha_t(X_j)}(f) + \mu_j \delta^\star_{\alpha_t(X_j^*)} \delta_{\alpha_t(X_j^*)}(f)\right) \eta(t) dt \tag{3.8}$$

where the operation $\star$ is taking the adjoint with respect to the scalar product. In general, this generator is an unbounded operator defined on a suitable dense domain such that $\mathfrak{L}_j(f) \in \mathbb{L}_2$ and depends on the choice of $X_j$ and the state $\omega$.

The corresponding Markovian form is defined by

$$\Gamma_{1,\Lambda}(f) = \sum_{j \in \Lambda} \Gamma_{1,j}(f) = \frac{1}{2}\left(\mathfrak{L}_\Lambda(f^*f) - f^* \mathfrak{L}_\Lambda(f) - \mathfrak{L}_\Lambda(f^*)f\right)$$

and for the above generator $\Gamma_{1,j}(f)$ can be given by

$$2\Gamma_{1,j}(f) = \Bigg\{ \int \left(\nu_j \delta^\star_{\alpha_t(X_j)} \delta_{\alpha_t(X_j)}(f^*f) + \mu_j \delta^\star_{\alpha_t(X_j^*)} \delta_{\alpha_t(X_j^*)}(f^*f)\right) \eta(t) dt$$
$$- f^* \left( \int \left(\nu_j \delta^\star_{\alpha_t(X_j)} \delta_{\alpha_t(X_j)}(f) + \mu_j \delta^\star_{\alpha_t(X_j^*)} \delta_{\alpha_t(X_j^*)}(f)\right) \eta(t) dt \right)$$
$$- \int \left(\nu_j \delta^\star_{\alpha_t(X_j)} \delta_{\alpha_t(X_j)}(f^*) + \mu_j \delta^\star_{\alpha_t(X_j^*)} \delta_{\alpha_t(X_j^*)}(f^*)\right) \eta(t) dt\, f \Bigg\}$$

Hence, we can write the following proposition.

**Proposition 3.2.** *Let $\mathfrak{L}$ be the Markov generator defined by the Dirichlet form. Then the corresponding Markovian form is given by*

$$\Gamma_1(f) = -\frac{1}{2} \int \left(|\delta_{\alpha_{t-i/4}(X^*)}(f)|^2 + |\delta_{\alpha_{t-i/4}(X)}(f))|^2\right)(\eta(t+i/4) + \eta(t-i/4))\, dt$$

*for all operators $f, f^* \in \mathcal{D}(\mathfrak{L})$.*

*In the infinite dimensional case on a lattice, it is necessary to have the finite speed of propagation of information to secure the dense domain.*

*Proof.* We use the fact that $\mathfrak{L}$ is given by (3.8). Using the Leibniz rule (3.3) for the derivation for the first part of the integrant, (the second will be analogous), we have

$$\delta^\star_{\alpha_t(X)}\delta_{\alpha_t(X)}(f^*f) = \delta^\star_{\alpha_t(X)}\left(\delta_{\alpha_t(X)}(f^*)f + f^*\delta_{\alpha_t(X)}(f)\right)$$

Using the formulae of Proposition 3.3 for modified Leibniz property of the adjoint, for $f, g \in \mathcal{D}_\varepsilon$, we have

$$\delta^\star_Y(fg) = \delta^\star_Y(f)g - f\delta_{\alpha_{-i/2}(Y^*)}(g)$$
$$= f\delta^\star_Y(g) - \delta_{\alpha_{i/2}(Y^*)}(f)g,$$

hence

$$\delta^\star_{\alpha_t(X)}\delta_{\alpha_t(X)}(f^*f) = \left(\delta^\star_{\alpha_t(X)}\delta_{\alpha_t(X)}(f^*)\right)f - \delta_{\alpha_t(X)}(f^*) \cdot \delta_{\alpha_{t-i/2}(X^*)}(f)$$
$$+ f^*\left(\delta^\star_{\alpha_t(X)}\delta_{\alpha_t(X)}(f)\right) - \delta_{\alpha_{t+i/2}(X^*)}(f^*) \cdot \delta_{\alpha_t(X)}(f)$$

Similarly for the term with $X^*$ replacing $X$, we get

$$\delta^\star_{\alpha_t(X^*)}\delta_{\alpha_t(X^*)}(f^*f) = \left(\delta^\star_{\alpha_t(X^*)}\delta_{\alpha_t(X^*)}(f^*)\right)f - \delta_{\alpha_t(X^*)}(f^*) \cdot \delta_{\alpha_{t-i/2}(X)}(f)$$
$$+ f^*\left(\delta^\star_{\alpha_t(X^*)}\delta_{\alpha_t(X^*)}(f)\right) - \delta_{\alpha_{t+i/2}(X)}(f^*) \cdot \delta_{\alpha_t(X^*)}(f)$$

Adding together both formulas we have

$$\begin{aligned}\mathfrak{L}(f^*f) &= \int \left(\delta^\star_{\alpha_t(X)}\delta_{\alpha_t(X)}(f^*f) + \delta^\star_{\alpha_t(X^*)}\delta_{\alpha_t(X^*)}(f^*f)\right)\eta(t)dt \\ &= f^*\mathfrak{L}(f) + \mathfrak{L}(f^*)f \\ &+ \int \left(\delta_{\alpha_t(X)}(f^*) \cdot \delta_{\alpha_{t-i/2}(X^*)}(f)\right)\eta(t)dt \\ &+ \int \left(\delta_{\alpha_{t+i/2}(X^*)}(f^*) \cdot \delta_{\alpha_t(X)}(f)\right)\eta(t)dt \\ &+ \int \left(\delta_{\alpha_t(X^*)}(f^*) \cdot \delta_{\alpha_{t-i/2}(X)}(f)\right)\eta(t)dt \\ &+ \int \left(\delta_{\alpha_{t+i/2}(X)}(f^*) \cdot \delta_{\alpha_t(X^*)}(f)\right)\eta(t)dt\end{aligned} \quad (\text{G.1})$$

Shifting the integration variable in first and third term by $-i/4$ and in the second and fourth by $-i/4$,

we get

$$\mathfrak{L}(f^*f) = \int \left( \delta^\star_{\alpha_t(X)} \delta_{\alpha_t(X)}(f^*f) + \delta^\star_{\alpha_t(X^*)} \delta_{\alpha_t(X^*)}(f^*f) \right) \eta(t) dt$$

$$= f^*\mathfrak{L}(f) + \mathfrak{L}(f^*)f$$

$$+ \int \left( \delta_{\alpha_{t+i/4}(X)}(f^*) \cdot \delta_{\alpha_{t-i/4}(X^*)}(f) \right) \eta(t + i/4) dt$$

$$+ \int \left( \delta_{\alpha_{t+i/4}(X^*)}(f^*) \cdot \delta_{\alpha_{t-i/4}(X)}(f) \right) \eta(t - i/4) dt$$

$$+ \int \left( \delta_{\alpha_{t+i/4}(X^*)}(f^*) \cdot \delta_{\alpha_{t-i/4}(X)}(f) \right) \eta(t + i/4) dt$$

$$+ \int \left( \delta_{\alpha_{t+i/4}(X)}(f^*) \cdot \delta_{\alpha_{t-i/4}(X^*)}(f) \right) \eta(t - i/4) dt$$

Now the terms in the brackets can be written as squares of operators. After adding corresponding terms with the same weight, we obtain

$$\mathfrak{L}(f^*f) = \int \left( \delta^\star_{\alpha_t(X)} \delta_{\alpha_t(X)}(f^*f) + \delta^\star_{\alpha_t(X^*)} \delta_{\alpha_t(X^*)}(f^*f) \right) \eta(t) dt$$

$$= f^*\mathfrak{L}(f) + \mathfrak{L}(f^*)f$$

$$+ \int \left( \left|\delta_{\alpha_{t-i/4}(X^*)}(f)\right|^2 + \left|\delta_{\alpha_{t-i/4}(X)}(f)\right|^2 \right) \eta(t + i/4) dt$$

$$+ \int \left( \left|\delta_{\alpha_{t-i/4}(X)}(f)\right|^2 + \left|\delta_{\alpha_{t-i/4}(X^*)}(f)\right|^2 \right) \eta(t - i/4) dt$$

Since the brackets are the same in both integral expressions, finally this can be rearranged as follows

$$\mathfrak{L}(f^*f) = \int \left( \delta^\star_{\alpha_t(X)} \delta_{\alpha_t(X)}(f^*f) + \delta^\star_{\alpha_t(X^*)} \delta_{\alpha_t(X^*)}(f^*f) \right) \eta(t) dt$$

$$= f^*\mathfrak{L}(f) + \mathfrak{L}(f^*)f$$

$$+ \int \left( \left|\delta_{\alpha_{t-i/4}(X^*)}(f)\right|^2 + \left|\delta_{\alpha_{t-i/4}(X)}(f)\right|^2 \right) (\eta(t + i/4) + \eta(t - i/4)) \, dt$$

Hence we conclude

$$\begin{aligned} 2\Gamma_1(f) &= \mathfrak{L}(f^*f) - f^*\mathfrak{L}(f) - \mathfrak{L}(f^*)f \\ &= \int \left( \left|\delta_{\alpha_{t-i/4}(X^*)}(f)\right|^2 + \left|\delta_{\alpha_{t-i/4}(X)}(f)\right|^2 \right) (\eta(t + i/4) + \eta(t - i/4)) \, dt \end{aligned} \tag{G.2}$$

□

If $\alpha_{\pm i/4}(X) = e^{\pm \frac{1}{2}\xi}X$, then the formula (G.2) yields

$$\mathfrak{L}(f^*f) = \int \left( \delta^\star_{\alpha_t(X)} \delta_{\alpha_t(X)}(f^*f) + \delta^\star_{\alpha_t(X^*)} \delta_{\alpha_t(X^*)}(f^*f) \right) \eta(t) dt$$

$$= f^* \mathfrak{L}(f) + \mathfrak{L}(f^*) f \qquad (A.2)$$

$$+ 2 \int \left( e^\xi |\delta_{\alpha_t(X^*)}(f)|^2 + e^{-\xi} |\delta_{\alpha_t(X)}(f)|^2 \right) \eta(t) dt$$

Thus we get in this case

$$\frac{1}{C} \Gamma_1(f) = e^\xi |\delta_{X^*}(f)|^2 + e^{-\xi} |\delta_X(f)|^2 \qquad (G.3)$$

with a constant $C \equiv \int (\eta(t + i/4) + \eta(t - i/4)) \, dt$.

For example this will be the case of modular dynamics associated to $\omega_0$ and derivations associated with $A_{I,J}$, where

$$A(I, J) \equiv \prod_{i \in I} A_i \prod_{j \in J} A_j^*. \qquad (3.9)$$

which are operators defined for algebra of invariant derivations. These are analysed in Section 3.9.

There exists another way to define Dirichlet form (with modular automorphism $\alpha_\omega$ corresponding to the state $\omega$), see e.g.[CM17], as follows

$$\tilde{\mathcal{E}}_\Lambda(f) \equiv \sum_{j \in \Lambda} \left( \nu_j \langle \delta_{E_j}(f), \delta_{E_j}(f) \rangle_\omega + \mu_j \langle \delta_{E_j^*}(f), \delta_{E_j^*}(f) \rangle_\omega \right) \qquad (\tilde{\mathcal{E}})$$

where $E_j$ are the eigenvectors of the modular operator $\alpha_\omega(\pm \frac{i}{2})$, associated to the state $\omega$ at time $\pm \frac{i}{2}$, such that

$$\alpha_\omega(\pm \tfrac{i}{2})(E_j) = e^{\pm \xi_j} E_j, \qquad (3.10)$$

for some $\xi_j \in \mathbb{R}$, and $\nu_j, \mu_j \in (0, \infty)$.

Now, the Markov generator corresponding to the second Dirichlet form is formally given by

$$\langle f, -\tilde{\mathfrak{L}}_\Lambda f \rangle = \tilde{\mathcal{E}}_\Lambda(f).$$

The tilded generator is as follows

$$-\tilde{\mathfrak{L}}(f^*f) = \sum_{j\in\mathbb{Z}^d}\left(\kappa\delta^\star_{X_j}\delta_{X_j}(f^*f) + \mu\delta^\star_{X_j^*}\delta_{X_j^*}(f^*f)\right)$$

and respectively

$$\tilde{\Gamma}_1(f) = \sum_{j\in\mathbb{Z}^d}\left(\kappa e^{-\xi}|\delta_{X_j}(f)|^2 + \mu e^{\xi}|\delta_{X_j^*}(f)|^2\right)$$

In the context of infinite dimensional spaces, ensuring the density of the domain requires having finite speed of propagation of information. This is because even when dealing with localized operators, the eigenvectors are typically not localized.

We now give some properties of the $\Gamma_1$ function.

The modified Leibnitz rule for the adjoint of derivations 3.3 in $\mathbb{L}_2$

$$\delta^\star_Y(fg) = \delta^\star_Y(f)g - f\delta_{\alpha_{-i/2}(Y^*)}(g),$$

we have the following calculations

$$\delta^\star_{X_t}\delta_{X_t}(f^*f) = \delta^\star_{X_t}\left(\delta_{X_t}(f^*)f + f^*\delta_{X_t}(f)\right)$$
$$= \left(\delta^\star_{X_t}\delta_{X_t}(f^*)\right)f - \delta_{X_t}(f^*)\cdot\delta_{\alpha_{-i/2}(X_t^*)}(f) + f^*\left(\delta^\star_{X_t}\delta_{X_t}(f)\right) - \delta_{\alpha_{i/2}(X_t^*)}(f^*)\cdot\delta_{\alpha_t(X)}(f)$$

which implies

$$\mathfrak{L}(f^*f) = \mathfrak{L}(f^*)f - \int\left(\delta_{X_t}(f^*)\cdot\delta_{\alpha_{-i/2}(X_t^*)}(f)\right)\eta_t dt + f^*\mathfrak{L}(f) - \int\left(\delta_{\alpha_{i/2}(X_t^*)}(f^*)\cdot\delta_{\alpha_t(X)}(f)\right)\eta_t dtf.$$

The expression provided assumes that the operator $f$ satisfies suitable conditions for the operations on the right-hand side to make sense. These conditions ensure that the actions involving $f$ and its adjoint $f^*$, as well as the composition of operators such as $P_s$ and $P_{t-s}$, are well-defined within the mathematical framework being considered. In particular $f^*f \in \mathcal{D}(\mathfrak{L})$, for bounded operators $f \in \mathcal{D}(\mathfrak{L})$, if $\exists C \in (0,\infty)$ such that we have

$$\|\int\left(\delta_{\alpha_{i/2}(X_t^*)}(f^*) + \delta_{\alpha_{i/2}(X_t)}(f^*)\right)\eta_t dt\|^2 \leq 2\int\left(\|\delta_{\alpha_{\pm i/2}(X_t^*)}(f^*)\|_2^2 + \|\delta_{\pm\alpha_{i/2}(X_t)}(f^*)\|_2^2\right)\eta_t dt\| \leq C\mathcal{E}(f)$$

The last condition can be satisfied if $\eta$ is analytic in a strip $[-i/2, +i/2]$ and $|\eta(t \pm i/2)| \leq C\,\eta(t)$. In the previously considered models, the validity domain of the aforementioned relation may include local polynomials of elements within $\mathcal{D}_\varepsilon$. For a positivity preserving contraction semigroup in $\mathbb{L}_2$, for any $t > 0$ and $f \in \mathbb{L}_\infty$, we have $P_t f \in \mathcal{D}(\mathfrak{L}) \cap \mathbb{L}_\infty$. Thus for bounded operators, we can identify $\Gamma_1$ as follows: For $0 < s < t$

$$\frac{d}{ds}\left(P_s|P_{t-s}f|^2\right) = 2P_s\Gamma_1(P_{t-s}f) \geq 0.$$

Using this, we have

$$P_t f^2 - |P_t f|^2 = \int_0^t \frac{d}{ds}\left(P_s(P_{t-s}f)^2\right)ds = \int_0^t ds\, 2P_s\Gamma_1(P_{t-s}f) \geq 0.$$

Hence, under the above conditions, for Markov semigroup in $\mathbb{L}_2(\omega)$ the following strong positivity conditions holds.

**Proposition 3.3.** *For the diffusion semigroup $P_t \equiv e^{t\mathfrak{L}}$ in $\mathbb{L}_2(\omega)$, for any $t > 0$ and $f \in \mathbb{L}_2(\omega) \cap \mathbb{L}_\infty(\omega)$, the following Schwartz inequality is true.*

$$|P_t f|^2 \leq P_t |f|^2.$$

**Remark 3.2.** *The Schwartz inequality is proven for completely positive semigroups/maps and are not always true for all positive maps [Bha15].*

Next for a suitable function $\gamma(t) \geq 0$, we have the following property.

**Proposition 3.4.** *Suppose $\Gamma_1(\alpha_{-i/4}(f)) \in \mathbb{L}_1(\omega)$, then there exist an admissible function $\gamma(t) \geq 0$ such that*

$$\omega\,\Gamma_1(\alpha_{-i/4}(f)) = \int \sum_\sharp \nu_\sharp \langle \delta_{\alpha_t(X^\sharp)}(f), \delta_{\alpha_t(X^\sharp)}(f) \rangle \gamma(t)dt \equiv \mathcal{E}_\gamma(f).$$

*where $\nu_\sharp$ are some positive constants.*

*Proof.* Using the formula for $\Gamma_1$, in general with some positive constants $\nu_\sharp$, we have

$$\Gamma_1(\alpha_{-i/4}(f)) = \int \sum_\sharp \nu_\sharp \left(|\alpha_{-i/4}(\delta_{\alpha_t(X^\sharp)}(f)|^2\right) \gamma(t)dt$$

$$= \int \sum_\sharp \nu_\sharp \left(\rho^{-1/4}\left(\delta_{\alpha_t(X^\sharp)}(f)\right)^* \rho^{1/2} \left(\delta_{\alpha_t(X^\sharp)}(f)\right)\rho^{-1/4}\right) \gamma(t)dt$$

Taking the expectation with respect to $\omega(f) = Tr(\rho f)$, we obtain the required result.  □

In case when $X^\sharp$ are eigenvectors of modular operator to the power $\pm 1/4$, with some positive constants $\nu, \mu$, we have

$$\Gamma_1(\alpha_{-i/4}(f)) = \nu e^{-\xi/2}|\delta_X(\alpha_{-i/4}(f))|^2 + \mu e^{\xi/2}|\delta_{X^*}(\alpha_{-i/4}(f))|^2 = \nu|\alpha_{-i/4}(\delta_X(f))|^2 + \mu|\alpha_{-i/4}(\delta_{X^*}(f))|^2$$

and hence

$$\omega\tilde{\Gamma}_1(\alpha_{-i/4}(f)) = \nu\langle\delta_X(f), \delta_X(f)\rangle + \mu\langle\delta_{X^*}(f), \delta_{X^*}(f)\rangle \sim \mathcal{E}(f) \tag{3.11}$$

We have the following property.

**Proposition 3.5.** *For $f_t \equiv P_t f \equiv e^{t\mathcal{Q}} f$, if*

$$\Gamma_1(f_t) \leq e^{-2mt} Const \, \Gamma_1(f), \tag{3.12}$$

*then we have*

$$\mathcal{E}(f_t) \leq e^{-2mt} Const \, \mathcal{E}(f)$$

*and if Poincaré inequality, that is, for a constant $k$,*

$$k\|f - \omega(f)\|_2^2 \leq \mathcal{E}(f)$$

*holds, we get*

$$\|f_t - \omega(f)\|_2^2 \leq e^{-2mt}\|f - \omega(f)\|_2^2. \tag{3.13}$$

*Proof.* Given

$$\Gamma_1(f_t) \le e^{-2mt} Const\, \Gamma_1(f).$$

Using Proposition 3.4, consider $f$ such that $\alpha_s(\alpha_{-i/4})(f)$ is in domain of $\Gamma_1$. Taking the expectation with respect to $\omega$ both sides, we get

$$\omega\Gamma_1(\alpha_s(\alpha_{-i/4})(f_t)) \le e^{-2mt} Const\, \omega\Gamma_1(\alpha_s(\alpha_{-i/4})(f)),$$

Now multiplying $\gamma(t)$ and integrating both the sides, we obtain

$$\int \omega\Gamma_1(\alpha_s(\alpha_{-i/4})(f_t))\gamma(s)ds \le e^{-2mt} Const \int \omega\Gamma_1(\alpha_s(\alpha_{-i/4})(f))\gamma(s)ds,$$

and hence,

$$\mathcal{E}(f_t) \le e^{-2mt} Const\, \mathcal{E}(f).$$

Next if Poincaré inequality holds, that is,

$$k\|f - \omega(f)\|_2^2 \le \mathcal{E}(f)$$

and

$$k\|f_t - \omega(f_t)\|_2^2 \le \mathcal{E}(f_t)$$
$$\le e^{-2mt} Const\, \mathcal{E}(f)$$

Hence, using spectral theory, we obtain

$$\|f_t - \omega(f)\|_2^2 \le e^{-2mt}\|f - \omega(f)\|_2^2.$$

$\square$

**Remark 3.3.** *For obtaining the inequality* (3.13), *one can alternatively use the following standard*

*procedure using Poincaré inequality. Consider*

$$\frac{d}{ds}\|f_s - \omega(f)\|_2^2 = -2\mathcal{E}(f_s) \leq -2k\|f_s - \omega(f)\|_2^2$$

*which is a differential inequality and can be solved to obtain the required inequality.*

In classical diffusions, the inequalities of the form (3.12) are known as the Bakry-Émery condition [Bak04], which implies numerous coercive inequalities. In the quantum setup, similar inequalities were discussed in [CM17] for the eigenvectors of modular operator (3.10) and quantum harmonic oscillator.

## 3.7 Models and their Dirichlet Forms

In this section, we construct specific examples of models and discuss their properties.

### 3.7.1 Mean Field Models

**Example 3.1.** *Assume we have commuting copies of CCR in an infinite dimensional Hilbert space. Consider $[X_j, X_k^*] = \delta_{j,k}$. Define*

$$\mathbf{X}_\Lambda \equiv \frac{1}{\sqrt{|\Lambda|}} \sum_{k \in \Lambda} X_k$$

*We notice that*

$$[\mathbf{X}_\Lambda, \mathbf{X}_\Lambda^*] = \left(\frac{1}{\sqrt{|\Lambda|}}\right)^2 \sum_{k \in \Lambda}[X_k, \sum_{j \in \Lambda} X_j^*] = \left(\frac{1}{\sqrt{|\Lambda|}}\right)^2 \sum_{k \in \Lambda} id = \left(\frac{1}{\sqrt{|\Lambda|}}\right)^2 |\Lambda| = id.$$

*Consider the quadratic Hamiltonian given as follows*

$$U_\Lambda = \mathbf{X}_\Lambda^* \mathbf{X}_\Lambda = \frac{1}{|\Lambda|} \sum_{j,k \in \Lambda} X_j^* X_k$$

*Then using computations similar to those in Section 3.4, the corresponding modular dynamics of $\mathbf{X}_\Lambda$ and $\mathbf{X}_\Lambda^*$ associated to $U_\Lambda$ is given by*

$$\alpha_{t,\Lambda}(\mathbf{X}_\Lambda) = e^{i\beta t}\mathbf{X}_\Lambda \quad \text{and} \quad \alpha_{t,\Lambda}(\mathbf{X}_\Lambda^*) = e^{-i\beta t}\mathbf{X}_\Lambda^*$$

*We can then write the Dirichlet form given by the following result.*

**Proposition 3.6.** *The Dirichlet form defined with the derivations in the directions of $\mathbf{X}_\Lambda$ and $\mathbf{X}_\Lambda^*$ with respect to $U_\Lambda$ is given by*

$$\mathcal{E}_\Lambda(f) = \hat{\eta}(0)\left(\langle \delta_{\mathbf{X}_\Lambda}(f), \delta_{\mathbf{X}_\Lambda}(f)\rangle_\omega + \langle \delta_{\mathbf{X}_\Lambda^*}(f), \delta_{\mathbf{X}_\Lambda^*}(f)\rangle_\omega\right). \tag{3.14}$$

*where $\hat{\eta}(0) = \int_\mathbb{R} \eta(t)e^{ist}dt$. This form has the dense domain $D(\mathcal{E}_\Lambda) \supset \mathcal{D}_\Lambda$.*

*Proof: We compute the derivations*

$$\delta_{\alpha_t(\mathbf{X}_\Lambda)}(f) = i[\alpha_t(\mathbf{X}_\Lambda), f] = ie^{i\beta t}[\mathbf{X}_\Lambda, f]$$

$$(\delta_{\alpha_t(\mathbf{X}_\Lambda)}(f))^* = -i[f^*, \alpha_t(\mathbf{X}_\Lambda)^*] = -e^{-i\beta t}i[f^*, (\mathbf{X}_\Lambda)^*]$$

*Substituting these relations in the formula for the Dirichlet form gives us the result.* □

*The associated Markovian generator is given by*

$$-\mathfrak{L}_\Lambda(f) = \hat{\eta}(0)\left(-e^{-\frac{\beta}{2}}[\mathbf{X}_\Lambda, f]\mathbf{X}_\Lambda^* + e^{\frac{\beta}{2}}\mathbf{X}_\Lambda^*[\mathbf{X}_\Lambda, f] - e^{\frac{\beta}{2}}[\mathbf{X}_\Lambda^*, f]\mathbf{X}_\Lambda + e^{-\frac{\beta}{2}}\mathbf{X}_\Lambda[\mathbf{X}_\Lambda^*, f]\right) \tag{3.15}$$

**Theorem 3.2.** *The Markov operator (3.15) for the above model is the generator of quantum OU semigroup which maps symmetric polynomials in creation/annihilation operators into itself.*

*Hence, the Poincaré inequality holds as established in [CFL00]. Similarly, the Logarithmic Sobolev inequality is satisfied, as shown in [CS07] and in the sense of [OZ99].*

*Additionally, the limiting theory can be described within the framework of [RW98].*

**Example 3.2.** *(Mean Field Model 2) Again, consider $[X_j, X_k^*] = \delta_{j,k}$. Define*

$$\mathbf{X}_{n,\Lambda} \equiv \frac{1}{|\Lambda|^\varepsilon} \sum_{k \in \Lambda} X_k^n$$

*where $n$ is an integer such that $n > 1$ and $\varepsilon \in [0, 1]$.*

*Then we have*

$$\left[\mathbf{X}_{n,\Lambda}, \mathbf{X}_{n,\Lambda}^*\right] = \frac{1}{|\Lambda|^{2\varepsilon}} \sum_{k \in \Lambda} [X_k^n, \sum_{j \in \Lambda} X_j^{*n}] = \frac{1}{|\Lambda|^{2\varepsilon}} \sum_{k \in \Lambda} [X_k^n, X_k^{*n}] = \frac{1}{|\Lambda|^{2\varepsilon}} \sum_{k \in \Lambda} P_n(N_k);$$

*with some polynomial $P_n$.*

*Consider the following quadratic Hamiltonian*

$$U_\Lambda = \mathbf{X}_\Lambda^* \mathbf{X}_\Lambda = \frac{1}{|\Lambda|} \sum_{j,k \in \Lambda} X_j^* X_k.$$

*Now we want to obtain the modular dynamics of $\mathbf{X}_{n,\Lambda}$ and $\mathbf{X}_{n,\Lambda}^*$.*

*We know that*

$$\alpha_{t,\Lambda}(\mathbf{X}_{n,\Lambda}^\sharp) = e^{-i\beta t U_\Lambda} \mathbf{X}_{n,\Lambda}^\sharp e^{i\beta t U_\Lambda} = e^{-i\beta t U_\Lambda} \left(\frac{1}{|\Lambda|^\varepsilon} \sum_{k \in \Lambda} (X_k^\sharp)^n\right) e^{i\beta t U_\Lambda}.$$

*Using the automorphism property, we can write*

$$\alpha_{t,\Lambda}(\mathbf{X}_{n,\Lambda}^\sharp) = \frac{1}{|\Lambda|^\varepsilon} \sum_{k \in \Lambda} \left(\alpha_{t,\Lambda}(X_k^\sharp)\right)^n = \frac{1}{|\Lambda|^\varepsilon} \sum_{k \in \Lambda} \left(e^{\upsilon i \beta t} X_k^\sharp\right)^n$$

$$= \frac{1}{|\Lambda|^\varepsilon} \sum_{k \in \Lambda} e^{\upsilon n i \beta t} (X_k^\sharp)^n = e^{\upsilon n i \beta t} \mathbf{X}_{n,\Lambda}^\sharp,$$

*where $\upsilon = +1$ for $\mathbf{X}_{n,\Lambda}$ and $\upsilon = -1$ for $\mathbf{X}_{n,\Lambda}^*$. Hence we obtain the following proposition.*

**Proposition 3.7.** *The modular dynamics of $\mathbf{X}_{n,\Lambda}$ and $\mathbf{X}_{n,\Lambda}^*$ associated to $U_\Lambda$ is given by*

$$\alpha_{t,\Lambda}(\mathbf{X}_{n,\Lambda}) = e^{ni\beta t} \mathbf{X}_{n,\Lambda} \quad \text{and} \quad \alpha_{t,\Lambda}(\mathbf{X}_{n,\Lambda}^*) = e^{-ni\beta t} \mathbf{X}_{n,\Lambda}^*.$$

*Hence we conclude with the following result.*

**Theorem 3.3.** *The Dirichlet form in the directions of $\mathbf{X}_\Lambda$ and $\mathbf{X}_\Lambda^*$ with respect to $U_\Lambda$ is given by*

$$\mathcal{E}_\Lambda(f) = \int_\mathbb{R} \Big( \langle \delta_{\alpha_t(\mathbf{X}_{n,\Lambda})}(f), \delta_{\alpha_t(\mathbf{X}_{n,\Lambda})}(f) \rangle + \langle \delta_{\alpha_t(\mathbf{X}_{n,\Lambda}^*)}(f), \delta_{\alpha_t(\mathbf{X}_{n,\Lambda}^*)}(f) \rangle \Big) \eta(t) dt. \tag{3.16}$$

*is well defined on the dense domain $D(\mathcal{E}_\Lambda) \supset \mathcal{D}_\Lambda$ and closable, and hence defines a Markov generator.*

**Remark 3.4.** *It is an interesting open question whether or not the limits $t \to \infty$ and $\Lambda \to \mathbb{Z}^d$ are interchangeable.*

### 3.7.2 Non-diagonal Dirichlet forms

The next few examples we provide here discusses the dissipative dynamics defined by a Dirichlet form with nonlocal derivations, that is, the Dirichlet form is not just influenced by local interactions but also between the ones that are not adjacent.

**Example 3.3** (Z-type fields). *Let $\kappa \equiv \{\kappa_j \in \mathbb{C} : \sum_j |\kappa_j| < \infty\}$. Define*

$$Z_\kappa = \sum_j \kappa_j A_j \tag{3.17}$$

*such that the series is convergent in any $\mathbb{L}_{p,\omega_0}$ for $p \in [1, \infty)$. Consider two absolutely convergent sequences $\kappa$ and $\xi$, we obtain the following CCR relation*

$$[Z_\kappa, Z_\xi^*] = \sum_j \kappa_j \bar{\xi}_j. \tag{3.18}$$

*The right hand side of (3.18) is convergent because by assumption, the sequences $\kappa$ and $\xi$ are square summable.*
*The modular dynamics (associated to $\omega_0$) of $Z_\kappa$ is given*

$$\alpha_t(Z_\kappa) = e^{i\beta t} Z_\kappa \text{ and } \alpha_t(Z_\kappa^*) = e^{-i\beta t} Z_\kappa^*$$

*which can be shown using the linearity of the modular dynamics.*
*Consider a translation $(T_j \kappa) \equiv (\kappa_{l-j})_{l \in \mathbb{Z}^d}$ defined by shifting each l. We can then write the translation* of $Z_\kappa$ by $Z_{T_j\kappa}$.

*From now, we assume that $\kappa$ is not a zero vector $\boldsymbol{\theta}$.*

**Theorem 3.4.** *Suppose $\kappa, \xi \neq \boldsymbol{\theta}$. The Dirichlet form associated to the directions of $Z_{T_j\kappa}$ and $Z^*_{T_j\xi}$, $j \in \mathbb{Z}^d$, with respect to the state $\omega_0$ is given by*

$$\mathcal{E}(f) = \hat{\eta}(0) \sum_{j \in \mathbb{Z}^d} \left( \langle \delta_{Z_{T_j\kappa}}(f), \delta_{Z_{T_j\kappa}}(f) \rangle_\omega + \langle \delta_{Z^*_{T_j\xi}}(f), \delta_{Z^*_{T_j\xi}}(f) \rangle_\omega \right).$$

*with a dense domain $D(\mathcal{E}) \supset \mathcal{D}_\varepsilon$ is closable and hence defines a Markov generator.*

*Proof.* Since the vectors $\kappa$ and $\xi$ are summable, that is $\sum_j |\kappa_j| < \infty$ and $\sum_j |\xi_j| < \infty$, there exists a constant say $C_\mathcal{E}$ such that

$$\hat{\eta}(0) \sum_{j \in \mathbb{Z}^d} \left( \langle \delta_{Z_{T_j\kappa}}(f), \delta_{Z_{T_j\kappa}}(f) \rangle_\omega + \langle \delta_{Z^*_{T_j\xi}}(f), \delta_{Z^*_{T_j\xi}}(f) \rangle_\omega \right) \leq C_\mathcal{E} \hat{\eta}(0) \sum_{j \in \mathbb{Z}^d} \left( \langle \delta_{A_j}(f), \delta_{A_j}(f) \rangle_\omega + \langle \delta_{A^*_j}(f), \delta_{A^*_j}(f) \rangle_\omega \right)$$

We already showed that the domain of the Dirichlet form in the directions of $A_j$ and $A^*_j$ contains a dense set $\mathcal{D}_\varepsilon$ on which the adjoint operators are well defined, see 3.3. Then one can define the pre-Markov generator on this dense domain. We then invoke the Friedrichs extension that extends this densely defined, symmetric and positive operator to a self adjoint operator which ensures that the resulting operator is closed and hence a Markov generator. □

**Remark 3.5.** *Similar conclusion holds for the Dirichlet forms associated to the Gibbs states. Although in the Dirichlet form for Gibbs state, generally the integral with $\eta$ does not factorises, however, as we discussed earlier in Section 3.4, all the arguments go through thanks to finite speed of propagation of information and the fact that $\mathcal{D}_\varepsilon$ is a subset of $\mathbb{L}_{p,\omega}$ for $p \in [1, \infty)$.*

**Example 3.4.** *This example involves infinite set of CCRs which are not independent in general. Given $j \sim k$, $j, k \in \mathbb{Z}^d$ and $\kappa_j, \varepsilon_k \in \mathbb{C}$ define*

$$Z_{j,k} := \kappa_j A_j + \varepsilon_k A_k$$

*and consider the following Hamiltonian*

$$H_\Lambda = \sum_{\substack{j \sim k \\ j,k \in \Lambda}} Z^*_{j,k} Z_{j,k} = \sum_{\substack{j \sim k \\ j,k \in \Lambda}} (\bar{\kappa}_j A^*_j + \bar{\varepsilon}_k A^*_k)(\kappa_j A_j + \varepsilon_k A_k) = \sum_{\substack{j \sim k \\ j,k \in \Lambda}} |\kappa_j|^2 A^*_j A_j + \bar{\kappa}_j \varepsilon_k A^*_j A_k + \bar{\varepsilon}_k \kappa_j A^*_k A_j + |\varepsilon_k|^2 A^*_k A_k.$$

*We note that the following derivation*

$$\delta_{\mathbf{H}}(f) \equiv \lim_{\Lambda \to \mathbb{Z}^d} \delta_{H_\Lambda}(f).$$

*is well defined on all local polynomials in creators and annihilators $A_j^\sharp$, $j \in \mathbb{Z}^d$. In particular, for a fixed $l, m \in \Lambda$, we have*

$$i\delta_{\mathbf{H}}(Z_{l,m}) = [Z_{l,m}, \sum_{j \sim k} Z_{j,k}^* Z_{j,k}] = \sum_{j \sim k} [Z_{l,m}, Z_{j,k}^* Z_{j,k}] = \sum_{j \sim k} [Z_{l,m}, Z_{j,k}^*] Z_{j,k}$$

$$= \sum_{j \sim k} (\delta_{l,j}|\kappa_j|^2 + \delta_{l,k}\kappa_k \bar{\varepsilon}_k + \delta_{m,j}\varepsilon_j \bar{\kappa}_j + \delta_{m,k}|\varepsilon_k|^2) Z_{j,k}$$

*Hence we note that the infinite dimensional hamiltonian evolution $\alpha_t(Z_{l,m}) \equiv \alpha_{t,\mathbf{H}}(Z_{l,m}) \equiv e^{-t\beta\delta_{\mathbf{H}}} Z_{l,m}$ is well defined and satisfies the following relation*

$$\frac{d}{dt}\alpha_t(Z_{l,m}) = \frac{d}{dt} e^{-it\beta H}(Z_{l,m}) e^{it\beta H} = i\beta e^{-it\beta H}(Z_{l,m} H) e^{it\beta H} - i\beta e^{-it\beta H}(H Z_{l,m}) e^{it\beta H} = i\beta e^{-it\beta H}[Z_{l,m}, H] e^{it\beta H}$$

$$= i\beta e^{-it\beta H} (\sum_{j \sim k} (\delta_{l,j}|\kappa_j|^2 + \delta_{l,k}\kappa_k \bar{\varepsilon}_k + \delta_{m,j}\varepsilon_j \bar{\kappa}_j + \delta_{m,k}|\varepsilon_k|^2) Z_{j,k}) e^{it\beta H}$$

$$= i\beta (\sum_{j \sim k} (\delta_{l,j}|\kappa_j|^2 + \delta_{l,k}\kappa_k \bar{\varepsilon}_k + \delta_{m,j}\varepsilon_j \bar{\kappa}_j + \delta_{m,k}|\varepsilon_k|^2) \alpha_t(Z_{j,k})$$

*We now want to check that corresponding hamiltonian dynamics*

$$\alpha_t(f) = e^{-t\delta_{\mathbf{H}}} f = \lim_{\Lambda \to \mathbb{Z}^d} e^{-t\delta_{H_\Lambda}} f = \lim_{\Lambda \to \mathbb{Z}^d} e^{-itH_\Lambda} f e^{itH_\Lambda}$$

*is well defined on local polynomials in creators and annihilators, note that we have*

$$\frac{d}{dt}\alpha_{t,\Lambda}(A_l) = -i\alpha_{t,\Lambda}([H_\Lambda, A_l])$$

$$= -i \sum_{\substack{m \in \Lambda \\ m \sim l}} \left((|\kappa_l|^2 + |\varepsilon_l|^2)\alpha_{t,\Lambda}(A_l) + (\bar{\kappa}_l \varepsilon_m + \bar{\varepsilon}_l \kappa_m)\alpha_{t,\Lambda}(A_m)\right).$$

*The above equation can be solved in the algebra if mollified by dividing by a power of $(1 + \varepsilon N_\Lambda)^{-1}$*

with $N_\Lambda \equiv \sum_{l \in \Lambda} N_l$. This is because

$$\delta_{H_\Lambda}(N_\Lambda) = i[H_\Lambda, N_\Lambda] = \left[ \sum_{\substack{j \sim k \\ j,k \in \Lambda}} |\kappa_j|^2 A_j^* A_j + \bar{\kappa}_j \varepsilon_k A_j^* A_k + \bar{\varepsilon}_k \kappa_j A_k^* A_j + |\varepsilon_k|^2 A_k^* A_k, \sum_{l \in \Lambda} N_l \right].$$

And since

$$\left[ \sum_{k,j} A_j^* A_k, \sum_{l \in \Lambda} N_l \right] = \sum_j A_j^* \left[ \sum_k A_k, \sum_{l \in \Lambda} N_l \right] + \left[ \sum_j A_j^*, \sum_{l \in \Lambda} N_l \right] \sum_k A_k = 0,$$

we obtain

$$\delta_{H_\Lambda}(N_\Lambda) = 0.$$

One can then use iteration scheme in operator norm for mollified problem, otherwise one needs to study convergence in $\mathbb{L}_p(\omega_0)$ spaces, $p \in (1, \infty)$. From that should be clear (via arguments given in the previous sections) that the infinite dimensional limit can be performed and that we have finite speed of propagation of interaction in the system.

**Remark 3.6.** *We have a generalisation of the last relation directly to the infinite dimensions. It can be written in the sense of commutation of derivations as follows*

$$[\delta_\mathbf{H}, \delta_\mathbf{N}] = 0$$

*where*

$$\delta_\mathbf{N}(f) \equiv \lim_{\Lambda \to \mathbb{Z}^d} \delta_{\mathbf{N}_\Lambda}(f)$$

*with the limit on the right hand side on local polynomials in creators and annihilators (weak or strong in a Hilbert space) or in operator norm on the algebra generated by the mollified local polynomials (with local mollification provided by $N_O$ with $O \subset\subset \mathbb{Z}^d$).*

With the well defined hamiltonian dynamics for which finite speed of propagation of information property holds we have the following result.

**Theorem 3.5.** *The following Dirichlet form is well defined on the set of local polynomials*

$$\mathcal{E}(f) \equiv \sum_{\substack{j \sim k \\ j,k \in \mathbb{Z}^d}} \int \left( \langle \delta_{\alpha_{t,\mathbf{H}}(Z_{jk})}(f), \delta_{\alpha_{t,\mathbf{H}}(Z_{jk})}(f) \rangle_{\omega_0} + \langle \delta_{\alpha_{t,\mathbf{H}}(Z^*_{jk})}(f), \delta_{\alpha_{t,\mathbf{H}}(Z^*_{jk})}(f) \rangle_{\omega_0} \right) \eta(t) dt$$

*and its closure defines a Markov generator in $\mathbb{L}_2(\omega_0)$.*

In the following example, we use the mixed representation of CCRs.

**Example 3.5** (Y-type fields). *For some $\kappa, \xi \neq \theta$,*

$$Y_{\kappa,\xi} = Z_\kappa - Z^*_\xi.$$

*Then*

$$[Y_{\kappa,\xi}, Y^*_{\kappa,\xi}] = \left( |\kappa|^2_2 - |\xi|^2_2 \right) \mathbf{id}$$

*where $|\kappa|^2_2 \equiv \sum_j |\kappa_j|^2$ and similarly for $|\xi|^2_2$. We also note that*

$$[Y_{\kappa,\xi}, N_l] = \kappa_l A_l - \bar{\xi}_l A^*_l$$

*Using the above and 3.3, we have the following modular dynamics.*

**Lemma 3.1.** *The modular dynamics of $Y_{\kappa,\xi}$ and $Y^*_{\kappa,\xi}$ with respect to the infinite product state is given*

$$\alpha_t(Y_{\kappa,\xi}) = e^{i\beta t} Z_\kappa - e^{-i\beta t} Z^*_\xi$$

*Consider the following corresponding Dirichlet form*

$$\mathcal{E}(f) = \sum_{j \in \mathbb{Z}^d} \int \langle \delta_{\alpha_t(Y_{T_j\kappa, T_j\xi})}(f), \delta_{\alpha_t(Y_{T_j\kappa, T_j\xi})}(f) \rangle \eta(t) dt$$

*We have the following representation of this Dirichlet form.*

**Theorem 3.6.** *The Dirichlet form in the directions of $Y_{T_j\kappa, T_j\xi}$ and $Y^*_{T_j\kappa, T_j\xi}$, $j \in \mathbb{Z}^d$, associated to the* *product state $\omega_0$ is well defined on a dense set including $\mathcal{D}_\varepsilon$ and is given by*

$$\mathcal{E}(f) = \sum_{j\in\mathbb{Z}^d} \hat{\eta}(0)\langle\delta_{Z_{T_j\kappa}}(f), \delta_{Z_{T_j\kappa}}(f)\rangle_{\omega_0} + \sum_{j\in\mathbb{Z}^d} \hat{\eta}(0)\langle\delta_{Z_{T_j\xi}}(f), \delta_{Z_{T_j\xi}}(f)\rangle_{\omega_0}$$

$$+ \sum_{j\in\mathbb{Z}^d} \sqrt{2\pi}\hat{\eta}(-2\beta)\langle\delta_{Z_{T_j\kappa}}(f), \delta_{Z^*_{T_j\xi}}(f)\rangle_{\omega_0} + \sum_{j\in\mathbb{Z}^d} \sqrt{2\pi}\hat{\eta}(2\beta)\langle\delta_{Z^*_{T_j\xi}}(f), \delta_{Z_{T_j\kappa}}(f)\rangle_{\omega_0}$$

*On the domain $\mathcal{D}_\varepsilon$ it defines a pre-Markov generator.*

*Proof.* We compute the derivations in the direction of $\alpha_t(Y_{\kappa,\xi})$

$$\delta_{\alpha_t(Y_{\kappa,\xi})}(f) = i[\alpha_t(Y_{\kappa,\xi}), f] = ie^{i\beta t}[Z_\kappa, f] + ie^{-i\beta t}[Z^*_\xi, f]$$

$$(\delta_{\alpha_t(Y_{\kappa,\xi})}(f))^* = -i[f^*, \alpha_t(Y_{\kappa,\xi})^*] = -ie^{-i\beta t}[f^*, Z^*_\kappa] - ie^{i\beta t}[f^*, Z_\xi]$$

We can now obtain

$$\langle\delta_{\alpha_t(Y_{\kappa,\xi})}(f), \delta_{\alpha_t(Y_{\kappa,\xi})}(f)\rangle_{\omega_o} = \frac{1}{Z}Tr(e^{-\beta\sum_{j\in\Lambda} N_j/2}(\delta_{\alpha_t(Y_{\kappa,\xi})}(f))^* e^{-\beta\sum_{j\in\Lambda} N_j/2}\delta_{\alpha_t(Y_{\kappa,\xi})}(f))$$

$$= \frac{1}{Z}Tr(e^{-\beta\sum_{j\in\Lambda} N_j/2}\left(-ie^{-i\beta t}[f^*, Z^*_\kappa] - ie^{i\beta t}[f^*, Z_\xi]\right) e^{-\beta\sum_{j\in\Lambda} N_j/2}\left(ie^{i\beta t}[Z_\kappa, f] + ie^{-i\beta t}[Z^*_\xi, f]\right))$$

$$= \frac{1}{Z}Tr(e^{-\beta\sum_{j\in\Lambda} N_j/2}\left(-ie^{-i\beta t}[f^*, Z^*_\kappa]\right) e^{-\beta\sum_{j\in\Lambda} N_j/2}\left(ie^{i\beta t}[Z_\kappa, f]\right))$$

$$+ \frac{1}{Z}Tr(e^{-\beta\sum_{j\in\Lambda} N_j/2}\left(-ie^{i\beta t}[f^*, Z_\xi]\right) e^{-\beta\sum_{j\in\Lambda} N_j/2}\left(ie^{i\beta t}[Z_\kappa, f]\right))$$

$$+ \frac{1}{Z}Tr(e^{-\beta\sum_{j\in\Lambda} N_j/2}\left(-ie^{-i\beta t}[f^*, Z^*_\kappa]\right) e^{-\beta\sum_{j\in\Lambda} N_j/2}\left(ie^{-i\beta t}[Z^*_\xi, f]\right))$$

$$+ \frac{1}{Z}Tr(e^{-\beta\sum_{j\in\Lambda} N_j/2}\left(-ie^{i\beta t}[f^*, Z_\xi]\right) e^{-\beta\sum_{j\in\Lambda} N_j/2}\left(ie^{-i\beta t}[Z^*_\xi, f]\right))$$

$$= \langle\delta_{\alpha_t(Z_\kappa)}(f), \delta_{\alpha_t(Z_\kappa)}(f)\rangle_\omega + e^{2i\beta t}\langle\delta_{\alpha_t(Z^*_\xi)}(f), \delta_{\alpha_t(A_j)}(f)\rangle_\omega$$

$$+ e^{-2i\beta t}\langle\delta_{\alpha_t(Z_\kappa)}(f), \delta_{\alpha_t(Z^*_\xi)}(f)\rangle_\omega + \langle\delta_{\alpha_t(Z^*_\xi)}(f), \delta_{\alpha_t(Z^*_\xi)}(f)\rangle_\omega$$

We can now find the Dirichlet form using the translation $(T_j\kappa) = (\kappa_{\mathbf{l}-\mathbf{j}})_{l\in\mathbb{Z}^d}$,

$$\mathcal{E}(f) = \sum_{j\in\mathbb{Z}^d} \int \langle\delta_{\alpha_t(Y_{T_j\kappa, T_j\xi})}(f), \delta_{\alpha_t(Y_{T_j\kappa, T_j\xi})}(f)\rangle \eta(t)dt$$

$$\mathcal{E}(f) = \sum_{j\in\mathbb{Z}^d} \hat{\eta}(0)\langle \delta_{Z_{T_{j^\kappa}}}(f), \delta_{Z_{T_{j^\kappa}}}(f)\rangle_{\omega_0} + \sum_{j\in\mathbb{Z}^d} \hat{\eta}(0)\langle \delta_{Z_{T_{j^\xi}}}(f), \delta_{Z_{T_{j^\xi}}}(f)\rangle_{\omega_0}$$
$$+ \sum_{j\in\mathbb{Z}^d} \sqrt{2\pi}\hat{\eta}(-2\beta)\langle \delta_{Z_{T_{j^\kappa}}}(f), \delta_{Z^*_{T_{j^\xi}}}(f)\rangle_{\omega_0} + \sum_{j\in\mathbb{Z}^d} \sqrt{2\pi}\hat{\eta}(2\beta)\langle \delta_{Z^*_{T_{j^\xi}}}(f), \delta_{Z_{T_{j^\kappa}}}(f)\rangle_{\omega_0}$$

□

**Remark 3.7.** *For the specific example of $\eta(t) = \frac{e^{ibt}}{cosh(8\pi t)}$, $b \in \mathbb{R}$, we have $\hat{\eta}(s) = [8cosh((s+b)/16)]^{-1}$. Then*

$$\hat{\eta}(0) = [8cosh(b/16)]^{-1}, \quad \hat{\eta}(2\beta) = [8cosh((\pm 2\beta + b)/16)]^{-1}.$$

*Playing with 8 in the cosh we can define number of eigenvectors along the idea of [CZ24] and hence have a possibility of introducing more complicated Dirichlet forms of type ($\mathcal{E}'$).*

**Example 3.6.** *Consider the weak monomials of creation and annihilation operators*

$$W_{j,k} \equiv W_{j,k}^{(n,m)} \equiv A_j^{*n} A_k^m.$$

*Then the modular dynamics corresponding to $\mathbf{N}_\Lambda \equiv \sum_{l\in\Lambda} N_l$ with $j, k \in \Lambda$ is as follows*

$$\alpha_t(W_{j,k}) = \lim_{\Lambda\to\mathbb{Z}^d} e^{-i\beta t \mathbf{N}_\Lambda} W_{j,k} e^{i\beta t \mathbf{N}_\Lambda} = \alpha_t(A_j^{*n})\alpha_t(A_k^m) = e^{i(m-n)\beta t} W_{j,k}$$

**Theorem 3.7.** *The Dirichlet form with derivations generated by directions $W_{j,k}^{(n,m)}$ given as follows*

$$\mathcal{E}(f) = \hat{\eta}(0) \sum_{\substack{j,k\in\mathbb{Z}^d \\ j\sim k}} \left( \|\delta_{W_{j,k}}(f)\|_{2,\omega}^2 + \|\delta_{W^*_{j,k}}(f)\|_{2,\omega}^2 \right)$$

*is well defined on the algebra of local polynomials in creators and annihilators. Its closure defines a Markov generator.*

*The proof is similar to previous examples.*

*If we add an additional term in $W_{j,k}$ such that it is a self adjoint operator*

$$W_{j,k} \equiv A_j^{*n} A_k^m + A_k^{*m} A_j^n = W_{j,k}^*.$$

*Then, the modular dynamics we obtain is as follows*

$$\alpha_t(W_{j,k}) = e^{i(m-n)\beta t} A_j^{*n} A_k^m + e^{-i(m-n)\beta t} A_k^{*m} A_j^n$$

*and hence a more complicated Dirichlet form given as follows*

$$\mathcal{E}(f) = \sum_{j,k} 2(\hat{\eta}(0) \langle \delta_{A_j^{*n} A_k^m}(f), \delta_{A_j^{*n} A_k^m}(f) \rangle_\omega + \hat{\eta}(2(n-m)\beta) \langle \delta_{A_j^{*n} A_k^m}(f), \delta_{A_k^{*m} A_j^n}(f) \rangle_\omega$$

$$+ \hat{\eta}(2(m-n)\beta) \langle \delta_{A_k^{*m} A_j^n}(f), \delta_{A_j^{*n} A_k^m}(f) \rangle_\omega + \hat{\eta}(0) \langle \delta_{A_k^{*m} A_j^n}(f), \delta_{A_k^{*m} A_j^n}(f) \rangle_\omega)$$

*The computations for this Dirichlet form is similar to 3.5.*

*Consider the special case when $m = n$. In this case, the time dependence factorises and for each term we get the same multiplier $\hat{\eta}(0)$ similar to previous examples. Moreover in the self adjoint case $W_{j,k} = W_{j,k}^*$ the Dirichlet form is not ergodic, i.e. it vanishes on nonzero elements.*

*In the case $m = 1 = n$, we have*

$$W_{j,k} \equiv A_j^* A_k + A_k^* A_j = W_{j,k}^*$$

*and they are invariant with respect to modular dynamics. Despite the general noncommutativity of such operators with different indices, in this special case we have the following nice commutation relations*

$$\begin{aligned}[W_{j,k}, W_{n,m}^*] &= [A_j^* A_k + A_k^* A_j, A_n^* A_m + A_m^* A_n] \\
&= \delta_{jm} A_n^* A_k + \delta_{kn} A_j^* A_m + \delta_{jn} A_m^* A_k + \delta_{km} A_j^* A_n \\
&\quad + \delta_{km} A_n^* A_j + \delta_{jn} A_k^* A_m + \delta_{kn} A_m^* A_j + \delta_{jm} A_k^* A_n \\
&= \delta_{jm} W_{k,n} + \delta_{kn} W_{j,m} + \delta_{jn} W_{k,m} + \delta_{km} W_{j,n}\end{aligned} \quad (3.19)$$

*To construct a Dirichlet form that is ergodic, one need to add terms with derivations related to*

$$W_{j,k} \equiv i(A_j^* A_k - A_k^* A_j).$$

**Example 3.7.** *Consider*

$$Z_{j,k} \equiv A_j^n - A_k^m$$

*With notation $A_j^{*n} \equiv (A_j^*)^n$, we have*

$$[Z_{j,k}, Z_{j,k}^*] = [A_j^n - A_k^m, A_j^{*n} - A_k^{*m}] = [A_j^n, A_j^{*n}] + [A_k^m, A_k^{*m}]$$

*Using formula (2.13) (or lemma 5.1 from [CZ24]),*

$$[A_j^n, A_j^{*n}] = (N+n)(N+n-1)\cdots(N+2)(N+1) - N(N-1)(N-2)\cdots(N-(n-1))$$

*Hence*

$$[Z_{j,k}, Z_{j,k}^*] = (N+n)(N+n-1)\cdots(N+2)(N+1) - N(N-1)(N-2)\cdots(N-(n-1))$$

$$+(N+m)(N+m-1)\cdots(N+2)(N+1) - N(N-1)(N-2)\cdots(N-(m-1))$$

**Lemma 3.2.** *The modular dynamics in the directions of $Z_{j,k}$ and $Z_{j,k}^*$ with respect to the product state of a system of quantum harmonic oscillators is given by*

$$\alpha_t(Z_{j,k}) = e^{in\beta t} A_j^n - e^{im\beta t} A_k^m \text{ and } \alpha_t(Z_{j,k}^*) = e^{-in\beta t} A_j^{*n} - e^{-im\beta t} A_k^{*m}$$

*Proof.* We know that for $j, k \in \Lambda$ we have

$$[Z_{j,k}, \sum_{j\in\Lambda} N_j] = [A_j^n - A_k^m, \sum_{j\in\Lambda} N_j] = [A_j^n, \sum_{j\in\Lambda} N_j] - [A_k^m, \sum_{j\in\Lambda} N_j]$$

$$= nA_j^n - mA_k^m$$

And

$$[Z_{j,k}^*, \sum_{j\in\Lambda} N_j] = -nA_j^{*n} + mA_k^{*m}$$

Then $\alpha_t(Z_{j,k}) = \alpha_t(A_j^n) - \alpha_t(A_k^m) = e^{in\beta t} A_j^n - e^{im\beta t} A_k^m$ and similarly $\alpha_t(Z_{j,k}^*) = \alpha_t(A_j^{*n}) - \alpha_t(A_k^{*m}) = e^{-in\beta t} A_j^{*n} - e^{-im\beta t} A_k^{*m}$. □

**Theorem 3.8.** *The Dirichlet form in the directions of $Z_{j,k}$ and $Z_{j,k}^*$ with respect to the product state of*

*a system of quantum harmonic oscillators given by*

$$\mathcal{E}_\Lambda(f) = \sum_{j \sim k} \hat{\eta}(0)\langle \delta_{A_j^n}(f), \delta_{A_j^n}(f)\rangle_\omega + \hat{\eta}(0)\langle \delta_{A_k^m}(f), \delta_{A_k^m}(f)\rangle_\omega + \hat{\eta}((m-n)\beta)\langle \delta_{A_j^n}(f), \delta_{A_k^m}(f)\rangle_\omega +$$

$$\hat{\eta}((n-m)\beta)\langle \delta_{A_k^m}(f), \delta_{A_j^n}(f)\rangle_\omega + \hat{\eta}(0)\langle \delta_{A_j^{*n}}(f), \delta_{A_j^{*n}}(f)\rangle_\omega + \hat{\eta}(0)\langle \delta_{A_k^{*m}}(f), \delta_{A_k^{*m}}(f)\rangle_\omega$$

$$- \hat{\eta}((n-m)\beta)\langle \delta_{A_j^{*n}}(f), \delta_{A_k^{*m}}(f)\rangle_\omega - \hat{\eta}((m-n)\beta)\langle \delta_{A_k^{*m}}(f), \delta_{A_j^{*n}}(f)\rangle_\omega$$

*with a dense domain $D(\mathcal{E}_\Lambda) \supset \mathcal{D}$ is closable and hence defines a Markov generator.*

*Proof: We compute the derivations in the direction of $\alpha_t(Z_{j,k})$*

$$\delta_{\alpha_t(Z_{j,k})}(f) = i[\alpha_t(Z_{j,k}), f] = ie^{in\beta t}[A_j^n, f] - ie^{im\beta t}[A_k^m, f]$$

$$(\delta_{\alpha_t(Z_{j,k})}(f))^* = -i[f^*, \alpha_t(Z_{j,k})^*] = -ie^{-in\beta t}[f^*, A_j^{*n}] + ie^{-im\beta t}[f^*, A_k^{*m}]$$

*And hence*

$$\int \langle \delta_{\alpha_t(Z_{j,k})}(f), \delta_{\alpha_t(Z_{j,k})}(f)\rangle_\omega \eta(t) dt = \frac{1}{Z} Tr(e^{-\beta \sum_{j \in \Lambda} N_j/2}(\delta_{\alpha_t(X_j)}(f))^* e^{-\beta \sum_{j \in \Lambda} N_j/2} \delta_{\alpha_t(X_j)}(f))$$

$$= \frac{1}{Z} Tr(-e^{-\beta \sum_{j \in \Lambda} N_j/2}(-ie^{-in\beta t}[f^*, A_j^{*n}] + ie^{-im\beta t}[f^*, A_k^{*m}])e^{-\beta \sum_{j \in \Lambda} N_j/2}(ie^{in\beta t}[A_j^n, f] - ie^{im\beta t}[A_k^m, f]))$$

$$= \hat{\eta}(0)\langle \delta_{A_j^n}(f), \delta_{A_j^n}(f)\rangle_\omega + \hat{\eta}(0)\langle \delta_{A_k^m}(f), \delta_{A_k^m}(f)\rangle_\omega + \hat{\eta}((m-n)\beta)\langle \delta_{A_j^n}(f), \delta_{A_k^m}(f)\rangle_\omega + \hat{\eta}((n-m)\beta)\langle \delta_{A_k^m}(f), \delta_{A_j^n}(f)\rangle_\omega$$

*Similarly*

$$\delta_{\alpha_t(Z_{j,k}^*)}(f) = i[\alpha_t(Z_{j,k}^*), f] = ie^{-in\beta t}[A_j^{*n}, f] - ie^{-im\beta t}[A_k^{*m}, f]$$

$$(\delta_{\alpha_t(Z_{j,k}^*)}(f))^* = -i[f^*, \alpha_t(Z_{j,k}^*)^*] = -ie^{in\beta t}[f^*, A_j^n] + ie^{im\beta t}[f^*, A_k^m]$$

*And hence*

$$\int \langle \delta_{\alpha_t(Z_{j,k}^*)}(f), \delta_{\alpha_t(Z_{j,k}^*)}(f)\rangle_\omega \eta(t) dt$$

$$= \hat{\eta}(0)\langle \delta_{A_j^{*n}}(f), \delta_{A_j^{*n}}(f)\rangle_\omega + \hat{\eta}(0)\langle \delta_{A_k^{*m}}(f), \delta_{A_k^{*m}}(f)\rangle_\omega - \hat{\eta}((n-m)\beta)\langle \delta_{A_j^{*n}}(f), \delta_{A_k^{*m}}(f)\rangle_\omega - \hat{\eta}((m-n)\beta)\langle \delta_{A_k^{*m}}(f), \delta_{A_j^{*n}}(f)\rangle_\omega$$

*Then*

$$\mathcal{E}_\Lambda(f) = \sum_{j \sim k} \int_{-\infty}^{\infty} \left( \langle \delta_{\alpha_t(Z_{j,k})}(f), \delta_{\alpha_t(Z_{j,k})}(f) \rangle_\omega + \langle \delta_{\alpha_t(Z^*_{j,k})}(f), \delta_{\alpha_t(Z^*_{j,k})}(f) \rangle_\omega \right) \eta(t) dt$$

$$= \hat{\eta}(0) \langle \delta_{A^n_j}(f), \delta_{A^n_j}(f) \rangle_\omega + \hat{\eta}(0) \langle \delta_{A^m_k}(f), \delta_{A^m_k}(f) \rangle_\omega + \hat{\eta}((m-n)\beta) \langle \delta_{A^n_j}(f), \delta_{A^m_k}(f) \rangle_\omega + \hat{\eta}((n-m)\beta) \langle \delta_{A^m_k}(f), \delta_{A^n_j}(f) \rangle_\omega +$$

$$\hat{\eta}(0) \langle \delta_{A^{*n}_j}(f), \delta_{A^{*n}_j}(f) \rangle_\omega + \hat{\eta}(0) \langle \delta_{A^{*m}_k}(f), \delta_{A^{*m}_k}(f) \rangle_\omega - \hat{\eta}((n-m)\beta) \langle \delta_{A^{*n}_j}(f), \delta_{A^{*m}_k}(f) \rangle_\omega - \hat{\eta}((m-n)\beta) \langle \delta_{A^{*m}_k}(f), \delta_{A^{*n}_j}(f) \rangle_\omega$$

**Example 3.8.** *Consider*

$$Y_{j,k} = A^n_j - A^{*m}_k$$

*then*

$$[Y_{j,k}, Y^*_{j,k}] = [A^n_j - A^{*m}_k, A^{*n}_j - A^m_k] = [A^n_j, A^{*n}_j] + [A^{*m}_k, A^m_k]$$

$$= n A^{n-1}_j A^{*n-1}_j - m A^{m-1}_k A^{*m-1}_k$$

*and*

$$[Y_{j,k}, \sum_j N_j] = [A^n_j - A^{*m}_k, \sum_j N_j] = n A^n_j + m A^{*m}_k$$

**Lemma 3.3.** *The modular dynamics of $Y_{j,k}$ and $Y^*_{j,k}$ with respect to the infinite product state $\omega_0$ of a system of quantum harmonic oscillators is given*

$$\alpha_t(Y_{j,k}) = e^{in\beta t} A^n_j - e^{-im\beta t} A^{*m}_k \quad \text{and} \quad \alpha_t(Y^*_{j,k}) = e^{-in\beta t} A^{*n}_j - e^{im\beta t} A^m_k$$

*Proof.* Using linearity of modular dynamics

$$\alpha_t(Y_{j,k}) = \alpha_t(A^n_j) - \alpha_t(A^{*m}_k) = e^{in\beta t} A^n_j - e^{-im\beta t} A^{*m}_k$$

Similarly

$$\alpha_t(Y^*_{j,k}) = e^{-in\beta t} A^{*n}_j - e^{im\beta t} A^m_k$$

□

**Theorem 3.9.** *The Dirichlet form in the directions of $Y_{j,k}$ and $Y^*_{j,k}$ with respect to the product state $\omega_0$*

*is given as follows*

$$\mathcal{E}_\Lambda(f) = \sum_{j \sim k} \hat{\eta}(0)\langle \delta_{A_j^n}(f), \delta_{A_j^n}(f)\rangle_\omega + \hat{\eta}(0)\langle \delta_{A_k^{*m}}(f), \delta_{A_k^{*m}}(f)\rangle_\omega - \hat{\eta}((-m-n)\beta)\langle \delta_{A_j^n}(f), \delta_{A_k^{*m}}(f)\rangle_\omega$$

$$- \hat{\eta}((n+m)\beta)\langle \delta_{A_k^{*m}}(f), \delta_{A_j^n}(f)\rangle_\omega + \hat{\eta}(0)\langle \delta_{A_j^{*n}}(f), \delta_{A_j^{*n}}(f)\rangle_\omega + \hat{\eta}(0)\langle \delta_{A_k^m}(f), \delta_{A_k^m}(f)\rangle_\omega$$

$$- \hat{\eta}((n+m)\beta)\langle \delta_{A_j^{*n}}(f), \delta_{A_k^m}(f)\rangle_\omega - \hat{\eta}((-m-n)\beta)\langle \delta_{A_k^m}(f), \delta_{A_j^{*n}}(f)\rangle_\omega$$

*Proof:* We compute the derivations in the direction of $\alpha_t(Y_{j,k})$

$$\delta_{\alpha_t(Y_{j,k})}(f) = i[\alpha_t(Y_{j,k}), f] = ie^{in\beta t}[A_j^n, f] - ie^{-im\beta t}[A_k^{*m}, f]$$

$$(\delta_{\alpha_t(Y_{j,k})}(f))^* = -i[f^*, \alpha_t(Y_{j,k})^*] = -ie^{-in\beta t}[f^*, A_j^{*n}] + ie^{im\beta t}[f^*, A_k^m]$$

*And hence*

$$\int \langle \delta_{\alpha_t(Y_{j,k})}(f), \delta_{\alpha_t(Y_{j,k})}(f)\rangle_\omega \eta(t)dt = \frac{1}{Z}Tr(e^{-\beta \sum_{j\in\Lambda} N_j/2}(\delta_{\alpha_t(Y_{j,k})}(f))^* e^{-\beta \sum_{j\in\Lambda} N_j/2}\delta_{\alpha_t(Y_{j,k})}(f))$$

$$= \frac{1}{Z}Tr(e^{-\beta \sum_{j\in\Lambda} N_j/2}(-ie^{-in\beta t}[f^*, A_j^{*n}] + ie^{im\beta t}[f^*, A_k^m])e^{-\beta \sum_{j\in\Lambda} N_j/2}(ie^{in\beta t}[A_j^n, f] - ie^{-im\beta t}[A_k^{*m}, f]))$$

$$= \hat{\eta}(0)\langle \delta_{A_j^n}(f), \delta_{A_j^n}(f)\rangle_\omega + \hat{\eta}(0)\langle \delta_{A_k^{*m}}(f), \delta_{A_k^{*m}}(f)\rangle_\omega - \hat{\eta}((-m-n)\beta)\langle \delta_{A_j^n}(f), \delta_{A_k^{*m}}(f)\rangle_\omega - \hat{\eta}((n+m)\beta)\langle \delta_{A_k^{*m}}(f), \delta_{A_j^n}(f)\rangle_\omega$$

*Similarly*

$$\delta_{\alpha_t(Y_{j,k}^*)}(f) = i[\alpha_t(Y_{j,k}^*), f] = ie^{-in\beta t}[A_j^{*n}, f] - ie^{im\beta t}[A_k^m, f]$$

$$(\delta_{\alpha_t(Y_{j,k}^*)}(f))^* = -i[f^*, \alpha_t(Y_{j,k}^*)^*] = -ie^{in\beta t}[f^*, A_j^n] + ie^{-im\beta t}[f^*, A_k^{*m}]$$

*And hence*

$$\int \langle \delta_{\alpha_t(Y_{j,k}^*)}(f), \delta_{\alpha_t(Y_{j,k}^*)}(f)\rangle_\omega \eta(t)dt$$

$$= \hat{\eta}(0)\langle \delta_{A_j^{*n}}(f), \delta_{A_j^{*n}}(f)\rangle_\omega + \hat{\eta}(0)\langle \delta_{A_k^m}(f), \delta_{A_k^m}(f)\rangle_\omega - \hat{\eta}((n+m)\beta)\langle \delta_{A_j^{*n}}(f), \delta_{A_k^m}(f)\rangle_\omega - \hat{\eta}((-m-n)\beta)\langle \delta_{A_k^m}(f), \delta_{A_j^{*n}}(f)\rangle_\omega$$

*Then*

$$\mathcal{E}_\Lambda(f) = \sum_{j \sim k} \int_{-\infty}^{\infty} \left(\langle \delta_{\alpha_t(Y_{j,k})}(f), \delta_{\alpha_t(Y_{j,k})}(f)\rangle_\omega + \langle \delta_{\alpha_t(Y_{j,k}^*)}(f), \delta_{\alpha_t(Y_{j,k}^*)}(f)\rangle_\omega\right)\eta(t)dt$$

$$= \sum_{j \sim k} \hat{\eta}(0) \langle \delta_{A_j^n}(f), \delta_{A_j^n}(f) \rangle_\omega + \hat{\eta}(0) \langle \delta_{A_k^{*m}}(f), \delta_{A_k^{*m}}(f) \rangle_\omega - \hat{\eta}((-m-n)\beta) \langle \delta_{A_j^n}(f), \delta_{A_k^{*m}}(f) \rangle_\omega - \hat{\eta}((n+m)\beta) \langle \delta_{A_k^{*m}}(f), \delta_{A_j^n}(f) \rangle_\omega$$

$$+ \hat{\eta}(0) \langle \delta_{A_j^{*n}}(f), \delta_{A_j^{*n}}(f) \rangle_\omega + \hat{\eta}(0) \langle \delta_{A_k^m}(f), \delta_{A_k^m}(f) \rangle_\omega - \hat{\eta}((n+m)\beta) \langle \delta_{A_j^{*n}}(f), \delta_{A_k^m}(f) \rangle_\omega - \hat{\eta}((-m-n)\beta) \langle \delta_{A_k^m}(f), \delta_{A_j^{*n}}(f) \rangle_\omega$$

**Remark 3.8.** *In Examples 3.7 and 3.8, the operators $Y_{j,k}$'s satisfy systems of CCRs but they are not necessarily commuting for different pairs of $j, k \in \Lambda$.*

**Remark 3.9.** *The Markov generators associated with these Dirichlet forms are defined on dense domains and can be extended via Friedrichs extension to self adjoint operators, denoted as $-\mathfrak{L}$. This extension leads to a strongly continuous semigroup in $\mathbb{L}_2$ space represented by $P_t = e^{t\mathfrak{L}}$.*

**Example 3.9** (**G**-Models). *For some $\kappa, \xi \neq 0$, let $Y_{\kappa,\xi} = \kappa A - \xi A^*$ and*

$$\mathcal{N} \equiv Y_{\kappa,\xi}^* Y_{\kappa,\xi} = (|\kappa|^2 + |\xi|^2) N + |\xi|^2 - \bar{\kappa}\xi A^{*2} - \bar{\xi}\kappa A^2.$$

*Consider*

$$G_{\kappa,\xi} \equiv \frac{1}{2} Y_{\kappa,\xi}^2 = \frac{1}{2} (\kappa A - \xi A^*)^2.$$

*Then, with $\mathcal{R} \equiv |\kappa|^2 - |\xi|^2$, we have*

$$\begin{aligned}
[G_{\kappa,\xi}, G_{\kappa,\xi}^*] &= \frac{1}{4} Y_{\kappa,\xi}[Y_{\kappa,\xi}, Y_{\kappa,\xi}^{*2}] + \frac{1}{4}[Y_{\kappa,\xi}, Y_{\kappa,\xi}^{*2}] Y_{\kappa,\xi} \\
&= \frac{1}{4} Y_{\kappa,\xi} Y_{\kappa,\xi}^*[Y_{\kappa,\xi}, Y_{\kappa,\xi}^*] + \frac{1}{4} Y_{\kappa,\xi}[Y_{\kappa,\xi}, Y_{\kappa,\xi}^*] Y_{\kappa,\xi}^* \\
&\quad + \frac{1}{4} Y_{\kappa,\xi}^*[Y_{\kappa,\xi}, Y_{\kappa,\xi}^*] Y_{\kappa,\xi} + \frac{1}{4}[Y_{\kappa,\xi}, Y_{\kappa,\xi}^*] Y_{\kappa,\xi}^* Y_{\kappa,\xi} \\
&= \frac{1}{4} 2\mathcal{R} \left( Y_{\kappa,\xi} Y_{\kappa,\xi}^* + Y_{\kappa,\xi}^* Y_{\kappa,\xi} \right) \\
&= \frac{1}{2} \mathcal{R}[Y_{\kappa,\xi}, Y_{\kappa,\xi}^*] + \mathcal{R} Y_{\kappa,\xi}^* Y_{\kappa,\xi} \\
&= \frac{1}{2} \mathcal{R}^2 + \mathcal{R} Y_{\kappa,\xi}^* Y_{\kappa,\xi} = \frac{1}{2} \mathcal{R}^2 + \mathcal{R} \mathcal{N}
\end{aligned}$$

*and*

$$[G_{\kappa,\xi}, \mathcal{N}] = 2\mathcal{R} G_{\kappa,\xi}, \qquad [G_{\kappa,\xi}^*, \mathcal{N}] = -2\mathcal{R} G_{\kappa,\xi}^*.$$

*We remark that in this model $\mathcal{R} \equiv |\kappa|^2 - |\xi|^2$ can take on positive as well negative values. It provides another example of type [SS67] in which the author talks about the quantum harmonic oscillator with hyperbolic phase space involving twisted transformations. We discuss this model and*

*its corresponding algebra in [MZ24b].*

*In the case of this model, we consider a product state $\omega_\theta$ associated to one particle interaction of the form $\mathcal{N}_j \equiv \mathcal{N}_j(\kappa_j, \xi_j)$, $j \in \mathbb{Z}^d$, as follows.*

$$\omega_\theta \equiv \lim_{\Lambda \to \mathbb{Z}^d} \frac{1}{Z_\Lambda} Tr_\Lambda e^{-\sum_{j\in\Lambda} \beta_j \mathcal{N}_j(\kappa_j, \xi_j)}$$

*where $\beta_j \in (0, \infty)$ and $0 < Z_\Lambda < \infty$ is the normalisation constant. This would be useful to introduce infinite dimensional models. The constants $\mathcal{R}_j$ in $\omega_\theta$ can vary with $j \in \mathbb{Z}^d$ on the integer lattice(or with the vertices of a triangulation of a manifold). In such cases, the modular and Hamiltonian automorphisms are defined by the following densely defined derivation, which is not necessarily inner in the following way*

$$ad_H(f) \equiv \lim_{\Lambda \to \mathbb{Z}^d} \sum_{j\in\Lambda} [\mathcal{N}_j, f]$$

*This setup includes a large family of eigenvectors of modular operator given as follows: For a finite set $J \subset\subset \mathbb{Z}^d$*

$$W_{J,O}(\kappa, \xi, \mathbf{n}) \equiv \prod_{j\in O} G^{n(j)}_{\kappa_j, \xi_j} \prod_{k\in J\setminus O} G^{*n(k)}_{\kappa_k, \xi_k}$$

*where indices $j, k$ indicate independent (commuting) copies of $G_{\kappa, \xi}$, and $\kappa \equiv (\kappa_j)_{j\in\mathbb{Z}^d}, \xi \equiv (\xi_j)_{j\in\mathbb{Z}^d}$, $\mathbf{n} \equiv \{n(l) \in \mathbb{N}, l \in \mathbb{Z}^d\}$.*

*One can then introduce a densely defined Dirichlet form for each of $W_{J,O}$ as follows*

$$\mathcal{E}_{J,O}(f) \equiv \sum_{j\in\mathbb{Z}^d} \left( \|\delta_{W_{J+j,O+j}} f\|_2 + \|\delta_{W^*_{J+j,O+j}} f\|_2 \right)$$

*where $J+j, O+j$ denote a shift of $J, O$ by a vector $j \in \mathbb{Z}^d$. We note that in case when $|n(O)| = |n(J\setminus O)|$, the corresponding $W_{J,O}$ is invariant with respect to modular operator, that is, $\alpha_t(W_{J,O}) = W_{J,O}$.*

*In the similar way, one can also explore Dirichlet forms that incorporate non-inner derivations*

$$\delta_{W(JO)} \equiv \sum_{j\in\mathbb{Z}^d} \delta_{W_{J+j,O+j}}.$$

## 3.8 No Spectral Gap Property

In this section, we generalise the results of commutative case obtained in [INZ12]. We will show that the quantum dissipative semigroups involving Dirichlet forms $\mathcal{E}$ defined with $Z_{j,k} = A_j - A_k$ and $Y_{j,k} = A_j - A_k^*$, for nearest neighbours' pairs $(j,k)$, decay to equilibrium only polynomially in time. We have the following result.

**Theorem 3.10.** *The Poincaré inequality does not hold for Dirichlet forms $\mathcal{E}$ defined with $Z_{j,k} = A_j - A_k$ as well as with $Y_{j,k} = A_j - A_k^*$.*

The only quantum example where Poincaré inequality failed was provided in [CFL00] for OU generator with equal coefficients for both directions of the derivations defined as the Quantum Brownian motion. In contrast to that example, an equilibrium state is present in our case. The concept behind the proof of the theorem aims to demonstrate the absence of a constant $m \in (0, \infty)$ such that

$$m\|f - \omega(f)\|_2^2 \leq \mathcal{E}(f).$$

This is accomplished by creating a series of operators denoted as $f_n$, which are confined within boxes $\Lambda_n$ of dimensions $2n + 1$, where the variance increases proportionally with the volume of the box, while the Dirichlet form grows proportionally to the surface area of the box.
The proof and hence the result generalises to all models in which the state $\omega$ has summable decay of correlations.

*Proof.* For $Z_{j,k} = \frac{1}{2}(A_j - A_k)$, consider an increasing sequence of a finite set $\Lambda_n \equiv [-n, n]^d$ and corresponding sequence of the following operators

$$F_n \equiv \sum_{k \in \Lambda_n} A_k^*.$$

Then we have

$$\delta_{Z_{j,k}^*}(F_n) = 0$$

and
$$\delta_{Z_{j,k}}(F_n) = i[\frac{1}{2}(A_j - A_k), \sum_{l\in[-n,n]^d} A_l^*] = \frac{i}{2} \sum_{l\in[-n,n]^d} (\delta_{j,l} - \delta_{k,l})$$

with
$$\sum_{l\in[-n,n]^d} (\delta_{j,l} - \delta_{k,l}) = 0 \text{ if } both\ j, k\ are\ outside\ or\ both\ inside\ \Lambda_n$$

and otherwise $\sum_{l\in[-n,n]^d} |(\delta_{j,l} - \delta_{k,l})| \leq 2d$. Hence we get

$$\mathcal{E}(F_n) = \hat{\eta}(0) \sum_{j,k\in\mathbb{Z}} \left( \|\delta_{Z_{j,k}}(F_n)\|_{2,\omega}^2 + \|\delta_{Z_{j,k}^*}(F_n)\|_{2,\omega}^2 \right) \leq Const|\partial\Lambda_n|$$

On the other hand, for the product state,

$$\|F_n - \langle F_n\rangle_\omega\|_{2,\omega}^2 \sim \sum_{j\in\Lambda_n} \|A_j^* - \langle A_j^*\rangle_\omega\|_{2,\omega}^2 \sim Vol(\Lambda_n)\|A_0^* - \langle A_0^*\rangle_\omega\|_{2,\omega}^2$$

Since the boundary $|\partial\Lambda_n|/Vol(\Lambda_n) \to_{n\to\infty} 0$ for a suitable sequence of sets $\Lambda$ invading the lattice, the Poincare inequality cannot hold. □

**Further examples**

A similar outcome applies to the Dirichlet form when it is defined using alternative derivations.

**Example 3.10.** *For $Z_{j,k} = A_j^m - A_k^m$, one can consider a sequence of the following operators*

$$F_n \equiv \sum_{k\in\Lambda_n} A_k^*.$$

*for a finite set $\Lambda_n$.*

*Then we have*

$$\delta_{Z_{j,k}}(F_n) = i[A_j^m - A_k^m, \sum_{l\in[-n,n]^d} A_l^*] = im \sum_{l\in[-n,n]^d} \left(\delta_{j,l}A_j^{m-1} - \delta_{k,l}A_k^{m-1}\right)$$

$$\sim \begin{cases} mA_j^{m-1} or -mA_k^{m-1} & if\ either\ j\ or\ k\ is\ outside\ \Lambda_n \\ 0 & if\ both\ are\ outside\ \Lambda_n \end{cases}$$

and

$$\delta_{Z^*_{j,k}}(F_n) = 0$$

Hence we conclude that

$$\mathcal{E}(F_n) \sim |\partial \Lambda_n|, \quad \text{and} \quad \|F_n - \langle F_n \rangle\|^2 \sim |\Lambda_n|.$$

which prevents the application of the Poincaré inequality.

Another important possibility of a sequence of test operators is provided by

$$F_n \equiv \sum_{k \in \Lambda_n} N_k.$$

In this case we have

$$\delta_{Z^\sharp_{j,k}}(F_n) = sign(\sharp)i[A_j^{\sharp m} - A_k^{\sharp m}, \sum_{l \in [-n,n]} N_l] = sign(\sharp)im\left(\delta_{j,l}A_j^{\sharp m} - \delta_{k,l}A_k^{\sharp m}\right)$$

where $sign(\sharp)$ equals $\pm 1$, respectively. Thus we reach the same conclusion.

**Example 3.11.** *For $Y_{j,k} = A_j + A_k^*$, consider a sequence*

$$F_n = \sum_{\substack{l,l' \in \Lambda_n \\ l \sim l'}} (A_l + A_{l'}^*)$$

**Remark** *This example can be extended to involve $Y_{\kappa,\xi}$, where $\kappa$ and $\xi$ are absolutely summable sequences that adhere to the condition that the sums of their elements across all indices equal zero:*

$$\sum_{j \in \mathbb{Z}^d} \kappa_j = 0, \quad \sum_{j \in \mathbb{Z}^d} \xi_j = 0$$

## 3.9   Algebra of Invariant Derivations

Consider the operators of the following form

$$A(I, J) \equiv \prod_{i \in I} A_i \prod_{j \in J} A_j^* \qquad (3.20)$$

and the setting associated to the product state $\omega_0$. In case $|I| = |J|$, we compute the modular dynamics

$$\alpha_t(A(I,J)) = \alpha_t(\prod_{i \in I} A_i \prod_{j \in J} A_j^*) = \prod_{i \in I} \alpha_t(A_i) \prod_{j \in J} \alpha_t(A_j^*) = \prod_{i \in I} A_i \prod_{j \in J} A_j^* = A(I,J)$$

Alternatively, one can use the relations (2.13) to show that such operators are invariant with respect to the modular dynamics when $|I| = |J|$. For $I = J$, the operator $A(I, J)$ is a polynomial in $N_j$, $j \in J$ using (2.13). Choosing $I \cap J = \emptyset$, we can define a Dirichlet form as follows

$$\mathcal{E}_{I,J}(f) \equiv \sum_{k \in \mathbb{Z}^d} \|\delta_{A(I+k,J+k)} f\|_{2,\omega_0}^2 \qquad (3.21)$$

for all operators $f$ for which the right hand side is finite. The domain is dense in $\mathbb{L}_2(\omega_0)$ because it includes all the local operators. We have the following general result.

**Theorem 3.11.** *The Dirichlet form $\mathcal{E}_{I,J}$ does not satisfy Poincaré inequality, that is, it has no spectral gap.*

*Proof.* We choose a sequence of operators

$$F_n \equiv \sum_{j \in \Lambda_n} N_j$$

and noticing that if $I + k, J + k \subset \Lambda_n$ where $\Lambda_n \equiv [-n, n]^d$ is an increasing sequence of a finite set, then

$$\delta_{A(I+k,J+k)}(F_n) = 0.$$

Hence

$$\mathcal{E}_{I,J}(F_n) \sim |\partial \Lambda_n|$$

while
$$\|F_n - \omega_0(F_n)\|^2_{2,\omega_0} \sim |\Lambda_n|.$$

As a result, the Poincaré inequality would fail for large values of $\Lambda_n$ regardless of the specific positive constant chosen. $\square$

We remark that there is a possibility to define an infinite number of Dirichlet forms which cannot satisfy Poincaré inequality. In the present setup we have an infinite dimensional algebra which is invariant with respect to the modular operator including the operators $A(I, J)$. Infact, it also includes more general operators, for example
$$A_I^{n(I)} \equiv \prod_{i \in I} A_i^{n(i)}$$
with multi-index $n(I) \equiv (n(i) \in \mathbb{N})_{i \in I}$, and setting $|n(I)| \equiv \sum_{i \in I} n(i)$, we can consider operators of the form
$$P_{I,J}(N_I, N_J)\left(A_I^{n(I)} \left(A_J^{n(J)}\right)^*\right)$$
with $|n(I)| = |n(J)|$ and a polynomial $P_{I,J}(N_I, N_J)$.

In addition to the inner derivations mentioned earlier, one can also explore exterior (or non-inner) derivations, forming an infinite dimensional Lie algebra. For instance, the given limit yields non-inner derivations that are effectively defined on a dense subset of local elements within $\mathbb{L}_{2,\omega_0}$.

$$\delta_{I,J}(f) \equiv \lim_{\Lambda \to \mathbb{Z}^d} \sum_{k \in \Lambda} \delta_{A(I+k,J+k)}(f).$$

Obtaining constructive examples of such algebra for Gibbs states obtained with a nontrivial interaction still remains an interesting open problem.

A new challenging area of noncommutative analysis is defining the Dirichlet forms with non-inner derivations.

## 3.10  Polynomial Decay to Equilibrium

In this section we generalise the commutative result of [INZ12] for the model with derivations in direction of $Z_{j,k} = A_j - A_k$, for $j \sim k$, $j,k \in \mathbb{Z}^d$ in the space associated to the product state $\omega_0$ describing a system of infinite number of quantum harmonic oscillators. The following result shows that the quantum system associated to this model decay only polynomially in time.

**Theorem 3.12.** *A quantum system described by the family of generators of the form as above decays to equilibrium algebraically in time in the sense that*

$$\sum_j |\delta_{A_j^\sharp}(e^{-tL}f)|^2 \to_{t\to\infty} 0$$

*with algebraic rate.*

*Proof.* **Z-Case**

For the model with derivations in direction of $Z_{j,k} = A_j - A_k$, for $j \sim k$, $j,k \in \mathbb{Z}^d$, set $f_t = P_t f$.

Writing the derivative of the scalar product

$$\frac{d}{dt}\langle \delta_{A_j}(P_t f), \delta_{A_j}(P_t f)\rangle_\omega = \langle \frac{d}{dt}\delta_{A_j}(P_t f), \delta_{A_j}(P_t f)\rangle_\omega + \langle \delta_{A_j}(P_t f), \frac{d}{dt}\delta_{A_j}(P_t f)\rangle_\omega \quad (3.22)$$

$$= \langle \delta_{A_j}L(P_t f), \delta_{A_j}(P_t f)\rangle_\omega + \langle \delta_{A_j}(P_t f), \delta_{A_j}L(P_t f)\rangle_\omega$$

$$= \langle L\delta_{A_j}(P_t f) + [\delta_{A_j}, L](P_t f), \delta_{A_j}(P_t f)\rangle_\omega + \langle \delta_{A_j}(P_t f), L\delta_{A_j}(P_t f) + [\delta_{A_j}, L](P_t f)\rangle_\omega$$

$$\leq \langle [\delta_{A_j}, L](P_t f), \delta_{A_j}(P_t f)\rangle_\omega + \langle \delta_{A_j}(P_t f), [\delta_{A_j}, L](P_t f)\rangle_\omega$$

Next we use the following fact.

**Lemma 3.4.**

$$[\delta_{A_j}, L](f) = 4\hat{\eta}(0)\sinh(\beta/2)\sum_{k\sim j}\left(\delta_{A_k}f - \delta_{A_j}f\right).$$

*Proof.* (*of Lemma*) Since

$$L = -\hat{\eta}(0)\sum_{j\sim k}\left(\delta^\star_{Z_{j,k}}\delta_{Z_{j,k}} + \delta^\star_{Z^*_{j,k}}\delta_{Z^*_{j,k}}\right)$$

and $A_l$ commute with $Z_{jk}$, for fixed $l$, we have

$$[\delta_{A_l}, L] = -\hat{\eta}(0) \sum_{j \sim k} \left( [\delta_{A_l}, \delta^\star_{Z_{j,k}}]\delta_{Z_{j,k}} + [\delta_{A_l}, \delta^\star_{Z^*_{j,k}}]\delta_{Z^*_{j,k}} \right).$$

Since, by Proposition 3.3, we have

$$\delta^\star_{Z_{j,k}}(g) = -\delta_{\alpha_{-i/2}(Z^*_{j,k})}(g) + i\left(\alpha_{-i/2}(Z^*_{j,k}) - \alpha_{i/2}(Z^*_{j,k})\right)g$$
$$= -e^{-\frac{\beta}{2}}\delta_{Z^*_{j,k}}(g) - 2i\sinh(\frac{\beta}{2})Z^*_{j,k}g$$

In our setup we have

$$[\delta_{A_l}, \delta_{Z^*_{j,k}}] = i\delta_{[A_l, Z^*_{j,k}]} = 0.$$

On the other hand for the left multiplication operator $l_{Z^*_{j,k}}$ by $Z^*_{j,k}$, using the Leibnitz rule 3.3 for the derivation

$$[\delta_{A_l}, L_{Z^*_{j,k}}] = \delta_{A_l}(Z^*_{j,k}) = \delta_{l,j} - \delta_{l,k}$$

Hence we obtain

$$[\delta_{A_l}, \delta^\star_{Z_{j,k}}](g) = [\delta_{A_l}, -e^{-\frac{\beta}{2}}\delta_{Z^*_{j,k}} - 2i\sinh(\frac{\beta}{2})l_{Z^*_{j,k}}](g) = [\delta_{A_l}, -2i\sinh(\frac{\beta}{2})l_{Z^*_{j,k}}](g)$$
$$= 2\sinh(\frac{\beta}{2})(\delta_{l,j} - \delta_{l,k})g$$

as $[A_l, Z^*_{j,k}] = \delta_{l,j} - \delta_{l,k}$. Thus we have

$$[\delta_{A_l}, \delta^\star_{Z_{j,k}}]\delta_{Z_{j,k}}(f) = 2\sinh(\beta/2)(\delta_{l,j} - \delta_{l,k})\delta_{Z_{j,k}}(f)$$

and hence for a fixed $l$,

$$[\delta_{A_l}, L](f) = -2\sinh(\beta/2)\hat{\eta}(0) \sum_{j \sim k}(\delta_{l,j} - \delta_{l,k})\delta_{Z_{j,k}}(f)$$
$$= -4\sinh(\beta/2)\hat{\eta}(0) \sum_{j \sim l}\delta_{Z_{l,j}}(f)$$
$$= -4\sinh(\beta/2)\hat{\eta}(0) \sum_{j \sim l}\left(\delta_{A_l}f - \delta_{A_j}f\right)$$

□

Using (3.22) and lemma 3.4, for any $j \in \mathbb{Z}^d$ we have

$$\frac{d}{dt}\langle \delta_{A_j}(P_t f), \delta_{A_j}(P_t f)\rangle_\omega \leq 8 \sinh(\beta/2)\hat{\eta}(0) \sum_{k\sim j} \mathfrak{Re}\langle \delta_{A_j}(P_t f), (\delta_{A_k}-\delta_{A_j})(P_t f)\rangle.$$

Using Cauchy-Schwartz inequality on the right hand side, this implies the following relation

$$\frac{d}{dt}\|\delta_{A_j}(P_t f)\| \leq 4\hat{\eta}(0)\sinh(\beta/2)\sum_{k\sim j}\left(\|\delta_{A_k}(P_t f)\|-\|\delta_{A_j}(P_t f)\|\right).$$

with the norm given by the scalar product.

Denoting by $\triangle$ the lattice Laplacian on $\mathbb{Z}^d$ and setting $F(t,j) = \|\delta_{A_j}(P_t f)\|$, with $C \equiv 4\hat{\eta}(0)\sinh(\beta/2)$, we get the following differential inequality

$$\frac{d}{dt}F(t,j) \leq C\,(\triangle F)(t,j)$$

Integrating both sides

$$\int_0^t dF(s) \leq \int_0^t C\Delta F(s)ds$$

$$F(t) \leq F(0) + \int_0^t C\Delta F(s)ds,$$

by iteration we arrive at the following bound

$$F(t) \leq e^{Ct\triangle}F(0)$$

This concludes that the system decays algebraically, see e.g. [INZ12]. The lattice Laplacian quantifies how the function value at a particular point differs from the average of its neighboring points. When the function $F(t)$ is acted upon by the lattice Laplacian $\Delta$, the Laplacian tends to "smooth out" the function by reducing the differences between neighboring points. As the Laplacian smooths out the function $F(t)$ over time, the sharp variations or gradients in the function diminish. Consequently, the rate of change of $F(t)$ with respect to time decreases, leading to a decay in the function's magnitude over time.

We obtain same algebraic bound for derivation with respect to $A_j^*$ using similar arguments.

Let us note that, similar to the case of functions, [INZ12], the space of linear combinations of $A_j, A_j^*$, $j \in \mathbb{Z}^d$, is mapped into itself by the generator $L$. Explicitly we have

$$LA_l = -\hat{\eta}(0) \sum_{j \sim k} \left( \delta^\star_{Z^*_{j,k}} \delta_{Z^*_{j,k}} \right) A_l$$

$$= i\hat{\eta}(0) \sum_{j \sim k} \left( \delta^\star_{Z^*_{j,k}} (\delta_{jl} - \delta_{kl}) \right) = 4\hat{\eta}(0) \sinh(\frac{\beta}{2}) \sum_{j \sim l} (A_j^* - A_l^*)$$

$$LA_l^* = -i\hat{\eta}(0) \sum_{j \sim k} \left( \delta^\star_{Z_{j,k}} (\delta_{jl} - \delta_{kl}) \right) = 4\hat{\eta}(0) \sinh(\frac{\beta}{2}) \sum_{j \sim l} (A_j - A_l)$$

where we used

$$\delta^\star_{Z_{j,k}}(g) = -e^{-\frac{\beta}{2}} \delta_{Z^*_{j,k}}(g) - 2i \sinh(\frac{\beta}{2}) Z^*_{j,k} g$$

$$\delta^\star_{Z^*_{j,k}}(g) = -e^{\frac{\beta}{2}} \delta_{Z_{j,k}}(g) + 2i \sinh(\frac{\beta}{2}) Z_{j,k} g$$

Hence, with $C \equiv 4\hat{\eta}(0) \sinh(\frac{\beta}{2})$, we get

$$L(A_j \pm A_j^*) = C \sum_{j \sim l} ((A_j \pm A_j^*) - (A_l \pm A_l^*)) \tag{3.23}$$

$$\equiv C(\triangle(A_\cdot \pm A_\cdot^*))_j$$

Thus for an operator linear in creators and annihilators

$$f \equiv \sum_j \kappa_j(t)(A_j \pm A_j^*)$$

we have using 3.23

$$\partial_t f = C \sum_j \dot{\kappa}_j(t)(A_j \pm A_j^*)$$

$$= C \sum_j \kappa_j(t)(\triangle(A_\cdot \pm A_\cdot^*))_j = C \sum_j (\triangle \kappa)_j(t)(A_j \pm A_j^{\cdot *})$$

which holds if and only if

$$\dot{\kappa}_j(t) = C \sum_j (\triangle \kappa)_j(t)$$

i.e.
$$\kappa(t) = e^{tC_\Delta}\kappa_0.$$

This provides explicit algebraic decay to equilibrium for operators linear in creators and annihilators. The equation $\kappa(t) = e^{tC_\Delta}\kappa_0$ provides a mathematical representation of how the coefficients $\kappa_j(t)$ decay over time under the specified dynamics. The action of the exponential operator $e^{tC_\Delta}$ smooths out the coefficients, resulting in algebraic decay towards equilibrium as time progresses. □

The $Y$-case where $Y_{j,k} = A_j - A_k^*$ is similar.

In this case, we have the following lemma.

**Lemma 3.5.** *Consider for fixed $l$*

$$[\delta_{A_l}, L] = \sum_{j,k} \hat{\eta}(0) 2\sinh(\beta/2)\left(\delta_{l,j}\delta_{A_j}(f) - \delta_{l,k}\delta_{A_j^*}(f)\right)$$
$$+ \hat{\eta}(-2\beta)2\sinh(\beta/2)\delta_{l,j}\delta_{A_k^*}(f) - \hat{\eta}(2\beta)2\sinh(\beta/2)\delta_{l,k}\delta_{A_k}(f)$$

*Proof.* Consider

$$[\delta_{A_l}, L] = \int [\delta_{A_l}, \sum_{j,k} \delta^\star_{\alpha_t(Y_{j,k})}\delta_{\alpha_t(Y_{j,k})} + \delta^\star_{\alpha_t(Y_{j,k}^*)}\delta_{\alpha_t(Y_{j,k}^*)}]\eta(t)dt$$

which can be written as

$$[\delta_{A_l}, L] = \int \sum_{j,k} [\delta_{A_l}, \delta^\star_{\alpha_t(Y_{j,k})}]\delta_{\alpha_t(Y_{j,k})} + \delta^\star_{\alpha_t(Y_{j,k})}[\delta_{A_l}, \delta_{\alpha_t(Y_{j,k})}] + [\delta_{A_l}, \delta^\star_{\alpha_t(Y_{j,k}^*)}]\delta_{\alpha_t(Y_{j,k}^*)} + \delta^\star_{\alpha_t(Y_{j,k}^*)}[\delta_{A_l}, \delta_{\alpha_t(Y_{j,k}^*)}]\eta(t)dt$$

We know

$$\delta^\star_{\alpha_t(Y_{j,k})}(f) = -ie^{-i\beta t}\left(e^{\beta/2}A_j^*f - fe^{-\beta/2}A_j^*\right) - ie^{i\beta t}\left(e^{-\beta/2}A_k f - fe^{\beta/2}A_k\right)$$

and similarly

$$\delta^\star_{\alpha_t(Y_{j,k}^*)}(f) = -ie^{i\beta t}\left(e^{-\beta/2}A_j f - fe^{\beta/2}A_j\right) - ie^{-i\beta t}\left(e^{\beta/2}A_k^*f - fe^{-\beta/2}A_k^*\right)$$

Since

$$[\delta_{A_l}, \delta_{\alpha_t(Y_{j,k})}](f) = \delta_{A_l}\delta_{\alpha_t(Y_{j,k})}(f) - \delta_{\alpha_t(Y_{j,k})}\delta_{A_l}(f) = -[Y_{j,k}, [A_l, f]] + [A_l, [Y_{j,k}, f]]$$

$$= -[f, [A_l, Y_{j,k}]] = -[f, [A_l, A_j - A_k^*]] = [f, \delta_{l,k}] = 0$$

and

$$[\delta_{A_l}, \delta_{\alpha_t(Y_{j,k}^*)}](f) = \delta_{A_l}\delta_{\alpha_t(Y_{j,k}^*)}(f) - \delta_{\alpha_t(Y_{j,k}^*)}\delta_{A_l}(f) = -[Y_{j,k}^*, [A_l, f]] + [A_l, [Y_{j,k}^*, f]]$$

$$= -[f, [A_l, Y_{j,k}^*]] = -[f, [A_l, A_j^* - A_k]] = -[f, \delta_{l,j}] = 0$$

second and fourth term is zero.

$$[\delta_{A_l}, \delta^\star_{\alpha_t(Y_{j,k})}](f) = \delta_{A_l}\delta^\star_{\alpha_t(Y_{j,k})}(f) - \delta^\star_{\alpha_t(Y_{j,k})}\delta_{A_l}(f)$$

$$= e^{-i\beta t}\left(e^{\beta/2}[A_l, A_j^*f] - e^{-\beta/2}[A_l, fA_j^*]\right) + e^{i\beta t}\left(e^{-\beta/2}[A_l, A_kf] - e^{\beta/2}[A_l, fA_k]\right)$$

$$- e^{-i\beta t}\left(e^{\beta/2}A_j^*[A_l, f] - [A_l, f]e^{-\beta/2}A_j^*\right) - e^{i\beta t}\left(e^{-\beta/2}A_k[A_l, f] - [A_l, f]e^{\beta/2}A_k\right)$$

$$= e^{-i\beta t}\left(e^{\beta/2} - e^{-\beta/2}\right)\delta_{l,j} = e^{-i\beta t}2\sinh(\beta/2)\delta_{l,j}$$

In a similar way

$$[\delta_{A_l}, \delta^\star_{\alpha_t(Y_{j,k}^*)}](f) = \delta_{A_l}\delta^\star_{\alpha_t(Y_{j,k}^*)}(f) - \delta^\star_{\alpha_t(Y_{j,k}^*)}\delta_{A_l}(f) = e^{i\beta t}\left(e^{-\beta/2} - e^{\beta/2}\right)\delta_{l,k} = e^{i\beta t}2\sinh(-\beta/2)\delta_{l,k}$$

Hence we obtain

$$[\delta_{A_l}, L](f) = \int \sum_{j,k}(e^{-i\beta t}2\sinh(\beta/2)\delta_{l,j})ie^{i\beta t}[A_j, f] + ie^{-i\beta t}[A_k^*, f] + (e^{i\beta t}2\sinh(-\beta/2)\delta_{l,k})ie^{-i\beta t}[A_j^*, f]$$

$$+ ie^{i\beta t}[A_k, f]\eta(t)dt$$

$$= \int \sum_{j,k} i2\sinh(\beta/2)\delta_{l,j}[A_j, f] + ie^{-2i\beta t}2\sinh(\beta/2)\delta_{l,j}[A_k^*, f] + i2\sinh(-\beta/2)\delta_{l,k}[A_j^*, f]$$

$$+ ie^{2i\beta t}2\sinh(-\beta/2)\delta_{l,k}[A_k, f]\eta(t)dt$$

$$= \sum_{j,k} \hat{\eta}(0)2\sinh(\beta/2)\delta_{l,j}\delta_{A_j}(f) + \hat{\eta}(-2\beta)2\sinh(\beta/2)\delta_{l,j}\delta_{A_k^*}(f) + \hat{\eta}(0)2\sinh(-\beta/2)\delta_{l,k}\delta_{A_j^*}(f)$$

$$+ \hat{\eta}(2\beta)2\sinh(-\beta/2)\delta_{l,k}\delta_{A_k}(f)$$

and we obtain the result. The rest of the proof is similar to the Z-Case. □

# Chapter 4

# Representations of Nilpotent Lie Algebras and Applications

## 4.1 Introduction and Background

In this chapter, we discuss the representations of Lie Algebras in terms of creation and annihilation operators. The generators of such representations can then be utilised to construct and analyse the dissipative dynamics. We provide examples where an interesting quantum stochastic analysis (see Section 2.8 for references) could be developed.

In the classical theory, given a family of noncommuting vector fields $\{X_j : j \in \mathcal{J}\}$ on an enveloping algebra $\mathcal{D}$, where $\mathcal{J}$ is a finite or countably infinite index set, a class of Markov generators of the form

$$\mathfrak{L} = \sum_{j \in \mathcal{J}} X_j^2 \qquad (4.1)$$

is extensively studied. If the Lie algebra (Definition 4.1) generated by $\{X_j : j \in \mathcal{J}\}$ spans $\mathcal{D}$, the operator $\mathfrak{L}$ of the above form is said to satisfy Hörmander rank condition. Such operators, instead of elliptic, are hypoelliptic in the sense of Hörmander. This means that the corresponding semigroups $P_t = e^{t\mathfrak{L}}$ still has a smooth density with respect to the Lebesgue measure but the mathematical techniques used for elliptic operators (like those by Bakry and Émery [Bak04]) do not work for these hypoelliptic operators.

By Lifting theorem [RS77], every operator $\mathfrak{L}$ given by (4.1) that satisfies the Hörmander rank condition can be approximated by a sub-Laplacian on a stratified Lie algebra. We give few basic definitions of Lie algebras. For more details, one can refer [BLU07].

**Definition 4.1.** *A Lie algebra is a vector space $\mathfrak{g}$ over a field F with an operation $[\cdot,\cdot] : \mathfrak{g} \times \mathfrak{g} \to \mathfrak{g}$ called Lie bracket such that it is bilinear, skew symmetric and satisfies Jacobi identity*

$$[x,[y,z]] + [y,[x,z]] + [z,[x,y]] = 0 \quad \text{for all } x, y, z \in \mathfrak{g}.$$

They describe the tangent space at the identity of a Lie group $\mathbb{G}$ (see Definition 1.2.1,[BLU07]) and provide information about infinitesimal transformations and symmetries.

Given a Lie algebra $\mathfrak{g}$, the lower central series of $\mathfrak{g}$ is a sequence of subalgebras $\mathfrak{g}_i$ defined as follows

$$\mathfrak{g}_1 = \mathfrak{g},$$

$$\mathfrak{g}_{i+1} = [\mathfrak{g}, \mathfrak{g}_i] \text{ for } i \geq 1,$$

A Lie algebra $\mathfrak{g}$ is called nilpotent if $\exists\, k \in \mathbb{N}$ such that $\mathfrak{g}_k = 0$.

Moreover, a Lie algebra $\mathfrak{g}$ is called a stratified Lie algebra if it has a vector space decomposition

$$\mathfrak{g} = \bigoplus_{j=1}^{\infty} V_j, \quad \text{such that} \quad [V_i, V_j] \subset V_{i+j}.$$

Furthermore, every element of $\mathfrak{g}$ can be expressed as a linear combination of iterated Lie brackets of elements of $V_1$.

A unitary representation of a group $\mathbb{G}$, if it exists, is a way of representing the group elements as unitary operators on a Hilbert space. The generators of unitary representations can be used to construct more complex operators that describe dissipative processes. For instance, the Hamiltonian can be related to the generators of the unitary representation of the symmetry group of the system.

In Section 4.2, we discuss unitary representations of some groups. In Section 4.3, we provide representations for Free nilpotent Lie algebras. Lastly, we provide some models for the Hamiltonians which are the combinations of so called Chevalley generators (4.11) and creation and annihilation operators in Section 4.4.

## 4.2 Quasi Invariance and Unitary Group Representations

Suppose, $\theta, \tau \in \mathbb{C}$ satisfy $|\tau|^2 - |\theta|^2 = 1$. Then the operators

$$a \equiv a(\tau, \theta) \equiv \tau A + \theta A^*, \qquad a^* \equiv a^*(\tau, \theta) \equiv \bar{\tau} A^* + \bar{\theta} A$$

satisfy CCR

$$[a, a^*] = id.$$

The transformation of this type is known in the mathematical/theoretical physics literature under the name of Bogolubov transformations.

We can consider here a class of transformations which includes the Lorenz group acting on two dimensional CCR vectors via matrices,

$$\mathbb{R} \ni t \longmapsto \begin{pmatrix} cosh(t) & sinh(t) \\ sinh(t) & cosh(t) \end{pmatrix}$$

(which can be extend to complex parameter). For a differentiable function $\mathbb{R} \ni s \to \{(\tau(s), \theta(s)) : |\tau(s)|^2 = |\theta(s)|^2 + 1\}$, we define

$$a_s \equiv a(\tau(s), \theta(s)), \ a_s^* \equiv a(\tau(s), \theta(s))^*,$$

with the initial condition $a_0 = A, a_0^* = A^*$. Given initial density $\rho \equiv \frac{1}{Z} e^{-U(A,A^*)}$ where $U$ is the interaction energy, we define transformed density as follows

$$\rho_s \equiv \frac{1}{Z_s} e^{-U(a_s, a_s^*)}$$

with normalisation factor $Z_s = \text{Tr } e^{-U(a_s, a_s^*)} = \text{Tr } e^{-U(A,A^*)}$. Then we have the following result.

**Theorem 4.1.** *For any polynomial function $f$, the following formula defines a unitary group repre-*

*sentation in* $\mathbb{L}_2(\rho)$

$$V_s(f(A, A^*)) \equiv \rho^{-\frac{1}{4}} \rho_s^{\frac{1}{4}} f(a_s, a_s^*) \rho_s^{\frac{1}{4}} \rho^{-\frac{1}{4}}.$$

*The generator of the group on polynomials is given by*

$$\partial_s V_s(f)(A, A^*)_{|s=0} = \partial_s f(a_s, a_s^*)_{|s=0} - \left\{ \frac{1}{4} \int_0^1 d\lambda \, \rho^{\frac{\lambda}{4}} \partial_s U(a_s, a_s^*)_{|s=0} \rho^{\frac{1-\lambda}{4}}, f(A, A^*) - \frac{1}{2} \frac{Z'}{Z} f(A, A^*) \right\}. \tag{4.2}$$

*where curly bracket denote anticommutator and* $Z' \equiv (\partial_s Z_s)_{|s=0}$.

*Proof.* Since the scalar product,

$$\begin{aligned}
\langle V_s(f(A, A^*)), V_s(f(A, A^*)) \rangle_\omega &= Tr\left( \rho^{\frac{1}{2}} (V_s(f(A, A^*)))^* \rho^{\frac{1}{2}} V_s(f(A, A^*)) \right) \\
&= Tr\left( \rho_s^{\frac{1}{2}} (f(a_s, a_s^*))^* \rho_s^{\frac{1}{2}} f(a_s, a_s^*) \right) \\
&= \langle f(a_s, a_s^*), f(a_s, a_s^*) \rangle_{\omega_s}
\end{aligned}$$

Since the commutation relations for the representations are same, the expectations will be same as with $\rho$. The generator (4.2) can be obtained using Fundamental Theorem of Calculus. □

To get to the representation of Lorenz group in higher space dimension one needs to consider higher order quantisation of space-time in a form of product space of many independent harmonic oscillators $(A_i, A_i^*)_{i=0,1..,n}$ in which we can consider the following CCR representation

$$(A_i, A_i^*)_{i=1,...,n} \mapsto S(\tau, \mathbf{x}) \equiv \tau \frac{1}{\sqrt{n}} \sum_{i=1,...,n} A_i + \sum_{i=1,...,n} x_i A_i^*.$$

(So in infinite dimensional limit time component is given by a "Gaussian random variable".) Then we have the following representation of the Minkowski product

$$[S, S^*] = |\tau|^2 - |\mathbf{x}|^2.$$

**Remark** *Given $n \in \mathbb{N}$, $n > 1$, CCR pairs $(A_i, A_i^*)_{i=0,1...n}$, we can define*

$$A_i(\boldsymbol{\gamma}, \boldsymbol{\kappa}) \equiv \sum_{j=1,...,n} \left(\gamma_{ij} A_j + \kappa_{ij} A_j^*\right)$$

$$A_i^*(\boldsymbol{\gamma}, \boldsymbol{\kappa}) \equiv \sum_{j=1,...,n} \left(\bar{\gamma}_{ij} A_j^* + \bar{\kappa}_{ij} A_j\right),$$

*such that they satisfy the CCR condition $[A_i(\boldsymbol{\gamma}, \boldsymbol{\kappa}), A_i^*(\boldsymbol{\gamma}, \boldsymbol{\kappa})] = \text{id}$ provided*

$$\sum_{j=1,...,n} |\gamma_{ij}|^2 - \sum_{j=1,...,n} |\kappa_{ij}|^2 = 1.$$

*Hence we can introduce a group of linear transformations*

$$\mathbf{T} \equiv (T, \tilde{T}) : (\boldsymbol{\gamma}, \boldsymbol{\kappa}) \mapsto \mathbf{T}(\boldsymbol{\gamma}, \boldsymbol{\kappa}) \equiv (T(\boldsymbol{\gamma}), \tilde{T}(\boldsymbol{\kappa}))$$

$$(\mathbf{T}(\boldsymbol{\gamma}))_i \equiv \sum_{j=1,...,n} \left(T_{ij} \gamma_{ij}\right), \qquad (\mathbf{T}(\boldsymbol{\kappa}))_i \equiv \sum_{j=1,...,n} \left(\tilde{T}_{ij} \kappa_{ij}\right)$$

*preserving the CCR condition, i.e. satisfying*

$$\sum_{j=1,...,n} |T(\boldsymbol{\gamma})_{ij}|^2 - \sum_{j=1,...,n} |\tilde{T}(\boldsymbol{\gamma})_{ij})|^2 = 1.$$

We can use the idea of the quasi-invariance of a state to obtain unitary representations of this extended group in a similar fashion.

## 4.3　Representations of Free Nilpotent Lie Algebras

In the classical theory, [GG90] developed an algorithm that produces vector fields say $E_1, E_2, \ldots, E_M$ in $\mathbf{R}^d$ with the property that they generate a Lie algebra isomorphic to a free nilpotent Lie algebra $\mathfrak{g}_{M,r}$. Here, $M$ is the number of generators and any iterated Lie bracket with more than $r$ elements vanishes. A free Lie algebra is defined as follows.

**Definition 4.2.** *(Definition 14.1.1, [BLU07]) For the fixed integers $M \geq 2$, $r \geq 1$, we say that $\mathfrak{g}_{M,r}$ is a free Lie algebra with $M$ generators $F_1, \ldots, F_M$ and nilpotent of step $r$ if:*

1. $\mathfrak{g}_{M,r}$ is a Lie algebra generated by its elements $F_1, \ldots, F_M$,

2. $\mathfrak{g}_{M,r}$ is nilpotent of step $r$,

3. for every Lie algebra $\mathfrak{n}$ nilpotent of step $r$ and for every map $\varphi$ from the set $\{F_1, \ldots, F_M\}$ to $\mathfrak{n}$, there is a (unique) homomorphism of Lie algebras from $\mathfrak{g}_{M,r}$ to $\mathfrak{n}$ which extends $\varphi$.

Given a hypoelliptic partial differential operator

$$L = \sum_{j=1}^{M} F_j^2.$$

By [RS77] discussed in Section 4.1, the vector fields $F_j$'s can be replaced by other vector fields say $\tilde{F}_j$'s in a larger space such that they are free at a given point. And finally these can be approximated by the generators $E_1, E_2, \ldots, E_M$ of a free nilpotent Lie algebra $\mathfrak{g}_{M,r}$. Then one can instead study the hypoellipticity of

$$L = \sum_{j=1}^{M} E_j^2.$$

Following Section 2, [GG90], we briefly discuss the construction of these generators given integers $r, M \geq 0$ such that they generate $\mathfrak{g}_{M,r}$. The construction utilises the concept of Hall basis which is a basis of $\mathfrak{g}_{M,r}$ and is defined below.

**Definition 4.3.** *(Definition 1.1, [GG90]) Each element of the Hall basis is a monomial in the generators and is defined recursively as follows. The generators $E_1, E_2, \ldots, E_M$ are elements of the basis and of length 1. If we have defined basis elements of lengths $1, \ldots, r-1$, they are ordered so that $E$ precedes $F$ written as $E < F$ if $\mathrm{length}(E) < \mathrm{length}(F)$. Moreover, if $\mathrm{length}(E) = s$ and $\mathrm{length}(F) = t$ and $r = s + t$, then $[E, F]$ is a basis elements of length $r$ if:*

1. *$E$ and $F$ are basis elements and $E > F$, and*

2. *if $E = [G, H]$, then $F \geq H$.*

Let $v$ be the dimension of $\mathfrak{g}_{M,r}$ and fix an element $E_i$ from the Hall basis $E_1, \ldots, E_v$. By the definition of Hall basis, $E_i = [E_{j_1}, E_{k_1}]$ where $j_1 > k_1$. Next, keeping $E_{k_1}$ fixed and repeating the process to

obtain

$$E_i = [[\cdots [[E_{j_n}, E_{k_n}], E_{k_{n-1}}], \cdots, E_{k_2}], E_{k_1}]$$

with

$$1 \leq k_n \leq j_n \leq M \quad \text{and} \quad k_{l+1} \leq k_l \quad \text{for} \quad 1 \leq l \leq n-1.$$

The maximal expansion of $E_i$ has $n$ commutations, and we write $d(i) = n$ and define $d(1) = \cdots = d(M) = 0$. One can also write a multi index $I(i) = (c_1^{(i)}, \cdots, c_v^{(i)})$ associated to $E_i$ where $c_s$ = cardinality of the set $\{t : k_t = s\}$. We note that $I(i) = (0, \cdots, 0)$ for $1 \leq i \leq M$. Then $E_i$ is the direct descendant of each $E_{j_l}$, that is, $j_l \prec i$.

For every pair $j \prec i$, we can define the polynomial $P_{j,i}$ by

$$P_{j,i} = \frac{-1^{(d(i)-d(j))}}{(I(i) - I(j))!} x^{(I(i)-I(j))}$$

where

$$x^{(I(i)-I(j))} = x_1^{c_1^{(i)}-c_1^{(j)}} x_2^{c_2^{(i)}-c_2^{(j)}} \ldots x_v^{c_v^{(i)}-c_v^{(j)}}.$$

Then the following result holds.

**Theorem 4.2.** *(Theorem 2.1, [GG90]) Let $r \geq 1, m \geq 2$, and let $v$ be the dimensions of $\mathfrak{g}_{M,r}$. The vector fields*

$$\begin{aligned}
E_1 &= \frac{\partial}{\partial x_1}, \\
E_2 &= \frac{\partial}{\partial x_2} + \sum_{j>2} P_{2,j} \frac{\partial}{\partial x_j}, \\
&\vdots \\
E_M &= \frac{\partial}{\partial x_M} + \sum_{j>M} P_{M,j} \frac{\partial}{\partial x_j},
\end{aligned} \quad (4.3)$$

*generates a Lie algebra isomorphic to $\mathfrak{g}_{M,r}$.*

In order to represent these generators in terms of $v$ independent copies of creation and annihilation operators satisfying (3.1), we first consider the substitutions

$$x_j \to A_j^* \quad \text{and} \quad \frac{\partial}{\partial x_j} \to A_j. \tag{4.4}$$

The expression

$$x^{(I(j)-I(l))} = x_1^{c_1^{(j)}-c_1^{(l)}} x_2^{c_2^{(j)}-c_2^{(l)}} \ldots x_v^{c_v^{(j)}-c_v^{(l)}}$$

in the construction can be written as

$$x^{(I(j)-I(l))} = (A_1^*)^{(c_1^{(j)}-c_1^{(l)})}(A_2^*)^{(c_2^{(j)}-c_2^{(l)})} \ldots (A_v^*)^{(c_v^{(j)}-c_v^{(l)})}.$$

Then, we state the following theorem.

**Theorem 4.3.** *Let $r \geq 1, m \geq 2$, and let $v$ be the dimensions of $\mathfrak{g}_{M,r}$. Consider $v$ independent copies of the operators $A_j, A_j^*$ such they satisfy* (3.1). *The vector fields*

$$\begin{aligned}
E_1 &= A_1, \\
E_2 &= A_2 + \sum_{j>2} \frac{-1^{(d(j)-d(2))}}{(I(j)-I(2))!}(A_1^*)^{(c_1^{(j)}-c_1^{(2)})}(A_2^*)^{(c_2^{(j)}-c_2^{(2)})} \ldots (A_v^*)^{(c_v^{(j)}-c_v^{(2)})} A_j, \\
&\vdots \\
E_M &= A_M + \sum_{j>M} \frac{-1^{(d(j)-d(M))}}{(I(j)-I(M))!}(A_1^*)^{(c_1^{(j)}-c_1^{(M)})}(A_2^*)^{(c_2^{(j)}-c_2^{(M)})} \ldots (A_v^*)^{(c_v^{(j)}-c_v^{(M)})} A_j,
\end{aligned} \tag{4.5}$$

*generates a Lie algebra isomorphic to $\mathfrak{g}_{M,r}$.*

*Proof.* The proof of classical case (Theorem 4.2) utilises the fact that $x_j$ and $\frac{\partial}{\partial x_k}$ satisfies CCR relations. Since the creation and annihilation operators considered satisfy the relations

$$[A_j, A_k^*] = \delta_{ik} \text{id}$$
$$[A_j, A_k] = 0 = [A_j^*, A_k^*],$$

a similar proof will follow. □

**Remark 4.1.** *The above model of representation is not unique, see Example 14.2.5, [BLU07].*

**Example 4.1.** *(Heisenberg algebra) First, we consider a simple example of Heinsenberg algebra where $[X, Y] = Z$. The classical fields are explicitly given by $X = \partial_{x_1} + 2x_2\partial_{x_3}$, $Y = \partial_{x_2} - 2x_1\partial_{x_3}$ and $Z = -4\partial_{x_3}$. We consider three independent copies of CCRs and write*

$$X = A_1 + 2A_2^*A_3,$$
$$Y = A_2 - 2A_1^*A_3, \quad (4.6)$$
$$Z = -4A_3.$$

*Then the following commutation relations*

$$[X, Y] = [A_1 + 2A_2^*A_3, A_2 - 2A_1^*A_3]$$
$$= [2A_2^*A_3, A_2 - 2A_1^*A_3] + [A_1, A_2 - 2A_1^*A_3]$$
$$= [2A_2^*A_3, A_2] + [A_1, -2A_1^*A_3] = -2A_3 - 2A_3$$
$$= -4A_3 = Z,$$
$$[X, Z] = [A_1 + 2A_2^*A_3, -4A_3]$$
$$= 0 = [Y, Z]$$

*holds.*

*Using the automorphism property, the modular dynamics for X, Y and Z with respect to the product state of quantum harmonic oscillator is given by*

$$\alpha_t(X) = e^{i\beta t}A_1 + A_2^*A_3, \quad \alpha_t(Y) = e^{i\beta t}A_2 - A_1^*A_3, \quad \alpha_t(Z) = 4e^{i\beta t}A_3.$$

*Then we can construct the Dirichlet form in the directions of generating fields X and Y as follows.*

**Theorem 4.4.** *The Dirichlet form in the directions of X and Y is given by*

$$\mathcal{E}(f) = \hat{\eta}(0)\langle \delta_{A_1}(f), \delta_{A_1}(f)\rangle_\omega + \hat{\eta}(\beta)\langle \delta_{A_2^*A_3}(f), \delta_{A_1}(f)\rangle_\omega + \hat{\eta}(-\beta)\langle \delta_{A_1}(f), \delta_{A_2^*A_3}(f)\rangle_\omega + \hat{\eta}(0)\langle \delta_{A_2^*A_3}(f), \delta_{A_2^*A_3}(f)\rangle_\omega$$
$$+ \hat{\eta}(0)\langle \delta_{A_2}(f), \delta_{A_2}(f)\rangle_\omega - \hat{\eta}(\beta)\langle \delta_{A_1^*A_3}(f), \delta_{A_2}(f)\rangle_\omega - \hat{\eta}(-\beta)\langle \delta_{A_2}(f), \delta_{A_1^*A_3}(f)\rangle_\omega + \hat{\eta}(0)\langle \delta_{A_1^*A_3}(f), \delta_{A_1^*A_3}(f)\rangle_\omega.$$

*on the domain where the right hand side is well defined.*

*Proof.* Using definitions in Section 2.5, we know that

$$\mathcal{E}(f) \equiv \int (\langle \delta_{\alpha_t(X)}(f), \delta_{\alpha_t(X)}(f) \rangle_\omega + \langle \delta_{\alpha_t(Y)}(f), \delta_{\alpha_t(Y)}(f) \rangle_\omega) \eta(t) dt.$$

First, we consider

$$\delta_{\alpha_t(X)}(f) = i[e^{i\beta t} A_1 + A_2^* A_3, f]$$

and hence

$$(\delta_{\alpha_t(X)}(f))^* = -i[f^*, e^{-i\beta t} A_1^* + A_3^* A_2].$$

Then we can compute

$$\begin{aligned}
\langle \delta_{\alpha_t(X)}(f), \delta_{\alpha_t(X)}(f) \rangle_\omega &= \frac{1}{Z} Tr(e^{-\beta \sum_{j\in\Lambda} N_j/2} (\delta_{\alpha_t(X)}(f))^* e^{-\beta \sum_{j\in\Lambda} N_j/2} \delta_{\alpha_t(X)}(f)) \\
&= \frac{1}{Z} Tr(e^{-\beta \sum_{j\in\Lambda} N_j/2} (-i[f^*, e^{-i\beta t} A_1^* + A_3^* A_2]) e^{-\beta \sum_{j\in\Lambda} N_j/2} (i[e^{i\beta t} A_1 + A_2^* A_3, f])) \\
&= \frac{1}{Z} Tr(e^{-\beta \sum_{j\in\Lambda} N_j/2} (-i[f^*, A_1^*]) e^{-\beta \sum_{j\in\Lambda} N_j/2} (i[A_1, f])) \\
&\quad + \frac{1}{Z} e^{i\beta t} Tr(e^{-\beta \sum_{j\in\Lambda} N_j/2} (-i[f^*, A_3^* A_2]) e^{-\beta \sum_{j\in\Lambda} N_j/2} (i[A_1, f])) \\
&\quad + \frac{1}{Z} e^{-i\beta t} Tr(e^{-\beta \sum_{j\in\Lambda} N_j/2} (-i[f^*, A_1^*]) e^{-\beta \sum_{j\in\Lambda} N_j/2} (i[A_2^* A_3, f])) \\
&\quad + \frac{1}{Z} Tr(e^{-\beta \sum_{j\in\Lambda} N_j/2} (-i[f^*, A_3^* A_2]) e^{-\beta \sum_{j\in\Lambda} N_j/2} (i[A_2^* A_3, f])) \\
&= \langle \delta_{A_1}(f), \delta_{A_1}(f) \rangle_\omega + e^{i\beta t} \langle \delta_{A_2^* A_3}(f), \delta_{A_1}(f) \rangle_\omega + e^{-i\beta t} \langle \delta_{A_1}(f), \delta_{A_2^* A_3}(f) \rangle_\omega \\
&\quad + \langle \delta_{A_2^* A_3}(f), \delta_{A_2^* A_3}(f) \rangle_\omega
\end{aligned}$$

Similarly we can write

$$\begin{aligned}
\langle \delta_{\alpha_t(Y)}(f), \delta_{\alpha_t(Y)}(f) \rangle_\omega &= \langle \delta_{A_2}(f), \delta_{A_2}(f) \rangle_\omega - e^{i\beta t} \langle \delta_{A_1^* A_3}(f), \delta_{A_2}(f) \rangle_\omega - e^{-i\beta t} \langle \delta_{A_2}(f), \delta_{A_1^* A_3}(f) \rangle_\omega \\
&\quad + \langle \delta_{A_1^* A_3}(f), \delta_{A_1^* A_3}(f) \rangle_\omega.
\end{aligned}$$

Multiplying the scalar products by $\eta(t)$ and integrating, we obtain the required result. □

**Example 4.2.** ($\mathfrak{g}_{4,2}$) *Consider a free nilpotent Lie algebra with 4 generators and nilpotent of step 2*

*such that the following commutation relations hold*

$$[E_2, E_1] = E_5, \quad [E_3, E_1] = E_6, \quad [E_4, E_1] = E_7,$$
$$[E_3, E_2] = E_8, \quad [E_4, E_2] = E_9, \quad [E_4, E_3] = E_{10}. \tag{4.7}$$

*Using the algorithm by [GG90], the generating vector fields can be written as*

$$E_1 = \frac{\partial}{\partial x_1},$$
$$E_2 = \frac{\partial}{\partial x_2} - x_1 \frac{\partial}{\partial x_5},$$
$$E_3 = \frac{\partial}{\partial x_3} - x_1 \frac{\partial}{\partial x_6} - x_2 \frac{\partial}{\partial x_8},$$
$$E_4 = \frac{\partial}{\partial x_4} - x_1 \frac{\partial}{\partial x_7} - x_2 \frac{\partial}{\partial x_9} - x_3 \frac{\partial}{\partial x_{10}}.$$

*For detailed explanation of the algorithm to obtain the above generators, one can refer Example 14.1.12, [BLU07]. We write these generators in terms of creation and annihilation operators using the substitutions* (4.4) *and obtain*

$$E_1 = A_1,$$
$$E_2 = A_2 - A_1^* A_5,$$
$$E_3 = A_3 - A_1^* A_6 - A_2^* A_8,$$
$$E_4 = A_4 - A_1^* A_7 - A_2^* A_9 - A_3^* A_{10}.$$

*Then using our assumption that pairs of $A_j$'s and $A_j^*$'s satisfy CCR relations, one can see that all the required commutation relations* (4.7) *are preserved.*

*Using the automorphism property, the modular dynamics for $E_1$, $E_2$, $E_3$ and $E_4$ with respect to the product state of quantum harmonic oscillator is given by*

$$\alpha_t(E_1) = e^{i\beta t} A_1, \quad \alpha_t(E_2) = e^{i\beta t} A_2 - A_1^* A_5,$$
$$\alpha_t(E_3) = e^{i\beta t} A_3 - A_1^* A_6 - A_2^* A_8, \quad \alpha_t(E_4) = e^{i\beta t} A_4 - A_1^* A_7 - A_2^* A_9 - A_3^* A_{10}.$$

*Then we can construct the Dirichlet form in the directions of generating fields as follows.*

**Theorem 4.5.** *The Dirichlet form in the directions of $E_1$, $E_2$, $E_3$ and $E_4$ is given by*

$$\mathcal{E}(f) = \hat{\eta}(0)\langle \delta_{A_1}(f), \delta_{A_1}(f)\rangle_\omega + \hat{\eta}(0)\langle \delta_{A_2}(f), \delta_{A_2}(f)\rangle_\omega - \hat{\eta}(\beta)\langle \delta_{A_1^*A_5}(f), \delta_{A_2}(f)\rangle_\omega - \hat{\eta}(-\beta)\langle \delta_{A_2}(f), \delta_{A_1^*A_5}(f)\rangle_\omega$$

$$+ \hat{\eta}(0)\langle \delta_{A_1^*A_5}(f), \delta_{A_1^*A_5}(f)\rangle_\omega + \hat{\eta}(0)\langle \delta_{A_3}(f), \delta_{A_3}(f)\rangle_\omega - \hat{\eta}(\beta)\langle \delta_{A_1^*A_6}(f), \delta_{A_3}(f)\rangle_\omega - \hat{\eta}(-\beta)\langle \delta_{A_3}(f), \delta_{A_1^*A_6}(f)\rangle_\omega$$

$$+ \hat{\eta}(0)\langle \delta_{A_1^*A_6}(f), \delta_{A_1^*A_6}(f)\rangle_\omega - \hat{\eta}(\beta)\langle \delta_{A_2^*A_8}(f), \delta_{A_3}(f)\rangle_\omega - \hat{\eta}(-\beta)\langle \delta_{A_3}(f), \delta_{A_2^*A_8}(f)\rangle_\omega + \hat{\eta}(0)\langle \delta_{A_2^*A_8}(f), \delta_{A_2^*A_8}(f)\rangle_\omega$$

$$+ \hat{\eta}(0)\langle \delta_{A_4}(f), \delta_{A_4}(f)\rangle_\omega - \hat{\eta}(\beta)\langle \delta_{A_1^*A_7}(f), \delta_{A_4}(f)\rangle_\omega - \hat{\eta}(-\beta)\langle \delta_{A_4}(f), \delta_{A_1^*A_7}(f)\rangle_\omega + \hat{\eta}(0)\langle \delta_{A_1^*A_7}(f), \delta_{A_1^*A_7}(f)\rangle_\omega +$$

$$- \hat{\eta}(\beta)\langle \delta_{A_2^*A_9}(f), \delta_{A_4}(f)\rangle_\omega - \hat{\eta}(-\beta)\langle \delta_{A_4}(f), \delta_{A_2^*A_9}(f)\rangle_\omega + \hat{\eta}(0)\langle \delta_{A_2^*A_9}(f), \delta_{A_2^*A_9}(f)\rangle_\omega - \hat{\eta}(\beta)\langle \delta_{A_3^*A_{10}}(f), \delta_{A_4}(f)\rangle_\omega$$

$$- \hat{\eta}(-\beta)\langle \delta_{A_4}(f), \delta_{A_3^*A_{10}}(f)\rangle_\omega + \hat{\eta}(0)\langle \delta_{A_3^*A_{10}}(f), \delta_{A_3^*A_{10}}(f).\rangle_\omega$$

*on the domain where the right hand side is well defined.*

*The computations are similar to the ones in previous example.*

Alternatively, one can consider the substitutions

$$x_j \to A_j, \quad \frac{\partial}{\partial x_j} \to -A_j^* \tag{4.8}$$

for which a similar result holds with generating fields given by

$$\begin{aligned}
\tilde{E}_1 &= -A_1^*, \\
\tilde{E}_2 &= -A_2^* - \sum_{2<j} \frac{-1^{(d(j)-d(2))}}{(I(j)-I(2))!} (A_1^{(c_1^{(j)}-c_1^{(2)})} A_2^{(c_2^{(j)}-c_2^{(2)})} \ldots A_v^{(c_d^{(j)}-c_d^{(2)})}) A_j^*, \\
&\vdots \\
\tilde{E}_M &= -A_M^* - \sum_{M<j} \frac{-1^{(d(j)-d(M))}}{(I(j)-I(M))!} (A_1^{(a_1^{(j)}-a_1^{(M)})} A_2^{(a_2^{(j)}-a_2^{(M)})} \ldots A_v^{(a_d^{(j)}-a_d^{(M)})}) A_j^*.
\end{aligned} \tag{4.9}$$

**Example 4.3.** *Again, for the classical fields $X = \partial_{x_1} + 2x_2\partial_{x_3}$, $Y = \partial_{x_2} - 2x_1\partial_{x_3}$ and $Z = -4\partial_{x_3}$, we consider three independent copies of CCR and using 4.8, we write*

$$X = -A_1^* - 2A_2A_3^*$$

$$Y = -A_2^* + 2A_1A_3^*$$

$$Z = 4A_3^*.$$

*Then the following commutation relations are preserved*

$$[X, Y] = [-A_1^* - 2A_2A_3^*, -A_2^* + 2A_1A_3^*] = 4A_3^* = Z$$

$$[X, Z] = [Y, Z] = 0.$$

Similarly, the preservation of commutation relations can be verified for Example 4.2.

Given the representations of the fields in terms of creation and annihilation operators, one can then write the inner derivations with respect to the fields. Since for any operators $A, B$,

$$[\delta_A, \delta_B] = i\delta_{[A,B]}, \tag{4.10}$$

the respective inner derivations would satisfy same commutation relations. For example, in the case of Heisenberg algebra, using 4.1, one can write the derivations as follows

$$\delta_{A_1+2A_2^*A_3}, \quad \delta_{A_2-2A_1^*A_3}, \quad i\delta_{-4A_3}.$$

This allows us to define the corresponding subgradient as follows

$$\nabla = (\delta_{E_1}, \ldots, \delta_{E_M})$$

which includes elements of the first strata of the stratified Lie algebra, $V_1$ since it includes the primary generating vectors of the algebra.

The noncommutative analogue of the Hörmander rank condition is satisfied if for any polynomial or any element of enveloping algebra say $f \in \mathcal{D}$, the condition

$$\delta_{E_j}(f) = 0$$

for every $E_j \in V_1$ implies that operator $f$ belongs to the centre of the algebra, $Z(\mathcal{D})$.

While the hypoellipticity theory in classical case is well developed, in noncommutative spaces this is more complicated since the tangent space possibly contains some non inner derivations.

**Example 4.4.** *(Heinsenberg algebra) For the fields given in 4.1 and 4.3, the generating derivations are $X, Y$ and $\delta_X(f), \delta_Y(f)$ vanishes for $f = Z$. Hence, the subgradient $\nabla$ vanishes only for $f = Z$ and $Z \in Z(\mathcal{D})$.*

For certain examples of Lie algebras, there is no guarantee that an algorithm exists to systematically provide representations in terms of creation and annihilation operators as in the case of free nilpotent lie algebras. However, we provide a few examples of representations that can be intuitively derived using the necessary commutation relations. We reiterate that these representations are not unique.

In the classification below, an algebra designated with $\mathfrak{n}$ is nilpotent, and the one with $\mathfrak{s}$ is solvable, but not nilpotent. The first subscript indicates the dimension, and the second index is for enumeration. So, $\mathfrak{s}_{4,11}$ is the 11th four-dimensional, solvable, non-nilpotent Lie algebra in the classification. We recall that a Lie algebra is called solvable if its derived series eventually terminates in the zero subalgebra. The derived series is constructed by iteratively taking commutators of the algebra with itself, [BLU07].

| Algebra | Commutation relations | Noncommutative representation |
|---|---|---|
| $\mathfrak{s}_{2,1}$ | $[e_2, e_1] = e_1,$ | $e_2 \sim i\delta_{A^*A},\ e_1 \sim i\delta_{A^*}$ |
| $\mathfrak{n}_{3,1}$ | $[e_2, e_3] = e_1$ | $e_2 \sim i\delta_{A_2 - \frac{1}{2}A_3^*A_1},\ e_3 \sim i\delta_{A_3 + \frac{1}{2}A_2^*A_1},\ e_1 \sim i\delta_{A_1}$ |
| $\mathfrak{s}_{3,1,\alpha}$ | $[e_3, e_1] = e_1, [e_3, e_2] = \alpha e_2,$ $0 < |\alpha| \le 1,\ if\ |\alpha| = 1\ then\ arg(\alpha) \le \pi$ | $e_3 \sim i\delta_{N_1 + \alpha N_2},\ e_2 \sim i\delta_{A_2^*},\ e_1 \sim i\delta_{A_1^*}$ |
| $\mathfrak{s}_{3,2}$ | $[e_3, e_1] = e_1, [e_3, e_2] = e_1 + e_2,$ | $e_3 \sim -i\delta_{N_1 + A_2},\ e_1 \sim -i\delta_{A_1^*},\ e_2 \sim -i\delta_{A_1^*A_2^*}$ |
| $\mathfrak{n}_{4,1}$ | $[e_2, e_4] = e_1, [e_3, e_4] = e_2$ | $e_1 \sim i\delta_{2A_2^*},\ e_2 \sim i\delta_{-2A_1^*A_2^*},\ e_3 \sim i\delta_{A_1^{*2}A_2^*},\ e_4 \sim i\delta_{A_1}$ |
| $\mathfrak{s}_{4,1}$ | $[e_4, e_2] = e_1, [e_4, e_3] = e_3$ | $e_1 \sim i\delta_{A_3},\ e_2 \sim i\delta_{A_2^*A_3},\ e_3 \sim i\delta_{A_1},\ e_4 \sim i\delta_{-N_1 + A_2}$ |
| $\mathfrak{s}_{4,3,\alpha,\beta}$ | $[e_4, e_1] = e_1, [e_4, e_2] = \alpha e_2, [e_4, e_3] = \beta e_3,$ $0 < |\alpha| \le |\alpha| \le 1, (\alpha, \beta) \ne (-1, -1)$ | $e_4 \sim i\delta_{N_1 + \alpha N_2 + \beta N_3},\ e_2 \sim i\delta_{A_2^*},\ e_1 \sim i\delta_{A_1^*},\ e_3 \sim i\delta_{A_3^*}$ |
| $\mathfrak{s}_{4,4,\alpha}$ | $[e_4, e_1] = e_1, [e_4, e_2] = e_1 + e_2, [e_4, e_3] = \alpha e_3,$ $\alpha \ne 0$ | $e_4 \sim i\delta_{N_1 + A_2 + \alpha N_3},\ e_2 \sim i\delta_{A_1^*A_2^*},\ e_1 \sim i\delta_{A_1^*},\ e_3 \sim i\delta_{A_3^*}$ |
| $\mathfrak{s}_{4,6}$ | $[e_2, e_3] = e_1, [e_4, e_2] = e_2, [e_4, e_3] = -e_3$ | $e_1 \sim i\delta_{A_4},\ e_2 \sim i\delta_{A_2 + A_3},\ e_3 \sim i\delta_{A_1 + A_3^*A_4},\ e_4 \sim i\delta_{N_1 - N_2}$ |
| $\mathfrak{s}_{4,11}$ | $[e_2, e_3] = e_1, [e_4, e_1] = e_1, [e_4, e_2] = e_2.$ | $e_1 \sim i\delta_{A_1^*},\ e_2 \sim i\delta_{A_2^*},\ e_3 \sim i\delta_{A_2 A_1^*},\ e_4 \sim i\delta_{N_1 + N_2}$ |

## 4.4 Chevalley-Serre relations

In this section, we discuss the foundational commutation relations known as Chevalley relations which occurs as a the generators of a simple complex Lie algebra.

We recall that a simple Lie algebra is a non-abelian Lie algebra $\mathfrak{g}$ that has no non-trivial ideals, where an ideal is a subspace $\mathfrak{a} \subseteq \mathfrak{g}$ such that for all $x \in \mathfrak{g}$ and $a \in \mathfrak{a}$, the commutator $[x, a] \in \mathfrak{a}$. This means the only ideals in $\mathfrak{g}$ are $\{0\}$ and $\mathfrak{g}$ itself. A semisimple Lie algebra is a Lie algebra $\mathfrak{g}$ that can be decomposed into a direct sum of simple Lie algebras. The Chevalley basis gives explicit construction of Lie algebra elements and their actions. It provides a standardized way to describe the Lie algebra, simplifying the comparison, classification, and study of different Lie algebras and their representations.

This systematic approach to organizing and describing the symmetries in Lie algebras is particularly useful in quantum mechanics research.

In case of general Lie algebras, the focus is on compact Lie algebras since they describe the symmetries of physical systems in quantum mechanics and their representation theory is well developed. Consider the operators $\{h_j, e_j, f_j\}_{j=1,\ldots,r}$ on a finite dimensional Hilbert space such that

$$
\begin{aligned}
&[h_i, h_j] = 0 \\
&[h_j, e_k] = \delta_{j,k} \lambda e_k \\
&[h_j, f_k] = -\delta_{j,k} \lambda f_k \\
&[e_j, f_k] = \delta_{jk} h_k \\
&ad(e_i)^{(1-\lambda)}(e_j) = 0 \\
&ad(f_i)^{(1-\lambda)}(f_j) = 0
\end{aligned}
\tag{4.11}
$$

where $\lambda \leq 0$ and the adjoint action notation $ad$ for elements $x$ and $y$ is defined as $ad(x)(y) = [x, y]$. Hence for any power $n$, $ad(x)^n(x) = [x, ad(x)^{n-1}(x)]$.

In the applications to theoretical physics, one can assume that $\lambda$ is real and $e_k^* = f_k$.

Since we assume a finite dimensional Hilbert space, the operators $\{h_j, e_j, f_j\}_{j=1,\ldots,r}$ are bounded. We construct and analyse the dissipative dynamics of some models of the systems involving Chevalley operators and creation/annihilation operators.

**Example 4.5.** *Consider the following Hamiltonian*

$$H_1 = \sum_j h_j.$$

*Using the relations* (4.11), *the corresponding modular dynamics of $e_k$ and $f_k$ associated to $H_1$ is given by*

$$\alpha_t(e_k) = e^{-i\beta t \lambda} e_k \quad \text{and} \quad \alpha_t(f_k) = e^{i\beta t \lambda} f_k$$

*and we have the following result.*

**Proposition 4.1.** *The Dirichlet form in the directions of $e_k$ and $f_k$ with respect to $H_1$ is given by*

$$\mathcal{E}(f) = \hat{\eta}(0) \sum_{k=1}^{r} \left( \langle \delta_{e_k}(f), \delta_{e_k}(f) \rangle_\omega + \langle \delta_{f_k}(f), \delta_{f_k}(f) \rangle_\omega \right). \tag{4.12}$$

*where $\hat{\eta}(0) = \int_\mathbb{R} \eta(t) e^{ist} dt$. This form has a dense domain such that the right hand side is well defined.*

*The computations of the Dirichlet form are similar to quantum harmonic oscillator. In this case, the Poincaré and Log Sobolev inequalities can be established; further elaboration will be provided in our forthcoming work [MZ24b].*

**Example 4.6.** *Suppose for the operators $e, f, h$, we have the following commutation relations*

$$[h, e] = \lambda e$$

$$[h, f] = -\lambda f$$

$$[e, f] = h.$$

*Assume $e^* = f$. Then we have $h^* = ([e, f])^* = f^* e^* - e^* f^* = ef - fe \equiv [e, f] = h$. Define a hamiltonian*

$$H_2 = N + A^* e + fA + h$$

*We have the following commutators*

$$\begin{aligned}
[e, H_2] &= [e, N + A^* e + fA + h] = hA - \lambda e \\
[f, H_2] &= [f, N + A^* e + fA + h] = -hA^* + \lambda f \\
[h, H_2] &= [h, N + A^* e + fA + h] = \lambda (eA^* - Af) \\
[A, H_2] &= [A, N + A^* e + fA + h] = A + e \\
[A^*, H_2] &= [A^*, N + A^* e + fA + h] = -A^* - f \\
[N, H_2] &= -(A^* + f)A + A^*(A + e) = eA^* - Af.
\end{aligned} \tag{4.13}$$

**Remark 4.2.** *In $\mathbb{L}_2(\omega)$, where $\omega$ is a state associated to $H_2$, for operators $F, G$ such that $\mathrm{ad}_{H_2}(F), \mathrm{ad}_{H_2}(G) \in$*

$\mathbb{L}_2(\omega)$, we have the following relationship

$$\langle [H_2, F], G \rangle_\omega = Tr(\rho^{\frac{1}{2}}([H_2, F])^* \rho^{\frac{1}{2}} G) = Tr(\rho^{\frac{1}{2}}(F)^* \rho^{\frac{1}{2}}[H_2, G]) = \langle F, [H_2, G] \rangle_\omega.$$

*Since the operator $\text{ad}_{H_2}$ is symmetric, it has a self adjoint extension. Then one can define a unitary group $\alpha_t$ in noncommutative $\mathbb{L}_2(\omega)$ space. Therefore, the modular dynamics exist, are well defined and hence can be used to write the Dirichlet form.*

*Formally, the modular dynamics would satisfy the following system of differential equations*

$$\begin{aligned}
\frac{d}{dt}\alpha_t(e) &= i\beta \left( \alpha_t(h)\alpha_t(A) - \lambda\alpha_t(e) \right) \\
\frac{d}{dt}\alpha_t(f) &= i\beta \left( -\alpha_t(h)\alpha_t(A^*) + \lambda\alpha_t(f) \right) \\
\frac{d}{dt}\alpha_t(A) &= i\beta \left( \alpha_t(A) + \alpha_t(e) \right) \\
\frac{d}{dt}\alpha_t(A^*) &= -i\beta \left( \alpha_t(A^*) + \alpha_t(f) \right) \\
\frac{d}{dt}\alpha_t(h) &= i\beta\lambda \left( \alpha_t(e)\alpha_t(A^*) - \alpha_t(A)\alpha_t(f) \right).
\end{aligned}$$

*Multiplying each differential equation by corresponding integrating factors, they can be rearranged as follows*

$$\begin{aligned}
\frac{d}{dt}e^{i\beta\lambda t}\alpha_t(e) &= i\beta e^{i\beta\lambda t}\alpha_t(h)\alpha_t(A) \\
\frac{d}{dt}e^{-i\beta\lambda t}\alpha_t(f) &= -i\beta e^{-i\beta\lambda t}\alpha_t(h)\alpha_t(A^*) \\
\frac{d}{dt}e^{-i\beta t}\alpha_t(A) &= i\beta e^{-i\beta t}\alpha_t(e) \\
\frac{d}{dt}e^{i\beta t}\alpha_t(A^*) &= -i\beta e^{i\beta t}\alpha_t(f) \\
\frac{d}{dt}\alpha_t(h) &= i\beta\lambda \left( \alpha_t(e)\alpha_t(A^*) - \alpha_t(A)\alpha_t(f) \right) = \lambda\frac{d}{dt}\alpha_t(N).
\end{aligned}$$

**Remark 4.3.** *We note that the last relation implies*

$$\frac{d}{dt}\alpha_t(\lambda N - h) = 0.$$

*This implies*

$$\alpha_t(e) = e^{-i\beta\lambda t}e + i\beta \int_0^t e^{-i\beta\lambda(t-s)}\alpha_s(h)\alpha_s(A)ds$$

$$\alpha_t(f) = e^{i\beta\lambda t}f - i\beta \int_0^t e^{i\beta\lambda(t-s)}\alpha_s(h)\alpha_s(A^*)ds$$

$$\alpha_t(A) = e^{i\beta t}A + i\beta \int_0^t e^{i\beta(t-s)}\alpha_s(e)ds$$

$$\alpha_t(A^*) = e^{-i\beta t}A^* - i\beta \int_0^t e^{i\beta(s-t)}\alpha_s(f)ds$$

$$\alpha_t(h) = h + i\beta\lambda \int_0^t (\alpha_s(e)\alpha_s(A^*) - \alpha_s(f)\alpha_s(A))\,ds.$$

*The Dirichlet form for this model is work in progress for our future work [MZ24b].*

**Remark 4.4.** *This model can then be extended involving more components of the operators* $\{h_j, e_j, f_j\}_{j=1,\ldots,r}$ *for the multicomponent Hamiltonian defined by*

$$H_2 = N + \sum_j e_j A^* + f_j A + \sum_j h_j$$

*where* $A, A^*$ *commute with* $\{h_j, e_j, f_j\}_{j=1,\ldots,r}$.

*Then, we have following commutation relations*

$$[e_k, H_2] = [e_k, N + \sum_j e_j A^* + f_j A + \sum_j h_j] = [e_k, \sum_j f_j A + \sum_j h_j] = h_k A - \lambda e_k$$

$$[f_k, H_2] = [f_k, N + \sum_j e_j A^* + f_j A + \sum_j h_j] = -h_k A^* + \lambda f_k$$

$$[A, H_2] = [A, N + \sum_j e_j A^* + f_j A + \sum_j h_j] = A + \sum_j e_j$$

$$[A^*, H_2] = [A^*, N + \sum_j e_j A^* + f_j A + \sum_j h_j] = -A^* - \sum_j f_j.$$

**Remark 4.5.** *We note that the third and sixth relations in (4.13) imply* $[\lambda N - h, H_2] = 0$.

*One can consider a Hamiltonian*

$$H_3 = \lambda N - h.$$

*Then the corresponding modular dynamics is given by*

$$\alpha_t(e) = e^{i\beta t\lambda}e, \quad \alpha_t(f) = e^{-i\beta t\lambda}f, \quad \alpha_t(A) = e^{i\beta t\lambda}A, \quad \alpha_t(A^*) = e^{-i\beta t\lambda}A^*.$$

*The corresponding Dirichlet form with respect to the Hamiltonian $H_3$ would be*

$$\mathcal{E}(f) = \hat{\eta}(0)\left(\langle \delta_e(f), \delta_e(f)\rangle_\omega + \langle \delta_f(f), \delta_f(f)\rangle_\omega + \langle \delta_A(f), \delta_A(f)\rangle_\omega + +\langle \delta_{A^*}(f), \delta_{A^*}(f)\rangle_\omega\right)$$

*defined for operators such that the right hand side is well defined.*

# Chapter 5

# Multivariate Graph supOU Processes

## 5.1 Introduction

In the noncommutative setting, the theory of Lévy processes has been significantly developed [Fra04]. In this chapter, we analyse a statistical model that is of independent interest but holds potential for generalization within the noncommutative framework. We approach this as a standalone chapter, since the setting and the questions addressed here are quite different to those of the preceding chapters. In the chapter, we continue to study generalisations of Ornstein-Uhlenbeck (OU) process but in the classical setup. Here instead of analysing the dynamics of OU processes in quantum setup, we focus on the application of classical OU process for dynamic networks. Since the considered process is multivariate, we utilise the theory of matrices which continue to follow various noncommutative properties similar to the previous chapters in this thesis.

Lévy-driven Ornstein-Uhlenbeck process are widely studied class of continuous-time models [Bar01] with numerous applications in finance, volatility modelling, neuroscience and electricity management. Such a mean-reverting process on a dynamic undirected graph known as Graph Ornstein-Uhlenbeck (GrOU) process has been formulated in [CV22]. This process provides the flexibility of continuous-time models and the sparsity of graphical models. The discrete time counterparts of the models on networks including Autoregressive models on Network (NAR) and Generalised NAR (GNAR) and GNAR-edge models have been extensively discussed in the recent works

[JLY23, Man+23].

The Lévy driven OU processes has exponentially decaying autocorrelation function. The presence of long-range dependence (or long memory) in a statistical model provides more flexibility and can potentially improve forecasting accuracy. Since these models consider a longer history of past values, they can better account for underlying trends and dependencies, leading to more accurate and reliable predictions. A generalisation of the Lévy driven OU process introduced in [BS11], [Bar01] called the superpositions of the OU processes (supOU) involves adding up independent OU processes. Such a generalisation allows for the process to exhibit long memory.

We extend the Graph OU model defined in [CV22] for the multivariate supOU processes. Since the modelling of the data requires estimation procedures, we utilise the Generalised Method of Moments (GMM) due to the unknown density of supOU processes. We perform the simulation study for the multivariate Graph supOU to include the possibility of long(er) memory. Note that we say long(er) memory instead of long memory since our technique for the proof of asymptotic normality for the GMM estimation of multivariate Graph supOU processes can only handle the short memory case, similar to [CS18]. We prove that a similar result is true in the case of multivariate Graph supOU processes.

Additionally, we provide a simulation study to estimate the parameters of the multivariate Graph supOU processes using a two step iterated GMM estimator. The moment based estimation of univariate supOU was introduced in [STW15]. On similar lines, we provide a novel algorithm for simulating multivariate supOU processes.

This chapter is organised as follows. In Section 5.2, we provide necessary background and definitions of the OU and supOU processes. In Section 5.3, we introduce the multivariate Graph supOU process with a case for long memory and a corresponding GMM estimator. Next, we prove the consistency and Central Limit theorem for multivariate Graph supOU process in Section 5.4 ending this chapter with the simulation study of these processes.

The simulation codes are written in Python and can be accessed through the github repository `https://github.com/shreyamehta31/Multivariate_Graph_supOU`.

## 5.2 Background and Preliminaries

### 5.2.1 Notation

We consider a filtered probability space $(\Omega, \mathcal{F}, P)$ endowed with a filtration $(\mathcal{F}_t, t \in \mathbb{R})$. The Lévy process, denoted as $(\mathbb{L}_t, t \in \mathbb{R})$, is a stochastic process possessing properties of stationary and independent increments, along with continuous probability distribution and càdlàg property.

We define various sets and operations related to matrices as follows. Let $M_{d,k}(\mathbb{R})$ represent the set of real $d \times k$ matrices. When $k = d$, we denote this set as $M_d(\mathbb{R})$. The linear subspace of $d \times d$ symmetric matrices is denoted by $\mathbb{S}_d$, the closed positive cone of symmetric matrices with non-negative real parts of their eigenvalues is denoted as $\mathbb{S}_d^+$, and the open positive definite cone of symmetric matrices with strictly positive real parts of eigenvalues is denoted as $\mathbb{S}_d^{++}$. The identity matrix of size $d \times d$ is represented by $\mathbf{I}_d$. The elements of a matrix $A \in M_{d,k}(\mathbb{R})$ is denoted by $A_{ij}$ and the adjoint is given by $A^*$. Moreover, the spectrum of a matrix consisting of all eigenvalues is denoted by $\sigma(\cdot)$.

Next, we define
$$M_d^- := \{X \in M_d(\mathbb{R}) : \sigma(X) \subset (-\infty, 0) + i\mathbb{R}\}$$

as the set of square $d \times d$ matrices with negative real parts of their eigenvalues and $\mathcal{B}_b(M_d^- \times \mathbb{R})$ to be the collection of bounded Borel sets of $M_d^- \times \mathbb{R}$.

In this chapter, we utilise the Kronecker (tensor) product of two matrices $A \in M_{d,n}(\mathbb{R})$ and $B$, denoted as $A \otimes B$. The vectorization transformation, which stacks the columns of a $d \times d$ matrix into a vector in $\mathbb{R}^{d^2}$, is represented as vec. Additionally, the half vectorization, which transforms the upper or lower triangular elements of a matrix into a vector by stacking the columns, is denoted as $\text{vec}_h$.

The norm of vectors or matrices are denoted by $\|\cdot\|$. The considered norm has no influence on the results since all the norms are equivalent. Although, it can be considered Euclidean norm or induced operator norm. The operator $\mathbf{1}_T$ stands for the indicator function of a set $T$.

Let $(X, \mathcal{H})$ and $(Y, \mathcal{G})$ be measurable spaces, where $\mathcal{H}$ and $\mathcal{G}$ are $\sigma$-algebras on sets $X$ and $Y$ respectively. A function $f : X \to Y$ is measurable if for every measurable set $B \in \mathcal{G}$, the set $\{x \in X : f(x) \in B\}$ is in $\mathcal{H}$. The Borel $\sigma$-algebras are denoted as $\mathcal{B}(\cdot)$ which is the smallest $\sigma$-algebra containing all open sets and let $\lambda$ denote the Lebesgue measure.

We will encounter the double integral $\int_A \int_B f(x,y) m(dx, dy)$, which represents the integral of the function $f$ over the set $A$ with respect to the variable $x$, and over the set $B$ with respect to the variable $y$, see Fubini's theory [Fub07].

### 5.2.2 The Lévy driven Ornstein-Uhlenbeck Process

We first define the d-dimensional OU process $\mathbb{X}_t = (X_t^{(1)}, \ldots, X_t^{(d)})^T$ for $t \geq 0$ satisfying the stochastic differential equation(SDE) for a dynamics matrix $\mathbf{Q} \in M_d(\mathbb{R})$,

$$d\mathbb{X}_t = \mathbf{Q}\mathbb{X}_{t-}dt + d\mathbb{L}_t, \tag{5.1}$$

for a $d-$dimensional Lévy process $\mathbb{L}_t = (L_t^{(1)}, \ldots, L_t^{(d)})^T$, where $\mathbb{X}_{t-} := \lim_{s\uparrow t} \mathbb{X}_s$ for any $t \in \mathbb{R}$. For the theory of stochastic differential equations, see [Øks03].

The Lévy process $\mathbb{L}$ is defined by the Lévy-Khintchine characteristic triplet $(\gamma, \Sigma, \nu)$ where $\gamma \in \mathbb{R}^d$, $\Sigma \in \mathbb{S}_d^+$, and $\nu$ is a Lévy measure on $\mathbb{R}^d$. The Lévy Khintchine representation for $\mathbb{L}_t$ is given as

$$\mathrm{E}\left(\exp(iu^*\mathbb{L}_t)\right) = \exp\left(t\left(iu^*\gamma - \frac{1}{2}u^*\Sigma u + \int_{\mathbb{R}^d} \left(e^{iu^*x} - 1 - iu^*x\mathbf{1}_{[0,1]}(\|x\|)\right) d\nu(x)\right)\right)$$

for all $u \in \mathbb{R}^d$, $t \in \mathbb{R}$ and $\nu$ is a Lévy measure on $\mathbb{R}^d$ such that $\int_{\mathbb{R}^d}(1 \wedge \|x\|^2)\nu(dx) < \infty$.

If $\mathrm{E}(\ln(\|\mathbb{L}_1\| \vee 1)) < \infty$ and $\mathbf{Q} \in M_d^-$, that is, all eigenvalues of $\mathbf{Q}$ have strictly negative real parts, we obtain the solution to (5.1)

$$\mathbb{X}_t = \int_{-\infty}^{t} e^{\mathbf{Q}(t-s)} d\mathbb{L}_s,$$

### 5.2.3 The Ornstein-Uhlenbeck Process on a Graph

We briefly recall the construction of Graph OU process. For the detailed description of the process and the estimation theory, we refer to [CV22]. The elements within the process $\mathbb{X}$ are understood as the vertices of a graph, connected to each other by a set of edges. These edges are represented by the adjacency matrix $\mathbf{A} = (a_{ij})$, where $a_{ii} = 0$ for $i \in \{1, \ldots, d\}$, and for $i \neq j$, $a_{ij} = 1$ for an edge between $i$ and $j$ and 0 otherwise. In what follows, we consider undirected graphs where $A$ is assumed to be symmetric. Additionally, $A$ is predetermined and constant in time.

We write the matrix $\bar{\mathbf{A}}$ as the row normalised adjacency matrix given by

$$\bar{\mathbf{A}} := diag(n_1^{-1}, \ldots, n_d^{-1})\mathbf{A}.$$

where $n_i := 1 \vee \sum_{j \neq i} a_{ij}$.

For two parameters $\theta = (\theta_1, \theta_2)^T \in \mathbb{R}^2$, define the matrix

$$\mathbf{Q}(\theta) = -\left(\theta_2 \mathbf{I}_{d \times d} + \theta_1 \bar{\mathbf{A}}\right). \tag{5.2}$$

The $\theta$ formulation of the GrOU process can be expressed as the solution of the SDE

$$d\mathbb{X}_t = \mathbf{Q}(\theta)\mathbb{X}_{t-}dt + d\mathbb{L}_t. \tag{5.3}$$

The set of two dimensional vectors $\theta = (\theta_1, \theta_2)^T$ where $\theta_1$ and $\theta_2$ represents the network and moment effect respectively and is such that

$$\theta = \left\{(\theta_1, \theta_2)^T : \theta_2 > 0, \theta_2 > |\theta_1|\right\}.$$

These conditions ensure $\mathbf{Q}(\theta) \in M_d^-$ and along with the condition $E(\ln(\|\mathbb{L}_1\| \vee 1)) < \infty$ the solution to (5.3) exists.

**Proposition 5.1.** *If $\theta_2 > 0, \theta_2 > |\theta_1|$, then $\mathbf{Q}(\theta) \in M_d^-$.*

*Proof.* We aim to show that all eigenvalues of $\mathbf{Q}(\theta)$ are strictly negative. By Geršgorin's circle theorem [Ger31], any eigenvalue of $\mathbf{Q}(\theta)$ lies within at least one Geršgorin disc. For the $i$-th row of $\mathbf{Q}(\theta)$, the center of the Geršgorin disc is the diagonal entry $Q_{ii}$, which is $-\theta_2$, and the radius is the sum of the absolute values of the off-diagonal entries in the $i$-th row. Since $\bar{A}$ is row-normalized, the sum of the absolute values of the non-diagonal entries in any row of $\bar{A}$ is 1. Thus, the radius of the Geršgorin disc for the $i$-th row is given by

$$\sum_{j \neq i} |Q_{ij}| = |\theta_1| \sum_{j \neq i} |\bar{A}_{ij}| = |\theta_1|.$$

Therefore, each Geršgorin disc for $Q(\theta)$ has a center at $-\theta_2$ and a radius $|\theta_1|$. Since $\theta_2 > |\theta_1|$, the disc centered at $-\theta_2$ with radius $|\theta_1|$ lies entirely in the left half-plane of the complex plane and does not intersect the imaginary axis.

Consequently, all eigenvalues of $\mathbf{Q}(\theta)$ must lie within this disc, which is entirely in the negative half-plane and does not include the origin. Thus, all eigenvalues are strictly negative. □

The $i$th component of (5.3) satisfies

$$dX_t^{(i)} = -\theta_2 X_{t-}^{(i)} dt - \theta_1 n_i^{-1} \sum_{j \neq i} a_{ij} X_{t-}^{(j)} dt + dL_t^{(i)}, \quad t \geq 0$$

which shows that the parameter $\theta_2$ acts on the $i$th node which measures the effect of the momentum of this node. On the other hand, the $\theta_1$ parameter corresponds to the effect of the neighbours of the $i$th node.

**Remark 5.1.** *The adjacency matrix A is defined for an undirected graph and hence, it is symmetric . However, this symmetry does not automatically imply that the matrix $\bar{A}$, and consequently the matrix $\mathbf{Q}$ is not symmetric. To ensure that $\mathbf{Q}$ is symmetric, the adjacency matrix A can be normalised in an alternative way [GB21]. We can write the scaled matrix $\bar{\bar{A}}$ as follows*

$$\bar{\bar{A}} = D^{-1/2} A^T D^{-1/2}$$

*where D is a d-dimensional normalising diagonal matrix with the ith diagonal entry defined as $\sum_{j=1}^n a_{ij} + \sum_{j=1}^n a_{ji}$ which is the sum of the in- and out- degrees of the ith vectex in the graph. This scaling would imply $\mathbf{Q} \in \mathbb{S}_d^+$. If A is assumed to be symmetric, then $A = A^T$ and $a_{ij} = a_{ji}$ and hence the ith diagonal entry of D would be $2 \sum_{j=1}^n a_{ij}$.*

For an illustration, consider the network in the figure below which includes four nodes.

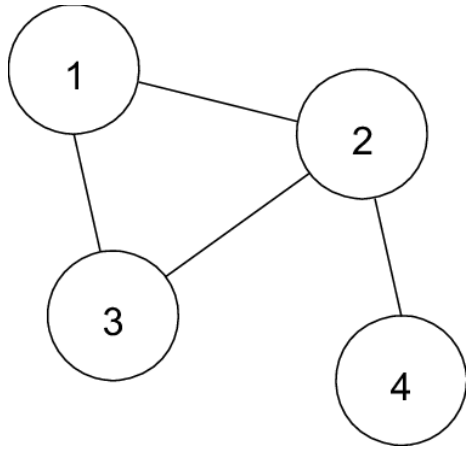

The adjacency matrix $A$ for this network is given by

$$A = \begin{bmatrix} 0 & 1 & 1 & 0 \\ 1 & 0 & 1 & 1 \\ 1 & 1 & 0 & 0 \\ 0 & 1 & 0 & 0 \end{bmatrix}.$$

We note that $n_1 = 2, n_2 = 3, n_3 = 2, n_4 = 1$. Then for the underlying Lévy process given by $\mathbb{L}_t = (L_t^{(1)}, L_t^{(2)}, L_t^{(3)}, L_t^{(4)})^T$, each component of the corresponding GrOU process would be the solution to each of the following equations

$$dX_t^{(1)} = -\theta_2 X_{t-}^{(1)} dt - \left(\frac{\theta_1}{2}\right)(X_{t-}^{(2)} + X_{t-}^{(3)})dt + dL_t^{(1)}, \quad t \geq 0$$

$$dX_t^{(2)} = -\theta_2 X_{t-}^{(2)} dt - \left(\frac{\theta_1}{3}\right)(X_{t-}^{(1)} + X_{t-}^{(3)} + +X_{t-}^{(4)})dt + dL_t^{(2)}, \quad t \geq 0$$

$$dX_t^{(3)} = -\theta_2 X_{t-}^{(3)} dt - \left(\frac{\theta_1}{2}\right)(X_{t-}^{(1)} + X_{t-}^{(2)})dt + dL_t^{(3)}, \quad t \geq 0$$

$$dX_t^{(4)} = -\theta_2 X_{t-}^{(4)} dt - \theta_1 X_{t-}^{(2)} dt + dL_t^{(4)}, \quad t \geq 0.$$

### 5.2.4 Lévy Bases

To define multivariate supOU processes with a random mean reversion parameter $\mathbf{Q}$, where $\mathbf{Q} \in M_d^-$, we generalise the driving Lévy process to Lévy bases. To this end, we briefly review the theory of Lévy bases. A $d$-dimensional Lévy process can be interpreted as an $\mathbb{R}^d$-valued random measure on the real numbers. For a $d$-dimensional Lévy process $\mathbb{L} = (\mathbb{L}_t)_{t \in \mathbb{R}}$, the measure over the set $(a, b]$ is given by $\mathbb{L}_b - \mathbb{L}_a$ for all $a, b \in \mathbb{R}, a < b$.

The Lévy bases is the infinitely divisible independently scattered random variables(i.d.i.s.r.m) defined as follows.

**Definition 5.1.** *A collection $\Lambda = \{\Lambda(B) : B \in \mathcal{B}_b(M_d^- \times \mathbb{R})\}$ of random variables taking values in $\mathbb{R}^d$ is defined as an $\mathbb{R}^d$-valued Lévy basis on $M_d^- \times \mathbb{R}$ if:*

1. *the probability distribution of $\Lambda(B)$ is infinitely divisible for every $B \in \mathcal{B}_b(M_d^- \times \mathbb{R})$,*

2. *for any natural number $l$ and pairwise disjoint sets $B_1, \ldots, B_l \in \mathcal{B}_b(M_d^- \times \mathbb{R})$, the random variables $\Lambda(B_1), \ldots, \Lambda(B_l)$ are independent, and*

3. *for any pairwise disjoint sets $B_i \in \mathcal{B}_b(M_d^- \times \mathbb{R})$ for $i \in \mathbb{N}$, where $\cup_{l \in \mathbb{N}} B_l \in \mathcal{B}_b(M_d^- \times \mathbb{R})$, the series $\sum_{l=1}^{\infty} \Lambda(B_l)$ converges almost surely, and $\Lambda(\cup_{l \in \mathbb{N}} B_l) = \sum_{l=1}^{\infty} \Lambda(B_l)$ almost surely.*

In the realm of supOU processes, emphasis is placed on Lévy bases, which exhibit homogeneity in time and can be decomposed into the effects of an underlying infinitely divisible distribution and a probability distribution on $M_d^-$. The characteristic function of such homogeneous Lévy bases is expressed by the Lévy Khintchine representation:

$$E(\exp(iu\Lambda(B))) = \exp(\phi(u)\Pi(B)) \qquad (5.4)$$

for all $u \in \mathbb{R}^d$ and $B \in \mathcal{B}_b(M_d^-(\mathbb{R}) \times \mathbb{R})$, where $\Pi = \pi \times \lambda$ is the product of a probability measure $\pi$ on $M_d^-(\mathbb{R})$ and the Lebesgue measure $\lambda$ on $\mathbb{R}$.

The function

$$\phi(u) = iu^*\gamma - \frac{1}{2}u^*\Sigma u + \int_{\mathbb{R}^d} \left(e^{iu^*x} - 1 - iu^*x\mathbf{1}_{[0,1]}(\|x\|)\right) \nu(dx)$$

represents the cumulant transform of an infinitely divisible distribution on $\mathbb{R}^d$ with Lévy Khintchine triplet $(\gamma, \Sigma, \nu)$, where $\gamma \in \mathbb{R}^d$, $\Sigma \in \mathbb{S}_d^+$, and $\nu$ is a Lévy measure. As described in [BS11], the distribution of the Lévy bases is fully determined by the "generating quadruple" $(\gamma, \Sigma, \nu, \pi)$.

We then define the Lévy bases $\mathbb{L}$ as

$$\mathbb{L}_t = \Lambda(M_d^- \times (0, t]) \quad \mathbb{L}_{-t} = \Lambda(M_d^- \times (-t, 0)) \quad t \in \mathbb{R}^+,$$

which possess the characteristic triplet $(\gamma, \Sigma, \nu)$, thus being termed "the underlying Lévy process".

Furthermore, Proposition 2.3 in [BS11] asserts that for an $\mathbb{R}^d$-valued Lévy basis with the aforementioned characteristic function, and a measurable function $f : M_d^- \times \mathbb{R} \to M_d(\mathbb{R})$, $f$ is $\Lambda$-integrable with respect to $\Lambda$ if and only if the following conditions are satisfied

$$\int_{M_d^-} \int_{\mathbb{R}} \left\| f(\mathbf{Q}, s)\gamma + \int_{\mathbb{R}^d} f(\mathbf{Q}, s)x(\mathbf{1}_{[0,1]}(\|f(\mathbf{Q}, s)x\|) - \mathbf{1}_{[0,1]}(\|x\|))\nu(dx) \right\| ds\pi(d\mathbf{Q}) < \infty, \tag{5.5}$$

$$\int_{M_d^-} \int_{\mathbb{R}} \|f(\mathbf{Q}, s)\Sigma f(\mathbf{Q}, s)^*\| ds\pi(d\mathbf{Q}) < \infty, \tag{5.6}$$

$$\int_{M_d^-} \int_{\mathbb{R}} \int_{\mathbb{R}^d} (1 \wedge \|f(\mathbf{Q}, s)x\|^2)\nu(dx)ds\pi(d\mathbf{Q}) < \infty. \tag{5.7}$$

If $f$ is $\Lambda$ integrable, then the distribution of the integral $\int_{M_d^-} \int_{\mathbb{R}^+} f(\mathbf{Q}, s)\Lambda(d\mathbf{Q}, ds)$ is infinitely divisible, with a characteristic function given by

$$\mathrm{E}\left(\exp\left(iu^* \int_{M_d^-} \int_{\mathbb{R}} f(\mathbf{Q}, s)\Lambda(d\mathbf{Q}, ds)\right)\right) = \exp\left(\int_{M_d^-} \int_{\mathbb{R}} \phi(f(\mathbf{Q}, s)^*u)ds\pi(d\mathbf{Q})\right).$$

Note that the integral over $s$ would be over $\mathbb{R}^+$ instead of $\mathbb{R}$ in the case of supOU processes defined in the next section.

### 5.2.5 Univariate and Multivariate Superposition of OU Process

In simple terms, superposition of OU(supOU) process are created by combining independent Ornstein-Uhlenbeck processes, each having its own mean reversion coefficient $\mathbf{Q}$.

We now give the construction of multivariate supOU processes as introduced in [BS11].

**Theorem 5.1.** *(Theorem 3.1, [BS11]) Let $\Lambda$ be an $\mathbb{R}^d$-valued Lévy basis on $M_d^- \times \mathbb{R}$ with generating quadruple $(\gamma, \Sigma, \nu, \phi)$ satisfying*

$$\int_{\|x\|>1} \ln(\|x\|)\nu(dx) < \infty \tag{5.8}$$

*and assume there exist measurable functions $\rho : M_d^- \to \mathbb{R}^+ \setminus \{0\}$ and $\kappa : M_d^- \to [1, \infty)$ such that*

$$\|e^{\mathbf{Q}s}\| \le \kappa(\mathbf{Q})e^{-\rho(\mathbf{Q})s} \quad \forall s \in \mathbb{R}^+, \pi - \text{almost surely}, \tag{5.9}$$

and

$$\int_{M_d^-} \frac{\kappa(\mathbf{Q})^2}{\rho(\mathbf{Q})} \pi(d\mathbf{Q}) < \infty. \tag{5.10}$$

Then the supOU process $(\mathbb{X}_t)_{t \in \mathbb{R}}$ is given by

$$\mathbb{X}_t = \int_{M_d^-} \int_{-\infty}^t e^{\mathbf{Q}(t-s)} \Lambda(d\mathbf{Q}, ds) \tag{5.11}$$

and is well defined for all $t \in \mathbb{R}$.

**Remark 5.2.** *In the univariate setting [Bar01], the definition simplifies in the following way. Given a real valued Lévy basis on $\mathbb{R} \times \mathbb{R}$ with generating quadruple $(\gamma, \Sigma, \nu, \phi)$ such that it satisfies*

$$\int_{|x|>1} \ln(|x|) \nu(dx) < \infty \quad \text{and} \quad \int_{-\infty}^0 \frac{1}{(-Q)} \pi(dQ) < \infty$$

*where $Q = \mathbf{Q} \in (-\infty, 0)$.*

*The supOU process $(X_t)_{t \in \mathbb{R}}$ is given by*

$$\mathbb{X}_t = \int_{\mathbb{R}} \int_{-\infty}^t e^{Q(t-s)} \Lambda(dQ, ds).$$

Furthermore, the necessary and sufficient condition for the supOU process $\mathbb{X}$ to have finite $r$th moments for $r \in (0, 2]$, that is, $E(\|\mathbb{X}_t\|^r) < \infty$ is

$$\int_{\|x\|>1} (\|x\|^r) \nu(dx) < \infty. \tag{5.12}$$

Hence, the moments for the supOU process is given by the following result.

**Theorem 5.2.** *(Theorem 3.11,[BS11]) The first and second moments of a stationary Lévy driven d-dimensional supOU exists given $\int_{\mathbb{R}^d} \|x\|^2 \nu(dx) < \infty$ holds. Then we have*

$$E(\mathbb{X}_0) = -\int_{M_d^-} \mathbf{Q}^{-1} \left(\gamma + \int_{|x|>1} x\nu(dx)\right) \pi(d\mathbf{Q}), \tag{5.13}$$

$$\text{var}(\mathbb{X}_0) = -\int_{M_d^-} (\mathcal{A}(\mathbf{Q}))^{-1} \left(\Sigma + \int_{\mathbb{R}^d} xx^* \nu(dx)\right) \pi(d\mathbf{Q}), \tag{5.14}$$

$$\text{cov}(\mathbb{X}_h, \mathbb{X}_0) = -\int_{M_d^-} e^{\mathbf{Q}h}(\mathcal{A}(\mathbf{Q}))^{-1}\left(\Sigma + \int_{\mathbb{R}^d} xx^*\nu(dx)\right)\pi(d\mathbf{Q}). \tag{5.15}$$

where $\mathcal{A}(\mathbf{Q}) : M_d(\mathbb{R}) \to M_d(\mathbb{R}), X \to \mathbf{Q}X + X\mathbf{Q}^*$, $\gamma + \int_{|x|>1} x\nu(dx) = \mu_{\mathbb{L}} = E(\mathbb{L}_1)$, $\Sigma + \int_{\mathbb{R}^d} xx^*\nu(dx) = \sigma_{\mathbb{L}}^2 = \text{var}(\mathbb{L}_1)$ and $h \in \mathbb{R}^+$.

## 5.3 Multivariate supOU Process on a Graph

Using the definitions of Graph OU and Multivariate supOU in previous sections, we define the Multivariate Graph supOU process with parameter $\theta = (\theta_1, \theta_2)^T \in \mathbb{R}^2$ in the following way

$$\mathbb{X}_t = \int_{M_d^-} \int_{-\infty}^t e^{\mathbf{Q}(\theta)(t-s)} \Lambda(d\mathbf{Q}(\theta), ds)$$

where $\Lambda$ is an $\mathbb{R}^d$-valued Lévy basis on $M_d^- \times \mathbb{R}$ with generating quadruple $(\gamma, \Sigma, \nu, \phi)$ and the matrix $\mathbf{Q}$ is given as

$$\mathbf{Q}(\theta) = -\left(\theta_2 \mathbf{I}_{d \times d} + \theta_1 \bar{\mathbf{A}}\right). \tag{5.16}$$

### 5.3.1 Specific Case for Possible Long Memory

When discussing the Graph supOU process in this section, we are establishing a novel model for its tendency to exhibit long memory effects. Here, long range dependence refers to the behavior where at least one part of the autocovariance function gradually decreases like $h^\alpha$, where $h$ represents the lag approaching infinity and $\alpha \in (0, 1)$.

Adapting Example 3.1 from [BS11], we assume that the mean reversion parameter $\mathbf{Q}$ is Gamma distributed. We consider the parametrisation

$$\theta_1 = c\theta_2, \quad \text{for} \quad \theta_2 > 0, \quad c \in \mathbb{R} \quad \text{with} \quad |c| < 1.$$

Then we can write

$$\mathbf{Q}(\theta) = -(\theta_2 \mathbf{I}_{d \times d} + \theta_1 \bar{\mathbf{A}}) = -\theta_2 \mathbf{I}_{d \times d} - c\theta_2 \bar{\mathbf{A}} = \theta_2(-\mathbf{I}_{d \times d} - c\bar{\mathbf{A}}) = \mathbf{Q}(c, \theta_2),$$

where $\theta_2 > |\theta_1| = |c|\theta_2$.

As a first attempt, we keep the parameter $c$ fixed and only randomise the parameter $\theta_2$. Let $\theta_2 \sim \Gamma(\alpha, \beta)$ with

$$\pi(d\theta_2) = \frac{\beta^\alpha}{\Gamma(\alpha)} \theta_2^{\alpha-1} e^{-\beta\theta_2} \mathbf{1}_{(0,\infty)}(\theta_2) d\theta_2, \quad (5.17)$$

where $\alpha > 1, \beta \in \mathbb{R}^+ \setminus \{0\}$.

Since c is fixed for now, we consider

$$\mathbf{Q}(\theta) = \theta_2(-\mathbf{I}_{d \times d} - c\bar{\mathbf{A}}) = \theta_2 K,$$

where $K = (-\mathbf{I}_{d \times d} - c\bar{\mathbf{A}}) \in M_d^-$ and $\theta_2$ has Gamma distribution (5.17).

The conditions (5.12) for the existence of supOU processes and finite moments can be written as

$$-\int_{M_d^-} \frac{1}{\max(\Re(\sigma(\mathbf{Q})))} \pi(d\mathbf{Q}) < \infty, \quad \int_{\|x\|>1} (\|x\|^r) \nu(dx) < \infty,$$

see Remark 3.10, [BS11]. Using the distribution (5.17), consider

$$-\int_{M_d^-} \frac{1}{\max(\Re(\sigma(\mathbf{Q})))} \pi(d\mathbf{Q}) = \frac{-\beta^\alpha}{\max(\Re(\sigma(K)))\Gamma(\alpha)} \int_{\mathbb{R}^+} \theta_2^{\alpha-2} e^{-\beta\theta_2} d\theta_2$$

$$= \frac{-\beta^\alpha}{\max(\Re(\sigma(K)))\Gamma(\alpha)} \cdot \frac{\Gamma(\alpha-1)}{\beta^{\alpha-1}} = \frac{-\beta}{\alpha \max(\Re(\sigma(K)))},$$

which is finite and hence the process $\mathbb{X}_t$ has finite second moments.

**Proposition 5.2.** *Let $\Lambda$ be a $d-$dimensional Lévy basis with generating quadruple $(\gamma, \Sigma, \nu, \pi)$ with $\pi$ defined as in (5.17). The moments of the process $\mathbb{X}_t = \int_{M_d^-} \int_{-\infty}^t e^{\mathbf{Q}(\theta)(t-s)} \Lambda(d\mathbf{Q}(\theta), ds)$ are given as follows*

$$\mathrm{E}(\mathbb{X}_0) = -\frac{\beta}{\alpha-1}(-\mathbf{I}_{d \times d} - c\bar{\mathbf{A}})^{-1}\left(\gamma + \int_{|x|>1} x\nu(dx)\right), \quad \alpha \neq 1, \quad (5.18)$$

$$\text{var}(\mathbb{X}_0) = -\frac{\beta}{\alpha - 1}(\mathcal{A}(-\mathbf{I}_{d\times d} - c\bar{\mathbf{A}}))^{-1}\left(\Sigma + \int_{\mathbb{R}^d} xx^*\nu(dx)\right), \quad \alpha \neq 1, \tag{5.19}$$

$$\text{cov}(\mathbb{X}_h, \mathbb{X}_0) = -\frac{\beta^\alpha}{\alpha - 1}(\beta\mathbf{I}_{d\times d} + \mathbf{I}_{d\times d}h + c\bar{\mathbf{A}}h)^{1-\alpha}(\mathcal{A}(-\mathbf{I}_{d\times d} - c\bar{\mathbf{A}}))^{-1}\left(\Sigma + \int_{\mathbb{R}^d} xx^*\nu(dx)\right), \tag{5.20}$$

where $\mathcal{A}(\mathbf{Q}) : M_d(\mathbb{R}) \to M_d(\mathbb{R}), X \to \mathbf{Q}X + X\mathbf{Q}^*$.

*Proof.* Using the formula (5.13) for mean of supOU processes and referring Example 3.1,[BS11],

$$E(\mathbb{X}_0) = -\int_{M_d^-} \mathbf{Q}^{-1}\left(\gamma + \int_{|x|>1} x\nu(dx)\right)\pi(d\mathbf{Q})$$

$$E(\mathbb{X}_0) = -\int_{M_d^-} (-\theta_2\mathbf{I}_{d\times d} - c\theta_2\bar{\mathbf{A}})^{-1}\left(\gamma + \int_{|x|>1} x\nu(dx)\right)\pi(d\theta_2)$$

$$= -\int_{\mathbb{R}^+} (-\theta_2\mathbf{I}_{d\times d} - c\theta_2\bar{\mathbf{A}})^{-1}\left(\gamma + \int_{|x|>1} x\nu(dx)\right)\frac{\beta^\alpha}{\Gamma(\alpha)}\theta_2^{\alpha-1}e^{-\beta\theta_2}\mathbf{1}_{(0,\infty)}(\theta_2)d\theta_2$$

$$= -\int_{\mathbb{R}^+} (-\theta_2\mathbf{I}_{d\times d} - c\theta_2\bar{\mathbf{A}})^{-1}\theta_2^{\alpha-1}e^{-\beta\theta_2}d\theta_2\left(\gamma + \int_{|x|>1} x\nu(dx)\right)\frac{\beta^\alpha}{\Gamma(\alpha)}$$

$$= -\int_{\mathbb{R}^+} (-\mathbf{I}_{d\times d} - c\bar{\mathbf{A}})^{-1}\theta_2^{\alpha-2}e^{-\beta\theta_2}d\theta_2\left(\gamma + \int_{|x|>1} x\nu(dx)\right)\frac{\beta^\alpha}{\Gamma(\alpha)}$$

$$= -\int_{\mathbb{R}^+} \theta_2^{\alpha-2}e^{-\beta\theta_2}d\theta_2\frac{\beta^\alpha}{\Gamma(\alpha)}(-\mathbf{I}_{d\times d} - c\bar{\mathbf{A}})^{-1}\left(\gamma + \int_{|x|>1} x\nu(dx)\right)$$

$$= -\Gamma(\alpha - 1)\beta^{(1-\alpha)}\frac{\beta^\alpha}{\Gamma(\alpha)}(-\mathbf{I}_{d\times d} - c\bar{\mathbf{A}})^{-1}\left(\gamma + \int_{|x|>1} x\nu(dx)\right).$$

Since $\Gamma(a) = (a - 1)!\Gamma(a - 1)$ for some $a \in \mathbb{R}^+$, we obtain

$$E(\mathbb{X}_0) = -\frac{\beta}{\alpha - 1}(-\mathbf{I}_{d\times d} - c\bar{\mathbf{A}})^{-1}\left(\gamma + \int_{|x|>1} x\nu(dx)\right), \quad \alpha \neq 1. \tag{5.21}$$

Now for variance, using (5.14)

$$\text{var}(\mathbb{X}_0) = -\int_{M_d^-} (\mathcal{A}(\mathbf{Q}))^{-1}\left(\Sigma + \int_{\mathbb{R}^d} xx^*\nu(dx)\right)\pi(d\mathbf{Q})$$

$$= -\int_{M_d^-} (\mathcal{A}(\theta_2(-\mathbf{I}_{d\times d} - c\bar{\mathbf{A}})))^{-1}\left(\Sigma + \int_{\mathbb{R}^d} xx^*\nu(dx)\right)\frac{\beta^\alpha}{\Gamma(\alpha)}\theta_2^{\alpha-1}e^{-\beta\theta_2}\mathbf{1}_{(0,\infty)}(\theta_2)d\theta_2$$

$$= -\int_{\mathbb{R}^+} (\mathcal{A}(-\mathbf{I}_{d\times d} - c\bar{\mathbf{A}}))^{-1}\left(\Sigma + \int_{\mathbb{R}^d} xx^*\nu(dx)\right)\frac{\beta^\alpha}{\Gamma(\alpha)}\theta_2^{\alpha-2}e^{-\beta\theta_2}d\theta_2$$

$$= -\int_{\mathbb{R}^+} \frac{\beta^\alpha}{\Gamma(\alpha)}\theta_2^{\alpha-2}e^{-\beta\theta_2}d\theta_2(\mathcal{A}(-\mathbf{I}_{d\times d} - c\bar{\mathbf{A}}))^{-1}\left(\Sigma + \int_{\mathbb{R}^d} xx^*\nu(dx)\right)$$

$$= -\Gamma(\alpha - 1)\beta^{(1-\alpha)}\frac{\beta^\alpha}{\Gamma(\alpha)}(\mathcal{A}(-\mathbf{I}_{d\times d} - c\bar{\mathbf{A}}))^{-1}\left(\Sigma + \int_{\mathbb{R}^d} xx^*\nu(dx)\right)$$

where $\mathcal{A}(\mathbf{Q}) : M_d(\mathbb{R}) \to M_d(\mathbb{R}), X \to \mathbf{Q}X + X\mathbf{Q}^*$ and hence,

$$\text{var}(\mathbb{X}_0) = -\frac{\beta}{\alpha - 1}(\mathcal{A}(-\mathbf{I}_{d\times d} - c\bar{\mathbf{A}}))^{-1}\left(\Sigma + \int_{\mathbb{R}^d} xx^*\nu(dx)\right), \quad \alpha \neq 1. \quad (5.22)$$

Lastly, the autocovariance function can be computed as follows

$$\text{cov}(\mathbb{X}_h, \mathbb{X}_0) = -\int_{M_d^-} e^{\mathbf{Q}h}(\mathcal{A}(\mathbf{Q}))^{-1}\left(\Sigma + \int_{\mathbb{R}^d} xx^*\nu(dx)\right)\pi(d\mathbf{Q})$$

$$= -\int_{\mathbb{R}^+} e^{-\theta_2 \mathbf{I}_{d\times d}h - c\theta_2\bar{\mathbf{A}}h}(\mathcal{A}(-\mathbf{I}_{d\times d} - c\bar{\mathbf{A}}))^{-1}\left(\Sigma + \int_{\mathbb{R}^d} xx^*\nu(dx)\right)\frac{\beta^\alpha}{\Gamma(\alpha)}\theta_2^{\alpha-2}e^{-\beta\theta_2}\mathbf{1}_{(0,\infty)}(\theta_2)d\theta_2$$

Since $e^{-\beta\theta_2}.\mathbf{I}_{d\times d} = \sum_{l=0}^{\infty}\frac{(-\theta_2\beta)^l}{l!}.\mathbf{I}_{d\times d} = \sum_{l=0}^{\infty}\frac{(-\theta_2\beta\mathbf{I}_{d\times d})^l}{l!} = e^{-\theta_2\beta\mathbf{I}_{d\times d}}$, we can write

$$\text{cov}(\mathbb{X}_h, \mathbb{X}_0) = -\int_{\mathbb{R}^+} \theta_2^{\alpha-2}e^{\theta_2(-\beta\mathbf{I}_{d\times d} - \mathbf{I}_{d\times d}h - c\bar{\mathbf{A}}h)}d\theta_2 \frac{\beta^\alpha}{\Gamma(\alpha)}\mathcal{A}(-(\mathbf{I}_{d\times d} + c\bar{\mathbf{A}})))^{-1}\left(\Sigma + \int_{\mathbb{R}^d} xx^*\nu(dx)\right).$$

Since $\bar{A}$ is diagonalisable, let $U \in GL_d(\mathbb{C})$ and $\lambda_1, \lambda_2, \ldots, \lambda_d$ be eigenvalues of $(-\mathbf{I}_{d\times d} - c\bar{\mathbf{A}})$ with negative real part is such that

$$U(-\mathbf{I}_{d\times d} - c\bar{\mathbf{A}})U^{-1} = \text{diag}\{\lambda_1, \lambda_2, \ldots, \lambda_d\} = D$$

then we can write

$$e^{(-\mathbf{I}_{d\times d} - c\bar{\mathbf{A}})} = e^{U^{-1}DU}$$

$$= \sum_{l=0}^{\infty} \frac{(U^{-1}DU)^l}{l!}$$

$$= U^{-1}\mathbf{I}U + U^{-1}DU + \frac{1}{2}U^{-1}DUU^{-1}DU + \frac{1}{6}U^{-1}DUU^{-1}DUU^{-1}DU + \ldots$$

$$= U^{-1}\mathbf{I}U + U^{-1}DU + U^{-1}\frac{1}{2}D^2U + U^{-1}\frac{1}{6}D^3U + \cdots = U^{-1}e^DU$$

Hence, to compute the integral

$$\int_{\mathbb{R}^+} \theta_2^{\alpha-2}e^{-\theta_2(\beta\mathbf{I}_{d\times d} + \mathbf{I}_{d\times d}h + c\bar{\mathbf{A}}h)}d\theta_2$$

$$= U^{-1}\int_{\mathbb{R}^+}\theta_2^{\alpha-2}\exp(-\theta_2(\beta\mathbf{I}_{d\times d} + \text{diag}\{\lambda_1, \lambda_2, \ldots, \lambda_d\}))d\theta_2 U$$

$$= \Gamma(\alpha - 1)U^{-1}(\beta\mathbf{I}_{d\times d} + \text{diag}\{\lambda_1, \lambda_2, \ldots, \lambda_d\}h)^{1-\alpha}U$$

$$\text{cov}(\mathbb{X}_h, \mathbb{X}_0) = -\Gamma(\alpha - 1)(\beta \mathbf{I}_{d\times d} + \mathbf{I}_{d\times d} h + c\bar{\mathbf{A}} h)^{1-\alpha} \frac{\beta^\alpha}{\Gamma(\alpha)} \mathcal{A}(-(\mathbf{I}_{d\times d} + c\bar{\mathbf{A}})))^{-1} \left(\Sigma + \int_{\mathbb{R}^d} xx^* \nu(dx)\right).$$

We obtain

$$\text{cov}(\mathbb{X}_h, \mathbb{X}_0) = -\frac{\beta^\alpha}{\alpha - 1}(\beta \mathbf{I}_{d\times d} + \mathbf{I}_{d\times d} h + c\bar{\mathbf{A}} h)^{1-\alpha} \mathcal{A}(-(\mathbf{I}_{d\times d} + c\bar{\mathbf{A}})))^{-1} \left(\Sigma + \int_{\mathbb{R}^d} xx^* \nu(dx)\right). \quad (5.23)$$

$\square$

Evidently, the autocovariance function exhibits polynomial decay. Specifically, for values of $\alpha \in (1, 2)$, it suggests the presence of long memory.

Similar to the expression (5.4), we can write the characteristic function of the Graph supOU process $\mathbb{X}_t$ using the Lévy Khintchine representation

$$\mathrm{E}\left(\exp\left(\int_{M_d} \int_{-\infty}^{0} iu^T e^{\mathbf{Q}(h-s)} \Lambda(d\mathbf{Q}, ds)\right)\right) = \exp\left(\int_{M_d} \int_{-\infty}^{0} \phi((e^{\mathbf{Q}(h-s)})^T u) ds \pi(d\mathbf{Q})\right),$$

where

$$\phi(u) = iu^T \gamma - \frac{1}{2} u^T \Sigma u + \int_{\mathbb{R}^d} \left(e^{iu^T x} - 1 - iu^T x \mathbf{1}_{[0,1]}(\|x\|)\right) \nu(dx),$$

for all $u \in \mathbb{R}^d$.

**Remark 5.3.** *We can randomise the parameter c in the mean reversion parameter* $\mathbf{Q}$ *according to convenience, for example, considering discrete distribution where* $c = \{0, 1\}$ *with probabilites* $\pi(0)$ *and* $\pi(1)$ *where the computations will be trivial. For continuous distribution, for smaller computable dimension and for a variety of examples of networks, one can consider* $c \sim U(a, b)$ *where*

$$\pi(c) = \frac{c - a}{b - a} \mathbf{1}_{(a,b)} dc.$$

In order to implement these processes within practical or financial contexts, it is necessary to estimate the generating quadruple. Traditional methods such as maximum likelihood or similar approaches are not viable due to the unknown density of a supOU process. Therefore, similar to [STW15], we advocate for a moment-based estimation method, which relies on understanding the second-order structure of Graph supOU processes.

### 5.3.2 Weak Dependence Properties of supOU Processes

To explore the asymptotic characteristics of the parameter estimation of multivariate Graph supOU processes, we provide fundamental definitions and properties related to weak dependence in this section.

We note that supOU processes are a special case of the mixed moving average processes which are discussed in [CS18]. First, we recall that a stochastic process $X_t$ is said to be mixing if for any two events $A$ and $B$ in the sigma-algebra $\mathcal{F}$ of the process, the probability of both events occurring approaches the product of their individual probabilities as the separation between them increases:

$$\lim_{|k|\to\infty} \sup \{|P(A \cap B) - P(A)P(B)| : A \in \sigma(X_1,\ldots,X_k), B \in \sigma(X_{k+1}, X_{k+2},\ldots)\} = 0.$$

Mixing properties indicate how quickly a process forgets its initial conditions or previous states and approaches randomness or equilibrium. Processes with strong mixing properties tend to converge to equilibrium faster, while processes with weaker mixing take longer to do so.

Furthermore, for an $\mathbb{R}^d$-valued Lévy basis on $M_d^- \times \mathbb{R}$ and a 'kernel' function $f : M_d^- \times \mathbb{R} \to M_d(\mathbb{R})$ which is a $\mathcal{B}(M_d^- \times \mathbb{R})$-measurable function satisfying the conditions (5.5),(5.6) and (5.7) the mixed moving average (MMA) process can be written as

$$\mathbb{X}_t = \int_{M_d^-} \int_{-\infty}^{t} f(\mathbf{Q}, t-s) \Lambda(d\mathbf{Q}, ds).$$

Evidently, for $f(\mathbf{Q}, t-s) = e^{\mathbf{Q}(t-s)}$ this process is a supOU process.

The purpose of discussing this process is to study the weak dependence properties of a supOU process which would be useful in the later sections. Weak dependence in a process refers to a situation where the values of the process are not entirely independent, but they exhibit a level of correlation or association that diminishes as the time lag between observations increases. We give the definition of $\zeta$-weakly dependent process. Note that this was introduced as $\theta$-weakly dependence in [CS18]. Since we have $\theta$ as a parameter of interest in the graph notation, we call it as a $\zeta$-weakly dependent process.

**Definition 5.2.** *(Definition 3.2,[CS18]) A process $X = (X_t)_{t\in\mathbb{R}}$ taking values in $\mathbb{R}^d$ is called a $\zeta$-weakly dependent process if there exists a sequence $(\zeta(r)_{r\in\mathbb{R}^+})$ that converges to 0, and this sequence satisfies* *the following condition*

$$|\text{cov}(F(X_{i_1}, X_{i_2}, \ldots, X_{i_u}), G(X_{j_1}, X_{j_2}, \ldots, X_{j_v}))| \leq c(v\text{Lip}(G)\|F\|_\infty)\zeta(r),$$

for all $(u, v) \in \mathbb{N}^+ \times \mathbb{N}^+$, $r \in \mathbb{R}^+$, $(i_1, \ldots, i_u) \in \mathbb{R}^u$ and $(j_1, \ldots, j_v) \in \mathbb{R}^v$, with $i_1 \leq \cdots \leq i_u \leq i_u + r \leq j_1 \leq \cdots \leq j_v$, functions $F : (\mathbb{R}^d)^u \to \mathbb{R}$ and $G : (\mathbb{R}^d)^v \to \mathbb{R}$ respectively belonging to $\{f \in \mathcal{H}^* : \|f\|_\infty \leq 1\}$ and $\{f \in \mathcal{H} : \|f\|_\infty \leq 1\}$ where $\mathcal{H} = \cup_{u \in \mathbb{N}^+} \mathcal{H}_u$ with $\mathcal{H}_u$ be the class of bounded functions with a special condition, refer [CS18], where $\text{Lip}(G) = \sup_{x \neq y} \frac{|G(x) - G(y)|}{\|x_1 - y_1\| + \|x_2 - y_2\| + \cdots + \|x_n - y_n\|}$, where $c$ is a constant independent of $r$. We call $(\zeta(r))_{r \in \mathbb{R}^+}$ the sequence of the $\zeta$-coefficients.

In the case of integer valued processes, [Ben+23], $\zeta$- weakly dependence implies the strong mixing condition.

The general expression for the $\zeta$ coefficient (see Corollary 3.4, [CS18]) for a $\zeta$-weakly dependent MMA process

$$\zeta_X(r) = \left( \int_{M_d^-} \int_{-\infty}^{-r} \text{tr}(f(\mathbf{Q}, -s)\Sigma_L f(\mathbf{Q}, -s)^*) ds \pi(d\mathbf{Q}) + \| \int_{M_d^-} \int_{-\infty}^{-r} f(\mathbf{Q}, -s)\mu ds \pi(d\mathbf{Q}) \|^2 \right)^{\frac{1}{2}} \quad (5.24)$$

for all $r \geq 0$ where $\Sigma_L = \left( \Sigma + \int_{\mathbb{R}^d} xx^* \nu(dx) \right)$ and $\mu = \left( \gamma + \int_{|x| > 1} x\nu(dx) \right)$.

It has been shown in equation 3.26, [CS18] that the univariate supOU process is $\zeta$-weakly dependent with coefficients given by

$$\zeta_X(r) = \left( \text{cov}(X_0, X_{2r}) + \frac{4\mu^2}{\sigma^4} \text{cov}(X_0, X_r)^2 \right)^{\frac{1}{2}}. \quad (5.25)$$

We compute the $\zeta$-coefficients for multivariate supOU using (5.24) and $f(\mathbf{Q}, -s) = e^{-\mathbf{Q}s}$,

$$\zeta_X(r) = \left( \int_{M_d^-} \int_{-\infty}^{-r} \text{tr}(e^{-\mathbf{Q}s} \Sigma_L e^{-\mathbf{Q}^* s}) ds \pi(d\mathbf{Q}) + \| \int_{M_d^-} \int_{-\infty}^{-r} e^{-\mathbf{Q}s} \mu ds \pi(d\mathbf{Q}) \|^2 \right)^{\frac{1}{2}}$$

Since [BS11] typically use the Euclidean norm, we use the relation $\|A\|_F \leq \sqrt{g}\|A\|$ where $g$ is the rank of $A$ and $\|\cdot\|_F$ is the Frobenius Norm. Additionally, using the inequality $|\text{tr}(A)| \leq \sqrt{d}\|A\|_F$, triangle inequality for $a, b$ such that $|a + b|^{\frac{1}{2}} \leq |a|^{\frac{1}{2}} + |b|^{\frac{1}{2}}$,

$$\zeta_X(r) = \left( \int_{M_d^-} \int_{-\infty}^{-r} \text{tr}(e^{-\mathbf{Q}s} \Sigma_L e^{-\mathbf{Q}^*s}) ds \pi(d\mathbf{Q}) + \| \int_{M_d^-} \int_{-\infty}^{-r} e^{-\mathbf{Q}s} \mu ds \pi(d\mathbf{Q}) \|^2 \right)^{\frac{1}{2}}$$

$$\leq \left( \int_{M_d^-} \int_{-\infty}^{-r} |\text{tr}(e^{-\mathbf{Q}s} \Sigma_L e^{-\mathbf{Q}^*s})| ds \pi(d\mathbf{Q}) \right)^{\frac{1}{2}} + \left( \| \int_{M_d^-} \int_{-\infty}^{-r} e^{-\mathbf{Q}s} \mu ds \pi(d\mathbf{Q}) \|^2 \right)^{\frac{1}{2}} \quad (5.26)$$

$$\leq \left( \int_{M_d^-} \int_{-\infty}^{-r} \sqrt{gd} \| e^{-\mathbf{Q}s} \Sigma_L e^{-\mathbf{Q}^*s} \| ds \pi(d\mathbf{Q}) \right)^{\frac{1}{2}} + \| \int_{M_d^-} \int_{-\infty}^{-r} e^{-\mathbf{Q}s} \mu ds \pi(d\mathbf{Q}) \|$$

We replace $-s$ by $s$, resulting in a change of the limits of integration to $r$ to $\infty$. Using the bound (5.35), we obtain

$$\zeta_X(r) \leq \left( \int_{M_d^-} \int_r^\infty \sqrt{gd} \| e^{\mathbf{Q}s} \Sigma_L e^{\mathbf{Q}^*s} \| ds \pi(d\mathbf{Q}) \right)^{\frac{1}{2}} + \| \int_{M_d^-} \int_r^\infty e^{\mathbf{Q}s} \mu ds \pi(d\mathbf{Q}) \|$$

$$\leq \left( \int_{M_d^-} \int_r^\infty \sqrt{gd} \| \Sigma_L \| \kappa(\mathbf{Q})^2 e^{-2\rho(\mathbf{Q})s} ds \pi(d\mathbf{Q}) \right)^{\frac{1}{2}} + \|\mu\| \int_{M_d^-} \int_r^\infty \kappa(\mathbf{Q}) e^{-\rho(\mathbf{Q})s} ds \pi(d\mathbf{Q}) \quad (5.27)$$

$$\leq \left( \int_{M_d^-} \sqrt{gd} \| \Sigma_L \| \frac{\kappa(\mathbf{Q})^2}{2\rho(\mathbf{Q})} e^{-2\rho(\mathbf{Q})r} \pi(d\mathbf{Q}) \right)^{\frac{1}{2}} + \|\mu\| \int_{M_d^-} \frac{\kappa(\mathbf{Q})}{\rho(\mathbf{Q})} e^{-\rho(\mathbf{Q})r} \pi(d\mathbf{Q})$$

Using Remark 3.2, [BS11], if $\mathbf{Q}$ is diagonalisable, the function $\kappa(\mathbf{Q}) = \|U\| \|U\|^{-1}$ where $U$ is such that $U\mathbf{Q}U^{-1}$ is diagonal and $\rho(\mathbf{Q}) = -\max(\Re(\sigma(\mathbf{Q})))$. For the long memory specific case on a graph, when $\mathbf{Q} = \theta_2 K$, due to the definitions of the functions, $\kappa(\mathbf{Q}) = \theta_2 \kappa(K)$ and $\rho(\mathbf{Q}) = \theta_2 \rho(K)$. Additionally, $\pi$ has gamma distribution,

$$\zeta_X(r) \leq \left( \int_{\mathbb{R}^+} \sqrt{gd} \| \Sigma_L \| \theta_2 \frac{\kappa(K)^2}{2\rho(K)} e^{-2\theta_2 \rho(K) r} \frac{\beta^\alpha}{\Gamma(\alpha)} \theta_2^{\alpha-1} e^{-\beta \theta_2} \mathbb{I}_{(0,\infty)}(\theta_2) d\theta_2 \right)^{\frac{1}{2}}$$

$$+ \|\mu\| \int_{\mathbb{R}^+} \frac{\kappa(K)}{\rho(K)} e^{-\theta_2 \rho(K) r} \frac{\beta^\alpha}{\Gamma(\alpha)} \theta_2^{\alpha-1} e^{-\beta \theta_2} \mathbb{I}_{(0,\infty)}(\theta_2) d\theta_2$$

$$= \left( \sqrt{gd} \| \Sigma_L \| \int_{\mathbb{R}^+} \theta_2^\alpha e^{(-2\rho(K)r - \beta)\theta_2} d\theta_2 \frac{\kappa(K)^2}{2\rho(K)} \frac{\beta^\alpha}{\Gamma(\alpha)} \right)^{\frac{1}{2}} \quad (5.28)$$

$$+ \|\mu\| \int_{\mathbb{R}^+} e^{(-\rho(K)r - \beta)\theta_2} \theta_2^{\alpha-1} (\theta_2) d\theta_2 \frac{\kappa(K)}{\rho(K)} \frac{\beta^\alpha}{\Gamma(\alpha)}$$

$$= \left( \sqrt{gd} \| \Sigma_L \| \left( \frac{\Gamma(\alpha+1)}{(2\rho(K)r + \beta)^{\alpha+1}} \right) \frac{\kappa(K)^2}{2\rho(K)} \frac{\beta^\alpha}{\Gamma(\alpha)} \right)^{\frac{1}{2}} + \|\mu\| \left( \frac{\Gamma(\alpha)}{(\rho(K)r + \beta)^\alpha} \right) \frac{\kappa(K)}{\rho(K)} \frac{\beta^\alpha}{\Gamma(\alpha)}$$

$$= \left( \sqrt{gd} \| \Sigma_L \| \left( \frac{(\alpha+1)\beta^\alpha}{(2\rho(K)r + \beta)^{\alpha+1}} \right) \frac{\kappa(K)^2}{2\rho(K)} \right)^{\frac{1}{2}} + \|\mu\| \left( \frac{\beta^\alpha}{(\rho(K)r + \beta)^\alpha} \right) \frac{\kappa(K)}{\rho(K)}.$$

The first component of the coefficient $\zeta_X(r)$ is of order $O(r^{-\frac{(\alpha+1)}{2}})$ and the second component is of order $O(r^{-\alpha})$.

The $\zeta$-weakly dependent process has the 'hereditary property', that is, the level of dependence between random variables in a stochastic process not only weakens as the time lag or separation between observations increases but also remains weakly dependent when considering subsets or sub-processes of the original process.

**Proposition 5.3.** *(Proposition 3.4, [CS18]) Let $(X_t)_{t\in\mathbb{R}}$ be a stationary process with values in $\mathbb{R}^d$. Suppose there exists a constant $C > 0$ such that $\mathrm{E}[\|X_0\|^p]^{\frac{1}{p}} \leq C$, where $p \geq 1$. Let $h : \mathbb{R}^d \to \mathbb{R}^m$ be a function satisfying $h(0) = 0$ and $h(x) = (h_1(x), \ldots, h_m(x))$, with*

$$\|h(x) - h(y)\| \leq c\|x - y\|(1 + \|x\|^{a-1} + \|y\|^{a-1}),$$

*for $x, y \in \mathbb{R}^d$, $c > 0$, and $1 \leq a < p$. Define $(Y_t)_{t\in\mathbb{R}}$ as $Y_t = h(X_t)$. If $(X_t)_{t\in\mathbb{R}}$ is a $\zeta$-weakly dependent process such that for all $r \geq 0$,*

$$\zeta_Y(r) = C\zeta_X(r)^{\frac{p-a}{p-1}},$$

*where $C$ is a constant independent of $r$.*

In addition, we recall the Cramér Wold's device [CW36] which asserts that a random vector $X_n = (X_{n1}, \ldots, X_{nd})$ converges in distribution to $X = (X_1, \ldots, X_d)$ if and only if

$$\sum_{i=1}^{d} t_i X_{ni} \xrightarrow[n\to\infty]{D} \sum_{i=1}^{d} t_i X_i \tag{5.29}$$

for each $(t_1, \ldots, t_d) \in \mathbb{R}^d$.

### 5.3.3 Generalised Method of Moments (GMM) Estimator

We want to estimate the parameters for the case where the mean reversion parameter is Gamma distributed. Let $\mathbb{X} = (X_t)_{t\in\mathbb{R}}$ be the Graph supOU process with the mean reversion parameter

$$\mathbf{Q}(\theta) = \theta_2 K,$$

where $K \in M_d^-$ and $\theta_2 \sim \Gamma(\alpha,\beta)$.

Consider the equidistant sample of $N$ observations $\{\mathbb{X}_t : t = 1\Delta,\ldots,N\Delta\}$ where $\mathbb{X}_t = \begin{bmatrix} X_t^{(1)} \\ X_t^{(2)} \\ \vdots \\ X_t^{(d)} \end{bmatrix}$ and $\Delta = T/N > 0, T > 0, N \in \mathbb{N}$. To construct the moment function, we consider the mean, variance and autocovariance up to lag $m \geq 2$. We introduce the vector

$$\mathbb{X}_t^{(m)} := (\mathbb{X}_t,\ldots,\mathbb{X}_{t+m}) \quad t \in \{1,\ldots,N-m\}.$$

We want to estimate the parameter vector which contains the parameters of the moments of underlying Lévy process $\begin{bmatrix} \mathbb{L}_t^{(1)} \\ \mathbb{L}_t^{(2)} \\ \vdots \\ \mathbb{L}_t^{(d)} \end{bmatrix}$ and the parameters of $\pi$. Let the vector $\xi$ is in the parameter space $\mathfrak{W}$.

We recall the moments of the process

$$\mu(\zeta) = \mathrm{E}(\mathbb{X}_0) = -\frac{\beta}{\alpha-1}(-\mathbf{I}_{d\times d} - c\bar{\mathbf{A}})^{-1}\left(\gamma + \int_{|x|>1} x\nu(dx)\right), \quad \alpha \neq 1, \tag{5.30}$$

$$\mathrm{var}(\mathbb{X}_0) = \mathrm{cov}(\mathbb{X}_0,\mathbb{X}_0) = -\frac{\beta}{\alpha-1}(\mathcal{A}(-\mathbf{I}_{d\times d} - c\bar{\mathbf{A}}))^{-1}\left(\Sigma + \int_{\mathbb{R}^d} xx^*\nu(dx)\right), \quad \alpha \neq 1, \tag{5.31}$$

and for lag $k$,

$$\mathrm{cov}(\mathbb{X}_k,\mathbb{X}_0) = -\frac{\beta^\alpha}{\alpha-1}(\beta\mathbf{I}_{d\times d} + \mathbf{I}_{d\times d}h + c\bar{\mathbf{A}}h)^{1-\alpha}(\mathcal{A}(-\mathbf{I}_{d\times d} - c\bar{\mathbf{A}}))^{-1}\left(\Sigma + \int_{\mathbb{R}^d} xx^*\nu(dx)\right). \tag{5.32}$$

Define a measurable function $f : \mathbb{R}^{d(m+1)} \times \mathfrak{W} \to \mathbb{R}^{(d+\frac{(m+1)d(d+1)}{2})}$

$$f(\mathbb{X}_t^{(m)},\xi) = \begin{pmatrix} f_E(\mathbb{X}_t^{(m)},\xi) \\ f_0(\mathbb{X}_t^{(m)},\xi) \\ f_1(\mathbb{X}_t^{(m)},\xi) \\ \vdots \\ f_m(\mathbb{X}_t^{(m)},\xi) \end{pmatrix} = \begin{pmatrix} \mathrm{vec}(\mathbb{X}_t - \mathrm{E}(\mathbb{X}_0)) \\ \mathrm{vec}_h(\mathbb{X}_t^*\mathbb{X}_t - \mathrm{var}(\mathbb{X}_0) - \mathrm{E}(\mathbb{X}_0)(\mathrm{E}(\mathbb{X}_0))^T \\ \mathrm{vec}_h(\mathbb{X}_t^*\mathbb{X}_{(t+1)} - \mathrm{cov}(\mathbb{X}_1,\mathbb{X}_0) - \mathrm{E}(\mathbb{X}_0)(\mathrm{E}(\mathbb{X}_0))^T \\ \vdots \\ \mathrm{vec}_h(\mathbb{X}_t^*\mathbb{X}_{(t+m)} - \mathrm{cov}(\mathbb{X}_m,\mathbb{X}_0) - \mathrm{E}(\mathbb{X}_0)(\mathrm{E}(\mathbb{X}_0))^T \end{pmatrix}.$$

We can now define the sample moment function

$$g_{N,m}(\mathbb{X}_t, \xi) = \frac{1}{N-m} \sum_{t=1}^{N-m} f(\mathbb{X}_t^{(m)}, \xi) = \begin{pmatrix} \frac{1}{N-m} \sum_{t=1}^{N-m} f_E(\mathbb{X}_t^{(m)}, \xi) \\ \frac{1}{N-m} \sum_{t=1}^{N-m} f_0(\mathbb{X}_t^{(m)}, \xi) \\ \frac{1}{N-m} \sum_{t=1}^{N-m} f_1(\mathbb{X}_t^{(m)}, \xi) \\ \vdots \\ \frac{1}{N-m} \sum_{t=1}^{N-m} f_m(\mathbb{X}_t^{(m)}, \xi) \end{pmatrix}.$$

One can then estimate $\xi_0$ by minimising the objective function

$$\hat{\xi}_0^{N,m} = \arg\min g_{N,m}(\mathbb{X}_t, \xi)^T V g_{N,m}(\mathbb{X}_t, \xi), \tag{5.33}$$

where $V$ is the $(d + \frac{(m+1)d(d+1)}{2}) \times (d + \frac{(m+1)d(d+1)}{2})$ positive definite weight matrix.

## 5.4 Asymptotic Theory of GMM for Multivariate Graph supOU

In this section, we determine the asymptotic normality of GMM estimators of the multivariate Graph supOU process. We utilise the asymptotic theory provided in [CS18] and [Ben+23] for mixed moving average processes.

We want to prove the asymptotic normality of the estimator (5.33). As a first step, we prove the central limit theorem for the moment function $f(\mathbb{X}_t^{(m)}, \xi)$, which extends Theorem 6.1 in [CS18] to the case of multivariate Graph supOU processes.

**Theorem 5.3.** *Let $\Lambda$ be a real valued Lévy basis with generating quadruple $(\gamma, \Sigma, \nu, \pi)$ such that*

$$\int_{\|x\|>1} \ln(\|x\|)\nu(dx) < \infty \tag{5.34}$$

*and suppose that $\int_{\|x\|>1} \|x\|^{4+\delta} \nu(dx) < \infty$ for some $\delta > 0$. Also, assume that there exist measurable functions $\rho : M_d^- \to \mathbb{R}^+ \setminus \{0\}$ and $\kappa : M_d^- \to [1, \infty)$ such that*

$$\|e^{\mathbf{Q}s}\| \leq \kappa(\mathbf{Q})e^{-\rho(\mathbf{Q})s} \quad \forall s \in \mathbb{R}^+, \pi - \text{almost surely}, \tag{5.35}$$

and

$$\int_{M_d^-} \frac{\kappa(\mathbf{Q})^2}{\rho(\mathbf{Q})} \pi(d\mathbf{Q}) < \infty. \tag{5.36}$$

*Additionally, assume that the probability distribution $\pi$ is Gamma distributed along with the condition*

$$\alpha - 1 > \left(1 + \frac{1}{\delta}\right)\left(\frac{6+2\delta}{2+\delta}\right). \tag{5.37}$$

*Then $f(\mathbb{X}_t^{(m)}, \xi_0)$ is a $\zeta-$ weakly dependent process. The matrix*

$$F_\Sigma = \sum_{l \in \mathbb{Z}} \mathrm{cov}(f(\mathbb{X}_0, \xi_0), f(\mathbb{X}_l, \xi_0))$$

*is finite and positive definite. As $N \to \infty$,*

$$\sqrt{N} g_{N,m}(\mathbb{X}, \xi_0) \xrightarrow{d} N(0, F_\Sigma). \tag{5.38}$$

*Proof.* Since multivariate supOU processes are a special case for mixed moving average process and given $\int_{\|x\|>1} \|x\|^{4+\delta} \nu(dx) < \infty$. By Proposition 2.1 in [CS18], $4+\delta$ moments of the multivariate supOU process exists.

Let $C = \mathrm{E}(\mathbb{X}_0)$ and define a function $F(\mathbb{X}_t^{(m)}) : \mathbb{R}^{d(m+1)} \to \mathbb{R}^{(d+\frac{(m+1)d(d+1)}{2})}$ such that

$$F(\mathbb{X}_t^{(m)}) = f(\mathbb{X}_t^{(m)}, \xi_0) + \begin{pmatrix} \mathrm{vec}(\mathrm{E}(\mathbb{X}_0)) \\ \mathrm{vec}_h(\mathrm{var}(\mathbb{X}_0) + \mathrm{E}(\mathbb{X}_0)(\mathrm{E}(\mathbb{X}_0))^T) \\ \mathrm{vec}_h(\mathrm{cov}(\mathbb{X}_1, \mathbb{X}_0) + \mathrm{E}(\mathbb{X}_0)(\mathrm{E}(\mathbb{X}_0))^T) \\ \vdots \\ \mathrm{vec}_h(\mathrm{cov}(\mathbb{X}_m, \mathbb{X}_0) + \mathrm{E}(\mathbb{X}_0)(\mathrm{E}(\mathbb{X}_0))^T) \end{pmatrix} = \begin{pmatrix} \mathrm{vec}(\mathbb{X}_{t\Delta}) \\ \mathrm{vec}_h(\mathbb{X}_{t\Delta}\mathbb{X}_{t\Delta}^T) \\ \mathrm{vec}_h(\mathbb{X}_t \mathbb{X}_{(t+1)}^T) \\ \vdots \\ \mathrm{vec}_h(\mathbb{X}_t \mathbb{X}_{(t+m)}^T) \end{pmatrix}.$$

This function $F$ satisfies the conditions of 5.3, if $p = 4 + \delta, c = 1, a = 2$. Hence, $F(\mathbb{X}_t^{(m)})$ is a $\zeta$ weakly dependent process with coefficients $\zeta_F(r) = C(\mathcal{D}\zeta_X(r - m\Delta))^{\frac{2+\delta}{3+\delta}}$ for constants $C, \mathcal{D} > 0$ where $\theta_X$ is given by (5.28). This implies $f(\mathbb{X}_t^{(m)}, \xi_0)$ is a $\zeta$ weakly dependent process with above coefficients and zero mean. We obtain

$$\zeta_f(r) = C\mathcal{D}^{\frac{2+\delta}{3+\delta}}(\zeta_X(r - m\Delta))^{\frac{2+\delta}{3+\delta}}$$

where $\alpha - 1 > 2\left(1 + \frac{1}{\delta}\right)\left(\frac{2+\delta}{6+2\delta}\right)$ as in the hypothesis.

Then using Theorem 1 of Dedecker and Rio [DR00] described in the previous section, the function $\sqrt{N}f$ is normally distributed. This is because the moment condition in this theorem would hold for $\zeta$-weakly dependent process with given coefficients. Consequently, applying Cramér Wold device (5.29) we conclude that $\sqrt{N}g_{N,m}(X_t^{(m)})$ is normally distributed (5.38). □

Now we prove the central limit theorem (CLT) for the GMM estimator. We utilise the sufficient assumptions postulated by Mátyás [Mát99] to first show weak consistency and then CLT for the GMM estimator (5.33).

In order to demonstrate consistency, it is necessary to have an assumption that guarantees the accurate identification of the true value of $\xi$. For the univariate supOU processes, the identifiability is proved in Proposition 3.3 of [STW15]. For multivariate supOU processes, we make the following assumption.

**Assumption 1**

1. $\mathrm{E}(f(\mathbb{X}_t^{(m)}, \xi))$ exists and is finite for all $\xi \in \mathfrak{W}$ which holds by construction of $f(\mathbb{X}_t^{(m)}, \xi)$.

2. If $g_t^{(m)}(\xi) = \mathrm{E}(f(\mathbb{X}_t^{(m)}, \xi))$, there exits $\xi_0 \in \mathfrak{W}$ such that $g_t^{(m)}(\xi) = 0$ for all $t$ if and only if $\xi = \xi_0$.

The subsequent assumption concerns the convergence of sample moments to population moments.

**Assumption 2**

Every component of the vector $g_{N,m}(\mathbb{X}, \xi_0) - \mathrm{E}(f(\mathbb{X}_t^{(m)}, \xi))$ uniformly converges in probability to zero for each $\xi$ within $\mathfrak{W}$.

We will show that this assumption holds by verifying the sufficient conditions Assumptions 4-6 as defined later in this section.

Moreover, the following assumption regarding the weighting matrix is also required.

**Assumption 3**

There is a deterministic sequence of positive definite matrices denoted by $\bar{V}_{N,m}$ such that the difference between $V_{N,m}$ and $\bar{V}_{N,m}$ tends to zero in probability.

Then we have the following result.

**Theorem 5.4.** *The GMM estimator $\hat{\xi}_0^{N,m}$, as defined in equation* (5.33), *demonstrates weak consistency under the Assumptions 1-3.*

Proof: The weak consistency of $\hat{\xi}_0^{N,m}$ directly follows from Theorem 1 in [Mát99].

Further, we consider additional and alternate assumptions to prove asymptotic normality of the GMM estimator.

**Assumption 4**

The parameter space $\mathfrak{W}$ is both compact and sufficiently extensive to encompass the true parameter $\xi_0$.

It's worth noting that while the parameter space may not inherently be closed and bounded, it can be made compact by imposing appropriate constraints during optimization.

**Assumption 5**

For each component, the vector $g_{N,m}(\mathbb{X}, \xi_0) - \mathrm{E}(f(\mathbb{X}_t^{(m)}, \xi))$ converges pointwise in probability to zero within the parameter space $\mathfrak{W}$.

Since the supOU process is a specific instance of a mixed moving average process, it is inherently mixing and ergodic, thus satisfying assumption 5.

**Assumption 6**

Every component of $f(\mathbb{X}_t^{(m)}, \xi)$ exhibits stochastic equicontinuity.

This property can be achieved by imposing a stochastic Lipschitz-type assumption on each component of $f(\mathbb{X}_t^{(m)}, \xi)$. Let $\xi_i$ represent parameter vectors within $\mathfrak{W}$. Considering the first component:

$$|f_E(\mathbb{X}_t^{(m)}, \xi_1) - f_E(\mathbb{X}_t^{(m)}, \xi_2)| = |\mathrm{vec}(\mathbb{X}_{t\Delta} - \mu(\xi_1)) - \mathrm{vec}(\mathbb{X}_{t\Delta} - \mu(\xi_2))| = |\mathrm{vec}(-\mu(\xi_1) + \mu(\xi_2))|$$

$$= \left| -\frac{\beta_1}{\alpha_1 - 1}(\mathbf{I}_{d\times d} + c\bar{\mathbf{A}})^{-1} \left( \int_{|x|>1} x\nu_1(dx) \right) + \frac{\beta_2}{\alpha_2 - 1}(\mathbf{I}_{d\times d} + c\bar{\mathbf{A}})^{-1} \left( \int_{|x|>1} x\nu_2(dx) \right) \right|.$$

Here, terms involving $\mathbb{X}_t$ cancel out due to construction. Similar deductions hold for other components. This leaves us with a Lipschitz condition on non-random terms in each component of $f(\mathbb{X}_t^{(m)}, \xi)$.

Moreover, if we consider partial derivatives with respect to the model parameters, we find that these partial derivatives are bounded, indicating that the components are Lipschitz continuous. Thus, assumption 6 is satisfied.

**Assumption 7**

The moment function $f(\mathbb{X}_t^{(m)}, \xi)$ exhibits continuous differentiability with respect to $\xi$ across $\mathfrak{W}$. This assumption is inherent in the construction of the function $f$.

**Assumption 8**

A weak law of large numbers applies to the first derivative of $f(\mathbb{X}_t^{(m)}, \xi)$ in a neighbourhood of $\xi_0$. Specifically, considering

$$G_{N,m}(\mathbb{X}, \xi) = \frac{1}{N-m} \sum_{t=1}^{N-m} \frac{\partial f(\mathbb{X}_t^{(m)}, \xi)}{\partial \xi^T}.$$

We need to show that for a sequence $\xi_N^* \xrightarrow{p} \xi_0$, it follows that $G_{N,m}(\mathbb{X}, \xi_N^*) \xrightarrow{p} G_0$. Since the partial derivative matrix $\frac{\partial f(\mathbb{X}_t^{(m)}, \xi)}{\partial \xi^T}$ is independent of $\mathbb{X}_t^{(m)}$, $G_{N,m} = \mathbb{E}(\frac{\partial f(\mathbb{X}_t^{(m)}, \xi)}{\partial \xi^T})$ and $G_0 = \mathbb{E}[\frac{\partial f(\mathbb{X}_t^{(m)}, \xi)}{\partial \xi^T}]_{\xi=\xi_0}$. Therefore, by applying the continuous mapping theorem, this assumption is satisfied.

**Assumption 9**

The moment function $f(\mathbb{X}_t^{(m)}, \xi)$ conforms to a central limit theorem, as implied by Theorem 1.

Given that all the sufficient conditions hold for multivariate supOU processes, and following the same procedures as outlined in [Mát99], we deduce the following outcome for the estimator.

**Theorem 5.5.** *Let $\Lambda$ be a real valued Lévy basis with generating quadruple $(\gamma, \Sigma, \nu, \pi)$ and $\mathbb{X}$ a supOU process such that*

$$\int_{\|x\|>1} \ln(\|x\|)\nu(dx) < \infty,$$

*and suppose that $\int_{\|x\|>1} \|x\|^{4+\delta}\nu(dx) < \infty$ for some $\delta > 0$. Also, assume there exist measurable functions $\rho : M_d^- \to \mathbb{R}^+ \setminus \{0\}$ and $\kappa : M_d^- \to [1, \infty)$ such that*

$$\|e^{\mathbf{Q}s}\| \leq \kappa(\mathbf{Q})e^{-\rho(\mathbf{Q})s} \quad \forall s \in \mathbb{R}^+, \pi - almost\ surely, \tag{5.39}$$

*and*

$$\int_{M_d^-} \frac{\kappa(\mathbf{Q})^2}{\rho(\mathbf{Q})} \pi(d\mathbf{Q}) < \infty. \tag{5.40}$$

*Additionally, assume that the probability distribution $\pi$ is Gamma distributed along with the condition*

$$\alpha - 1 > \left(1 + \frac{1}{\delta}\right)\left(\frac{6 + 2\delta}{2 + \delta}\right).$$

*Moreover, Assumptions 6-9 hold. Then as $N \to \infty$,*

$$\sqrt{N}(\hat{\xi}_0^{N,m} - \xi_0) \xrightarrow{d} N(0, MF_\Sigma M'). \tag{5.41}$$

*where*

$$M = (G_0'AG_0)^{-1}G_0'A, \quad G_0 = \mathbb{E}[\frac{\partial f(\mathbb{X}_t^{(m)}, \xi)}{\partial \xi^T}]_{\xi=\xi_0}$$

$$\text{and} \quad F_\Sigma = \sum_{l \in \mathbb{Z}} \text{cov}(f(\mathbb{X}_0, \xi_0), f(\mathbb{X}_l, \xi_0)).$$

**Remark** It should be noted that the conditions outlined in these theorems necessitate

$$\alpha - 1 > \left(1 + \frac{1}{\delta}\right)\left(\frac{6 + 2\delta}{2 + \delta}\right).$$

This condition implies that $\alpha > 1$, thereby excluding the possibility of accommodating the long memory setting.

## 5.5 Setup and Methodology of Simulation Study

In this section, we discuss the simulation study to illustrate the generalised method of moments estimators for the multivariate Graph supOU process with the parametric framework discussed in 5.3.3. We adapt the simulation setup for univariate supOU processes as described in [RS72] and adapt it for the multivariate Graph supOU processes. The underlying Lévy process $\mathbb{L}$ of the supOU process in this simulation setup is a compound Poisson processes with positive jumps. The compound Poisson process considered will have many jumps and no restriction on the rate, hence making this model very flexible. We recall that the univariate supOU process $\mathbb{X}$ given by

$$\mathbb{X}_t = \int_{M_d} \int_{-\infty}^{t} e^{-(\theta_1 \mathbf{I}_{d \times d} + \theta_2 \bar{\mathbf{A}})(t-s)} \Lambda(d\mathbf{Q}, ds). \tag{5.42}$$

For a general underlying compound Poisson process with $\mu$ as a Poisson random measure, using the Lévy-Itô decomposition for $\mathbb{X}$ we get

$$\mathbb{X}_t = \int_{\mathbb{R}} \int_{\mathbb{R}_-} \int_{-\infty}^{t} e^{\mathbf{Q}(t-s)} x \mu(dx, d\mathbf{Q}, ds) \tag{5.43}$$

For the Graph supOU process, we have $\mathbf{Q} = \theta_2(-\mathbf{I}_{d \times d} - c\bar{\mathbf{A}})$.

The supOU process $\mathbb{X}$ can then be written as

$$\mathbb{X}_t = \sum_{i \geq 1, \tau \leq t} e^{\mathbf{Q}_i(t-\tau_i)} U_i + \sum_{i=1}^{\infty} e^{\mathbf{Q}_{-i}(t-\tau_{-i})} U_{-i}, \tag{5.44}$$

where

$$\tau_i := \sum_{j=1}^{i} T_j \text{ and } \tau_{-i} := \sum_{j=1}^{i} T_{-j}$$

and $(T_i)_{i \in \mathbb{Z} \setminus \{0\}}, (U_i)_{i \in \mathbb{Z} \setminus \{0\}}$ and $(Q_i)_{i \in \mathbb{Z} \setminus \{0\}}$ are independent i.i.d. random variables denoting the arrival times, jumps and mean reversion parameters.

To start with, we simulate the univariate Graph supOU processes in the next section. Later, we extend it for bivariate Graph supOU processes.

### 5.5.1 Univariate Graph supOU Simulation Study

For an illustrative example, we consider the same rates and parameters as in [STW15] where the compound Poisson say $\mathcal{N}_t$ has rate $0.1$ and the positive jumps for the simulations are $\Gamma(3, 20)$-distributed. In [STW15], the parameters were selected based on an examination of simulated paths to determine if they exhibit a reasonable shape for daily log returns in financial data.

The univariate Graph supOU process $\mathbb{X}$ with $\mathbf{Q} = \theta_2(-\mathbf{I}_{d \times d} - c\bar{\mathbf{A}})$ can be written as

$$\mathbb{X}_t = \sum_{i \geq 1, \tau \leq t} e^{\mathbf{Q}_i(t-\tau_i)} U_i + \sum_{i=1}^{\infty} e^{\mathbf{Q}_{-i}(t-\tau_{-i})} U_{-i}, \tag{5.45}$$

where
$$\tau_i := \sum_{j=1}^{i} T_j \text{ and } \tau_{-i} := \sum_{j=1}^{i} T_{-j}$$

and $(T_i)_{i\in\mathbb{Z}\setminus\{0\}}, (U_i)_{i\in\mathbb{Z}\setminus\{0\}}$ and $(Q_i)_{i\in\mathbb{Z}\setminus\{0\}}$ are independent i.i.d. random variables with $T_i \sim \exp(0.1), U_i \sim \Gamma(3, 20)$ and $\mathbf{Q}_i \sim B\Gamma(\alpha_n, 1)$ where $B \in \mathbb{R}^-$. Note that the paper mentions $\mathbf{Q}_i \sim -B\Gamma(\alpha_n, 1)$ with $B \in \mathbb{R}^-$ which leads to the error in the sign of $\mathbf{Q}_i$ later in the simulation.

The mean and variance of the considered compound Poisson considered would be given by

$$\mu = \left(\frac{1}{0.1}\right)\left(\frac{3}{20}\right) = 1.5 \quad \text{and} \quad \sigma^2 = \left(\frac{1}{0.1}\right)\left(\left(\frac{3}{20^2}\right) + \left(\frac{3}{20}\right)^2\right) = 0.3.$$

In the following univariate simulation, we simulate the process 200 times with both 500 and 1000 observations in each return. Since the infinite sum in 5.45 can be obtained approximately, we estimate the second infinite sum approximately by taking only 2000 observations and ignoring the jumps beyond -2000.

As the next step, we calculate the GMM estimators for each of these 1000 independent paths. Computing the estimators involves solving the optimization problem 5.33 for each path. Similar to the approach in [STW15], we utilise the 2-step iterated GMM estimation. In the first step of the optimisation process, the weight matrix $V$ is considered to be the identity matrix. In the second step, the weight matrix would be $V^{-1}$ where

$$V = \lim_{n \to \infty} \text{var}\left(\frac{1}{\sqrt{n}} \sum_{t=1}^{n} f(X_t^{(m)}, \xi_1)\right)$$

and $\xi_1$ is the estimation result of the first step. The estimation 5.33 simplifies to

$$\hat{V} = \frac{1}{n} \sum_{t=1}^{n} f(X_t^{(m)}, \xi_1) f(X_t^{(m)}, \xi_1)^T.$$

We recall that theoretical mean, variance and autocovariance of the univariate supOU are given by

$$\mathrm{E}(\mathbb{X}_0) = \frac{-\mu}{B(\alpha_n - 1)}, \quad \text{var}(\mathbb{X}_0) = \frac{-\sigma^2}{2B(\alpha_n - 1)}, \quad \text{cov}(\mathbb{X}_h, \mathbb{X}_0) = \frac{-\sigma^2(1 - Bh)^{1-\alpha_n}}{2\beta(\alpha_n - 1)}.$$

In the estimation of the parameter vector $\xi = (\mu, \sigma^2, \alpha_n, B)$ using two step GMM method, similar to [STW15], we use the mean, variance and first five lags of the autocovariance function. Since there are four parameters to be estimated, the number of moment conditions required needs to be atleast four. Hence we consider a number bigger than the minimum number of conditions required. We have $2 + m$ moment conditions for $m$ lags, and we consider $m = 5$ to overestimate the system. Starting with an initial parameter vector $\xi_0$ and the initial weighting matrix as identity matrix, the moment function of supOU (5.33) is minimised. In order to analyse the long memory case, we consider the initial parameter vector to be $\xi_0 = (\mu_0, \sigma_0^2, \alpha_{n,0}, B_0) = (1.5, 0.3, 1.95, -0.1)$. This is chosen randomly from a neighborhood of the true parameters. All simulations and estimations in this chapter have been carried out with Python.

### 5.5.2  Bivariate Graph supOU Simulation Study

In this section, we provide a novel algorithm for estimating the parameters for multivariate Graph supOU process. The considered graph would have two nodes and an undirected edge between them with the adjacency matrix $A = \begin{bmatrix} 0 & 1 \\ 1 & 0 \end{bmatrix}$.

The Graph supOU process $\mathbb{X}_t$ with the underlying Lévy process as the compound Poisson process given by

$$\mathbb{X}_t = \int_{\mathbb{R}} \int_{\mathbb{R}_-} \int_{-\infty}^{t} e^{\mathbf{Q}(t-s)} x \mu(dx, d\mathbf{Q}, ds) \tag{5.46}$$

is now bivariate and hence the process $\mathbb{X}_t$ will now be a vector value

$$\mathbb{X}_t = \begin{bmatrix} X_t^{(1)} \\ X_t^{(2)} \end{bmatrix}. \tag{5.47}$$

For the univariate graph supOU in the previous section, we had one compound Poisson say $\mathcal{N}_t$ with rate 0.1 and $\Gamma(3, 20)$-distributed jumps. In the bivariate case, we make a modelling assumption that there is one common joint process for all the corresponding independent coordinates, hence we write the two components of the compound Poisson as follows

$$\mathcal{N}^{(1)} = \mathcal{N}^{(\|)} + \mathcal{M}^{(1)},$$

$$\mathcal{N}^{(2)} = \mathcal{N}^{(\|)} + \mathcal{M}^{(2)},$$

where $\mathcal{N}^{(\|)}$ is the joint compound Poisson for both the components showing dependence between them. We consider joint jump vector to be $\begin{bmatrix} U_i^{\|(1)} \\ U_i^{\|(2)} \end{bmatrix}$ with different sizes for the two components. Then we can discretise (5.46) as follows

$$\begin{bmatrix} X_t^{(1)} \\ X_t^{(2)} \end{bmatrix} = \sum_{i\geq 1, \tau^{(\|)}\leq t} e^{\mathbf{Q}_i(t-\tau_i^{(\|)})} \begin{bmatrix} U_i^{\|(1)} \\ U_i^{\|(2)} \end{bmatrix} + \sum_{j\geq 1, \tau^{(1)}\leq t} e^{\mathbf{Q}_j(t-\tau_j^{(1)})} \begin{bmatrix} U_j^{(1)} \\ 0 \end{bmatrix} + \sum_{k\geq 1, \tau^{(2)}\leq t} e^{\mathbf{Q}_k(t-\tau_k^{(2)})} \begin{bmatrix} 0 \\ U_k^{(2)} \end{bmatrix}$$
$$+ \sum_i^{\infty} e^{\mathbf{Q}_{-i}(t-\tau_{-i}^{(\|)})} \begin{bmatrix} U_{-i}^{\|(1)} \\ U_{-i}^{\|(2)} \end{bmatrix} + \sum_j^{\infty} e^{\mathbf{Q}_{-j}(t-\tau_{-j}^{(1)})} \begin{bmatrix} U_{-j}^{(1)} \\ 0 \end{bmatrix} + \sum_k^{\infty} e^{\mathbf{Q}_{-k}(t-\tau_{-k}^{(2)})} \begin{bmatrix} 0 \\ U_{-k}^{(2)} \end{bmatrix}. \quad (5.48)$$

Since $\mathbf{Q} = \theta_2 K$, we obtain

$$\begin{bmatrix} X_t^{(1)} \\ X_t^{(2)} \end{bmatrix} = \sum_{i\geq 1, \tau^{(\|)}\leq t} e^{\theta_{2,i}^{(\|)} K(t-\tau_i^{(\|)})} \begin{bmatrix} U_i^{\|(1)} \\ U_i^{\|(2)} \end{bmatrix} + \sum_{j\geq 1, \tau^{(1)}\leq t} e^{\theta_{2,j}^{(1)} K(t-\tau_j^{(1)})} \begin{bmatrix} U_j^{(1)} \\ 0 \end{bmatrix} + \sum_{k\geq 1, \tau^{(2)}\leq t} e^{\theta_{2,k}^{(2)} K(t-\tau_k^{(2)})} \begin{bmatrix} 0 \\ U_k^{(2)} \end{bmatrix}$$
$$+ \sum_i^{\infty} e^{\theta_{2,-i}^{(\|)} K(t-\tau_{-i}^{(\|)})} \begin{bmatrix} U_{-i}^{\|(1)} \\ U_{-i}^{\|(2)} \end{bmatrix} + \sum_j^{\infty} e^{\theta_{2,-j}^{(1)} K(t-\tau_{-j}^{(1)})} \begin{bmatrix} U_{-j}^{(1)} \\ 0 \end{bmatrix} + \sum_k^{\infty} e^{\theta_{2,-k}^{(2)} K(t-\tau_{-k}^{(2)})} \begin{bmatrix} 0 \\ U_{-k}^{(2)} \end{bmatrix}, \quad (5.49)$$

where

$$\tau_i^{(k)} := \sum_{m=1}^{i} T_m^{(k)} \text{ and } \tau_{-i} := \sum_{m=1}^{i} T_{-m}^{(k)},$$

which is corresponding to the $k = 1, 2, \|$ compound Poisson process and $(T_i)_{i\in\mathbb{Z}\setminus\{0\}}^{(k)}, (U_i)_{i\in\mathbb{Z}\setminus\{0\}}^{(k)}$ are independent i.i.i.d. random variables with $T_i^{(k)} \sim \exp(0.1), U_i^{(k)} \sim \Gamma(3, 20)$ and $\theta_{2,i}^{(k)} \sim \Gamma(\alpha_n, 1)$ and $K = -\mathbf{I}_{2\times 2} - c\bar{\mathbf{A}} \in \mathcal{M}_2^-$ for $k = \|, 1, 2$. In this simulation study, we consider the distribution of $U_i^{\|(1)}$ and $U_i^{\|(2)}$ to be the same.

Since $K$ is diagonalisable in the graph considered, it can be written as $K = O(D)O^{-1}$ where $O$ is a matrix with columns as eigenvectors and $D$ is a diagonal matrix with eigenvalues on the diagonal. Then the exponential of $\theta_{2,-k}^{(2)} K(t - \tau_{-k}^{(2)})$ can be written using the expansion as

$$\exp(\theta_{2,-i}^{(2)} K(t - \tau_{-k}^{(2)})) = \sum_i \frac{(\theta_{2,-i}^{(2)} K(t - \tau_{-i}^{(2)}))^n}{n!} = O \sum_i \frac{(D)^n (\theta_{2,-i}^{(2)} (t - \tau_{-i}^{(2)}))^n}{n!} O^{-1}.$$

Let the eigenvalues of $K$ be $\lambda_1, \lambda_2$, then we can write

$$\exp(\theta_{2,-i}^{(2)} K(t - \tau_{-i}^{(2)})) = O \sum_i \left( \frac{\begin{bmatrix} \lambda_1^n (\theta_{2,i}^{(1)})^n (t - \tau_i^{(1)})^n & 0 \\ 0 & \lambda_2^n (\theta_{2,i}^{(1)})^n (t - \tau_i^{(1)})^n \end{bmatrix}}{n!} \right) O^{-1}$$

$$= O \sum_i \begin{bmatrix} \frac{\lambda_1^n (\theta_{2,i}^{(1)})^n (t-\tau_i^{(1)})^n}{n!} & 0 \\ 0 & \frac{\lambda_2^n (\theta_{2,i}^{(1)})^n (t-\tau_i^{(1)})^n}{n!} \end{bmatrix} O^{-1}$$

$$= O \begin{bmatrix} e^{\lambda_1 (\theta_{2,i}^{(1)})(t-\tau_i^{(1)})} & 0 \\ 0 & e^{\lambda_2 (\theta_{2,i}^{(1)})(t-\tau_i^{(1)})} \end{bmatrix} O^{-1}$$

The exponentials in (5.51) can be written in the above form for simulating the paths.

We consider $c = 0.5$, then $K = -\mathbf{I}_{d \times d} - c\bar{\mathbf{A}} = \begin{bmatrix} -1 & -0.5 \\ -0.5 & -1 \end{bmatrix}$ and $K^{-1} = \begin{bmatrix} -1.333 & 0.6667 \\ 0.6667 & -1.333 \end{bmatrix}$.

The mean of the bivariate compound Poisson process $\begin{bmatrix} \mathbb{L}_t^{(1)} \\ \mathbb{L}_t^{(2)} \end{bmatrix}$ can be computed in the following way

$$\mu_L = E\left( \begin{bmatrix} \mathbb{L}_t^{(1)} \\ \mathbb{L}_t^{(2)} \end{bmatrix} \right) = E\left( \sum_{i \geq 1, \tau^{(\parallel)} \leq t} \begin{bmatrix} U_i^{\parallel(1)} \\ U_i^{\parallel(2)} \end{bmatrix} + \sum_{j \geq 1, \tau^{(1)} \leq t} \begin{bmatrix} U_j^{(1)} \\ 0 \end{bmatrix} + \sum_{k \geq 1, \tau^{(2)} \leq t} \begin{bmatrix} 0 \\ U_k^{(2)} \end{bmatrix} \right)$$

$$= E\left( \sum_{i \geq 1, \tau^{(\parallel)} \leq t} \begin{bmatrix} U_i^{\parallel(1)} \\ U_i^{\parallel(2)} \end{bmatrix} \right) + E\left( \sum_{j \geq 1, \tau^{(1)} \leq t} \begin{bmatrix} U_j^{(1)} \\ 0 \end{bmatrix} \right) + E\left( \sum_{k \geq 1, \tau^{(2)} \leq t} \begin{bmatrix} 0 \\ U_k^{(2)} \end{bmatrix} \right)$$

$$= \frac{1}{0.1} \sum_{i \geq 1, \tau^{(\parallel)} \leq t} \begin{bmatrix} E(U_i^{\parallel(1)}) \\ E(U_i^{\parallel(2)}) \end{bmatrix} + \frac{1}{0.1} \sum_{j \geq 1, \tau^{(1)} \leq t} \begin{bmatrix} E(U_j^{(1)}) \\ 0 \end{bmatrix} + \frac{1}{0.1} \sum_{k \geq 1, \tau^{(2)} \leq t} \begin{bmatrix} 0 \\ E(U_k^{(2)}) \end{bmatrix}$$

$$= \frac{1}{0.1} \sum_{i \geq 1, \tau^{(\parallel)} \leq t} \begin{bmatrix} 3/20 \\ 3/20 \end{bmatrix} + \frac{1}{0.1} \sum_{j \geq 1, \tau^{(1)} \leq t} \begin{bmatrix} 3/20 \\ 0 \end{bmatrix} + \frac{1}{0.1} \sum_{k \geq 1, \tau^{(2)} \leq t} \begin{bmatrix} 0 \\ 3/20 \end{bmatrix} = \begin{bmatrix} 1.5 \\ 1.5 \end{bmatrix} + \begin{bmatrix} 1.5 \\ 0 \end{bmatrix} + \begin{bmatrix} 0 \\ 1.5 \end{bmatrix} = \begin{bmatrix} 3 \\ 3 \end{bmatrix}$$

Similarly, the variance of the compound Poisson with $\Sigma = 0$, $\left(\Sigma + \int_{\mathbb{R}^d} xx^* \nu(dx)\right)$ can be computed

$$\sigma_L^2 = \mathrm{var}\left(\begin{bmatrix}\mathbb{L}_t^{(1)} \\ \mathbb{L}_t^{(2)}\end{bmatrix}\right) = \mathrm{var}\left(\sum_{i\geq 1,\tau^{(\|)}\leq t}\begin{bmatrix}U_i^{\|(1)} \\ U_i^{\|(2)}\end{bmatrix} + \sum_{j\geq 1,\tau^{(1)}\leq t}\begin{bmatrix}U_j^{(1)} \\ 0\end{bmatrix} + \sum_{k\geq 1,\tau^{(2)}\leq t}\begin{bmatrix}0 \\ U_k^{(2)}\end{bmatrix}\right)$$

$$= \mathrm{var}\left(\sum_{i\geq 1,\tau^{(\|)}\leq t}\begin{bmatrix}U_i^{\|(1)} \\ U_i^{\|(2)}\end{bmatrix}\right) + \mathrm{var}\left(\sum_{j\geq 1,\tau^{(1)}\leq t}\begin{bmatrix}U_j^{(1)} \\ 0\end{bmatrix}\right) + \mathrm{var}\left(\sum_{k\geq 1,\tau^{(2)}\leq t}\begin{bmatrix}0 \\ U_k^{(2)}\end{bmatrix}\right)$$

First we compute

$$\mathrm{var}\left(\sum_{i\geq 1,\tau^{(\|)}\leq t}\begin{bmatrix}U_i^{\|(1)} \\ U_i^{\|(2)}\end{bmatrix}\right) = \frac{1}{0.1}\left(\mathrm{var}\left(\begin{bmatrix}U_i^{\|(1)} \\ U_i^{\|(2)}\end{bmatrix}\right) + \mathrm{E}\left(\begin{bmatrix}U_i^{\|(1)} \\ U_i^{\|(2)}\end{bmatrix}\right)\mathrm{E}\left(\begin{bmatrix}U_i^{\|(1)} \\ U_i^{\|(2)}\end{bmatrix}\right)^*\right)$$

$$= \frac{1}{0.1}\left(\begin{bmatrix}\mathrm{var}(U_i^{\|(1)}) & 0 \\ 0 & \mathrm{var}(U_i^{\|(2)})\end{bmatrix} + \begin{bmatrix}0.15 \\ 0.15\end{bmatrix}\begin{bmatrix}0.15 & 0.15\end{bmatrix}\right)$$

$$= \frac{1}{0.1}\left(\begin{bmatrix}0.0075 & 0 \\ 0 & 0.0075\end{bmatrix} + \begin{bmatrix}0.0225 & 0.0225 \\ 0.0225 & 0.0225\end{bmatrix}\right) = \begin{bmatrix}0.3 & 0.225 \\ 0.225 & 0.3\end{bmatrix}.$$

Next we compute

$$\mathrm{var}\left(\sum_{j\geq 1,\tau^{(1)}\leq t}\begin{bmatrix}U_j^{(1)} \\ 0\end{bmatrix}\right) = \frac{1}{0.1}\left(\mathrm{var}\left(\begin{bmatrix}U_j^{(1)} \\ 0\end{bmatrix}\right) + \mathrm{E}\left(\begin{bmatrix}U_j^{(1)} \\ 0\end{bmatrix}\right)\mathrm{E}\left(\begin{bmatrix}U_j^{(1)} \\ 0\end{bmatrix}\right)^*\right)$$

$$= \frac{1}{0.1}\left(\begin{bmatrix}\mathrm{var}(U_j^{(1)}) & 0 \\ 0 & 0\end{bmatrix} + \begin{bmatrix}0.15 \\ 0\end{bmatrix}\begin{bmatrix}0.15 & 0\end{bmatrix}\right)$$

$$= \frac{1}{0.1}\left(\begin{bmatrix}0.0075 & 0 \\ 0 & 0\end{bmatrix} + \begin{bmatrix}0.0225 & 0 \\ 0 & 0\end{bmatrix}\right) = \begin{bmatrix}0.3 & 0 \\ 0 & 0\end{bmatrix}.$$

Similarly,

$$\mathrm{var}\left(\sum_{k\geq 1,\tau^{(2)}\leq t}\begin{bmatrix}0 \\ U_k^{(2)}\end{bmatrix}\right) = \begin{bmatrix}0 & 0 \\ 0 & 0.3\end{bmatrix}$$

and hence we obtain

$$\sigma_L^2 = \mathrm{var}\left(\begin{bmatrix}\mathbb{L}_t^{(1)} \\ \mathbb{L}_t^{(2)}\end{bmatrix}\right) = \begin{bmatrix}0.6 & 0.225 \\ 0.225 & 0.6\end{bmatrix}.$$

Similar to the univariate case, we consider the initial parameter vector $\xi_0 = (\mu_{L,0}, \sigma^2_{L,0}, a_n)$. Using the above computations and the expressions (5.18),(5.19),(5.20), we obtain the moments of the bivariate graph supOU

$$E(\mathbb{X}_0) = \begin{bmatrix} 2.101 \\ 2.101 \end{bmatrix}, \text{var}(\mathbb{X}_0) = \begin{bmatrix} 0.3419 & -0.0526 \\ -0.0526 & 0.3419 \end{bmatrix} \tag{5.50}$$

and

$$\text{cov}(\mathbb{X}_h, \mathbb{X}_0) = \left(\begin{bmatrix} 1+h & 0.5h \\ 0.5h & 1+h \end{bmatrix}\right)^{-0.95} \begin{bmatrix} 0.3419 & -0.0526 \\ -0.0526 & 0.3419 \end{bmatrix}.$$

**Remark 5.4.** *The moment expressions for multivariate supOU (5.19), (5.20) includes computing the inverse of the function $\mathcal{A}(Q) : X \to QX + XQ^*$. This problem is similar to solving the Lyapunov equation (also known as Sylvester equation). We utilise the naive approach for solving Lyapunov equation [Jar17] by writing them as a system of linear equations*

$$(I \otimes A + A \otimes I)\text{vec}(X) = \text{vec}(W).$$

Similar to the univariate case, we simulate 200 independent paths for the bivariate graph supOU using 2 step iterated GMM and the estimator given by (5.33). First we consider 500 observations and then 1000 observations in each return. The initial weight matrix $V$ would be the identity matrix and first five lags of the autocovariance function are considered.

### 5.5.3 Extension to $d$ dimensions

For multivariate graph supOU, we extend the simulation model defined for bivariate case in the previous section. We retain our modelling assumption of having a common jump compound Poisson process for every coordinate and its corresponding independent compound Poisson. Hence we write the components of the compound Poisson as follows

$$\mathcal{N}^{(1)} = \mathcal{N}^{(\|)} + \mathcal{M}^{(1)},$$

$$\mathcal{N}^{(2)} = \mathcal{N}^{(\|)} + \mathcal{M}^{(2)},$$

$$\vdots$$

$$\mathcal{N}^{(d)} = \mathcal{N}^{(\|)} + \mathcal{M}^{(d)},$$

where $\mathcal{N}^{(\|)}$ is the joint compound Poisson for both the components showing dependence between them. The discretised multivariate graph supOU is given by

$$\begin{bmatrix} X_t^{(1)} \\ X_t^{(2)} \\ \vdots \\ X_t^{(d)} \end{bmatrix} = \sum_{i \geq 1, \tau^{(\|)} \leq t} e^{\theta_{2,i}^{(\|)} K(t-\tau_i^{(\|)})} \begin{bmatrix} U_i^{\|(1)} \\ U_i^{\|(2)} \\ \vdots \\ U_i^{\|(d)} \end{bmatrix} + \sum_{j \geq 1, \tau^{(1)} \leq t} e^{\theta_{2,j}^{(1)} K(t-\tau_j^{(1)})} \begin{bmatrix} U_j^{(1)} \\ 0 \\ \vdots \\ 0 \end{bmatrix} + \sum_{k \geq 1, \tau^{(2)} \leq t} e^{\theta_{2,k}^{(2)} K(t-\tau_k^{(2)})} \begin{bmatrix} 0 \\ U_k^{(2)} \\ \vdots \\ 0 \end{bmatrix}$$

$$+ \cdots + \sum_{l \geq 1, \tau^{(d)} \leq t} e^{\theta_{2,l}^{(2)} K(t-\tau_l^{(d)})} \begin{bmatrix} 0 \\ \vdots \\ 0 \\ U_l^{(d)} \end{bmatrix} + \sum_{i}^{\infty} e^{\theta_{2,-i}^{(\|)} K(t-\tau_{-i}^{(\|)})} \begin{bmatrix} U_{-i}^{\|(1)} \\ U_{-i}^{\|(2)} \\ \vdots \\ U_{-i}^{\|(d)} \end{bmatrix} + \sum_{j}^{\infty} e^{\theta_{2,-j}^{(1)} K(t-\tau_{-j}^{(1)})} \begin{bmatrix} U_{-j}^{(1)} \\ 0 \\ \vdots \\ 0 \end{bmatrix}$$

$$+ \sum_{k}^{\infty} e^{\theta_{2,-k}^{(2)} K(t-\tau_{-k}^{(2)})} \begin{bmatrix} 0 \\ U_{-k}^{(2)} \\ \vdots \\ 0 \end{bmatrix} + \cdots + \sum_{l}^{\infty} e^{\theta_{2,-l}^{(2)} K(t-\tau_{-l}^{(2)})} \begin{bmatrix} 0 \\ \vdots \\ 0 \\ U_{-l}^{(d)} \end{bmatrix}$$

(5.51)

where

$$\tau_i^{(k)} := \sum_{m=1}^{i} T_m^{(k)} \quad \text{and} \quad \tau_{-i} := \sum_{m=1}^{i} T_{-m}^{(k)}$$

which is corresponding to the $k = 1, 2, \ldots, d, \|$ compound Poisson process.

### 5.5.4 Results of the Simulation Study

**Univariate Graph supOU model**

The second plot of Figure 5.1 shows polynomially decaying autocorrelation function indicating long memory in a univariate supOU process. In Figure 5.2 and 5.3, we see the estimation results for this process using 200 independent paths with 500 and 1000 observations each respectively. In all the histograms, the dotted line represents the true parameter values. The estimates are mostly evenly distributed around the true parameter values with slight bias in parameter *B*. Moreover, Figure 5.4 and 5.5 presents normal QQ-plots of the obtained parameters. In the case of 1000 observations for each path, there is a short stagnation of the values in the QQ-plot for estimators $\mu$ and $\sigma$. But the QQ-plots of all of the parameters mostly indicates an asymptotic normality of the estimators with tails skewed away from the normal distribution. The estimates improve significantly with 1000 observations compared to 500 observations.

**Bivariate Graph supOU model**

In Figure 5.6, we see one of the 200 independent simulated paths for first and second component of the bivariate Graph supOU process. The next plot indicates the decaying autocorrelation function. For the first 20 lags, the autocorrelation function decays slowly indicating long(er) memory. Figures 5.7-5.10 shows the histograms and QQ-plots for the parameter estimators when using 200 paths with 500 and 1000 observations each. Again, most estimates are distributed around true parameters for both components. The QQ-plots for $\mu$ of first and second component and $a_n$ clearly indicate the asymptotic normality in the long memory case. The normality can be seen for the parameter $\sigma^2$ as well with slight deviation on the tails. Again, the estimates are more accurate with 1000 observations rather than 500 observations, due to the influence of the outliers. In the estimations with 500 observations each, we remove some outliers to plot the histograms and QQ-plots. For the estimates of mean of both components, we consider first 98.5 percentile of the data, excluding the top 1.5 percentile to eliminate outliers. Similarly, we considered first 93 percentile for variance parameters and first 97.5 percentile of data for $a_n$ parameter.

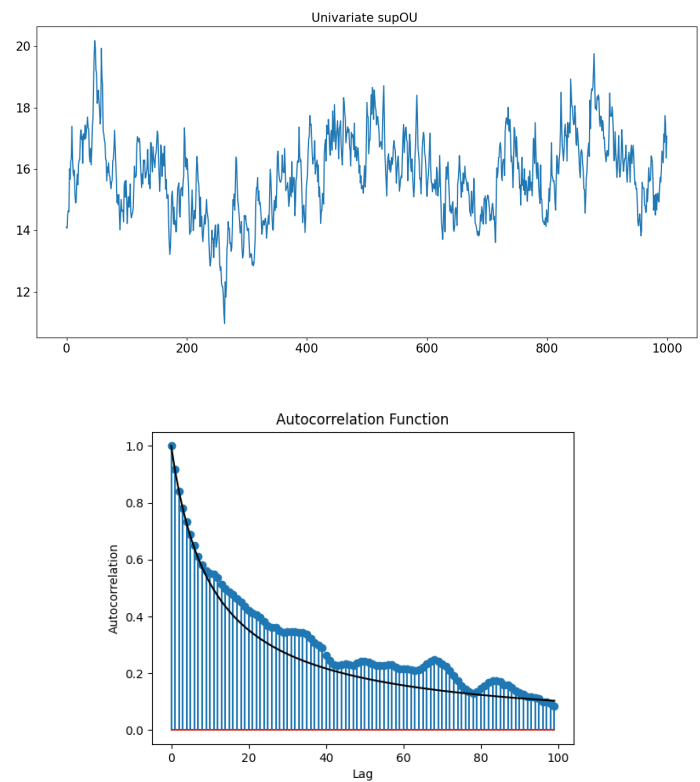

Figure 5.1: The first plot is a simulated path of univariate graph supOU. The next plot is the autocovariance plot showing long memory and the black line is the theoretical autocorrelation function.

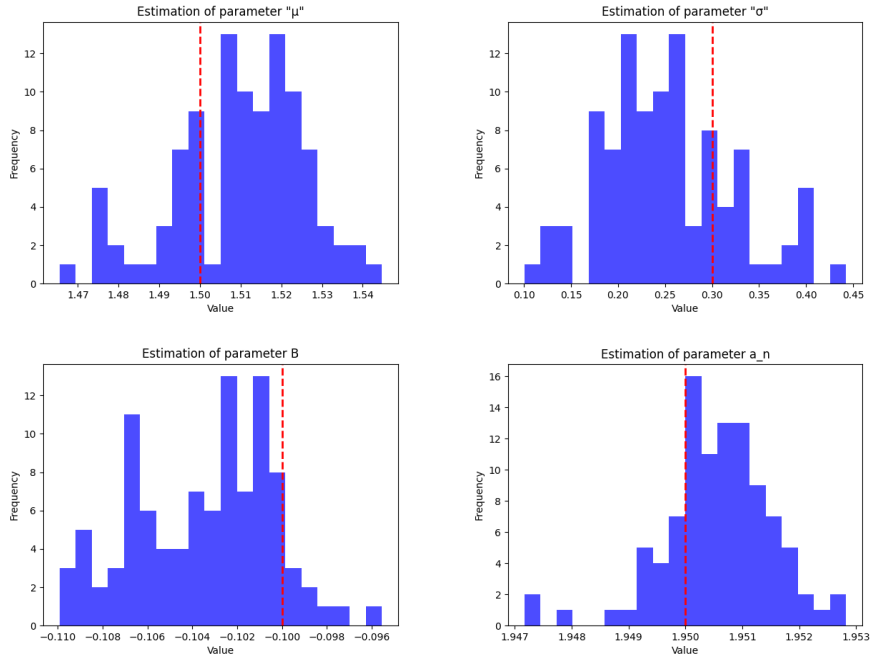

Figure 5.2: Histograms of parameter estimates of 200 paths of length 500 of a univariate graph supOU process with long memory. The true values are indicated by dotted lines.

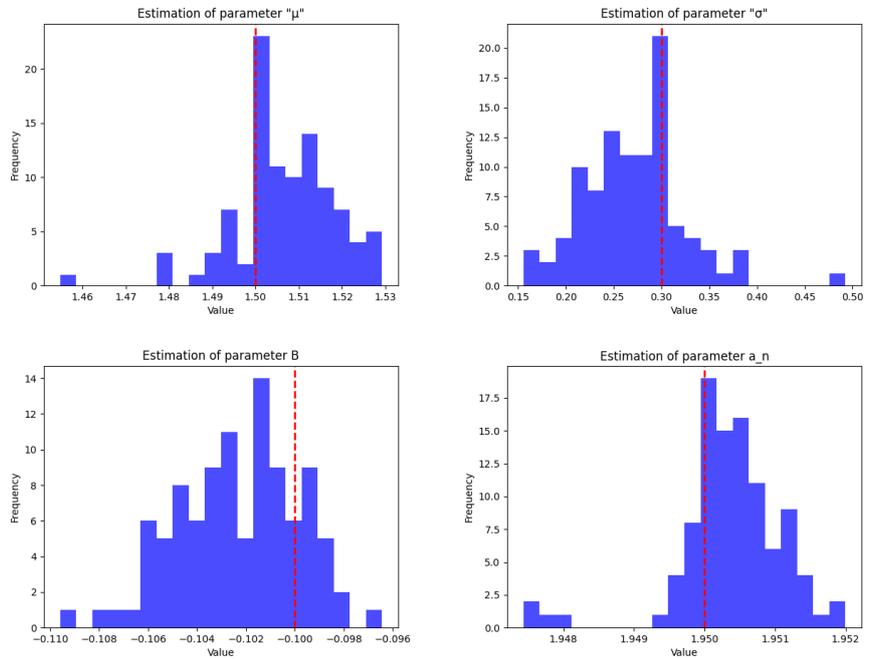

Figure 5.3: Histograms of parameter estimates of 200 paths of length 1000 of a univariate graph supOU process with long memory. The true values are indicated by dotted lines.

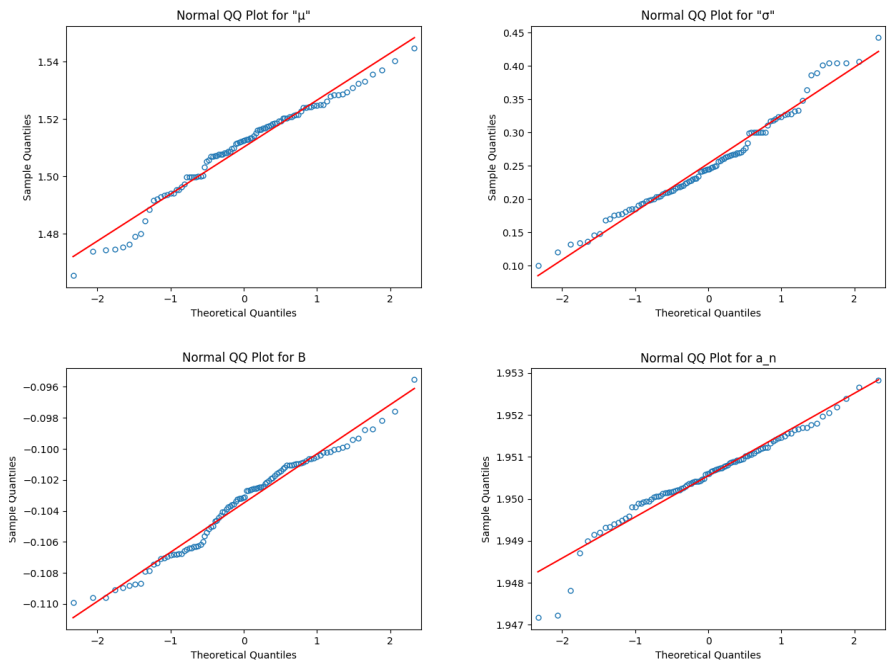

Figure 5.4: Normal QQ-plots of parameter estimates of 200 paths of length 500 of a univariate Graph supOU model

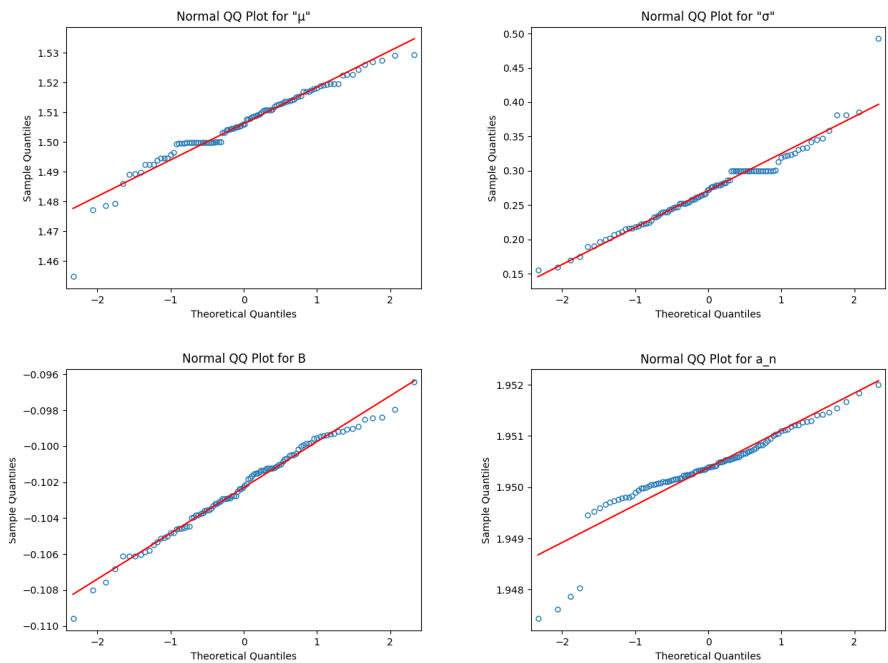

Figure 5.5: Normal QQ-plots of parameter estimates of 200 paths of length 1000 of a univariate Graph supOU model

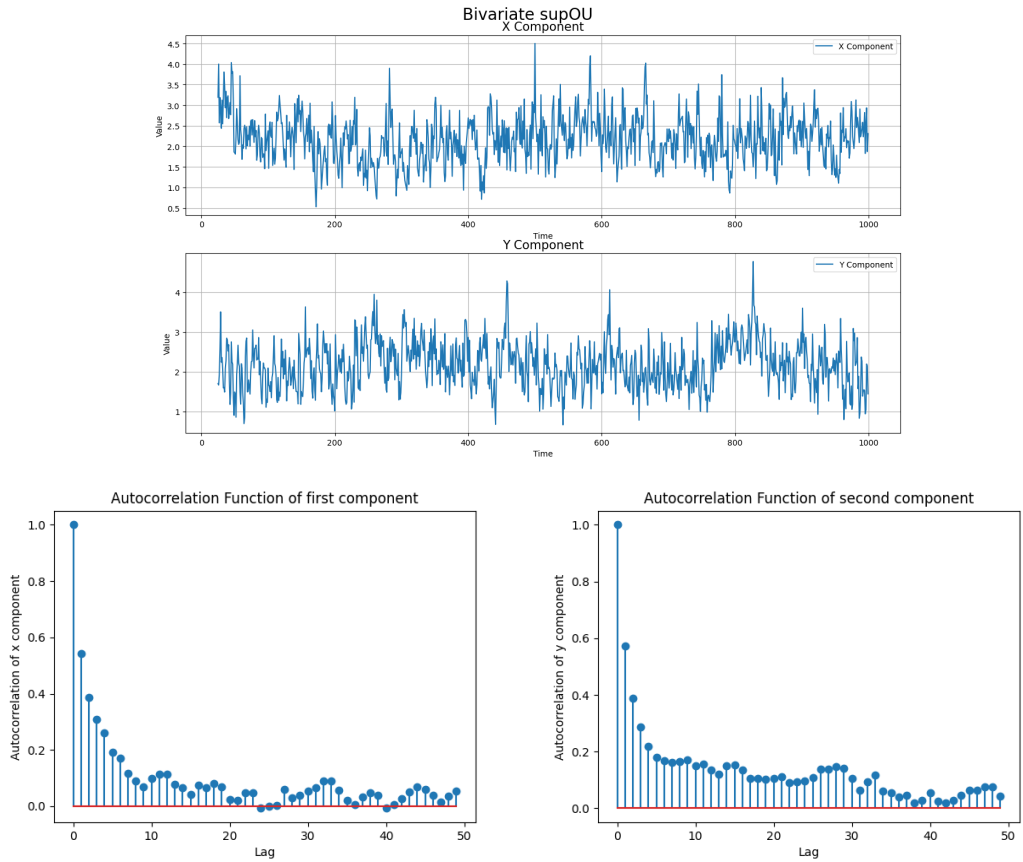

Figure 5.6: The plot at the top illustrates a simulated path of both components of a bivariate graph supOU. The next plot is the autocovariance plot of both components showing long(er) memory.

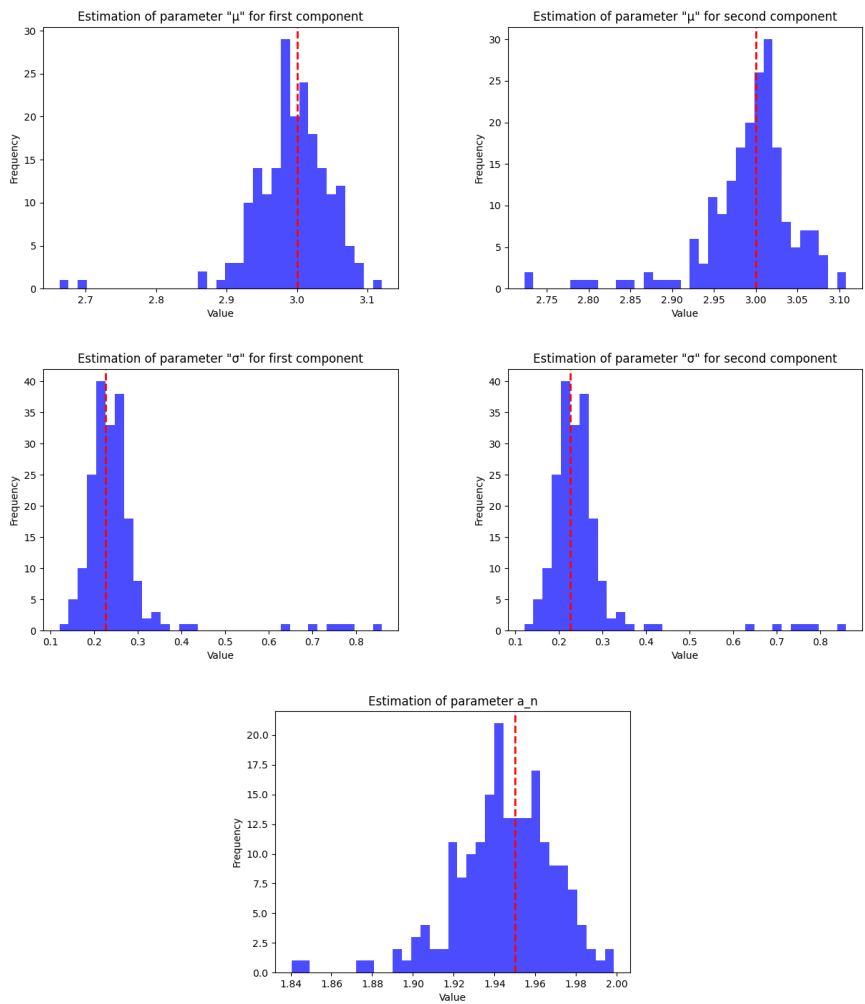

Figure 5.7: Histograms of parameter estimates of 200 paths of length 500 of a bivariate graph supOU process with long memory. The true values are indicated by dotted lines.

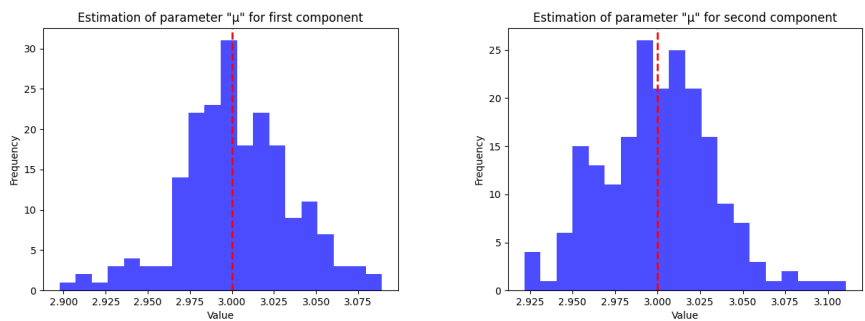

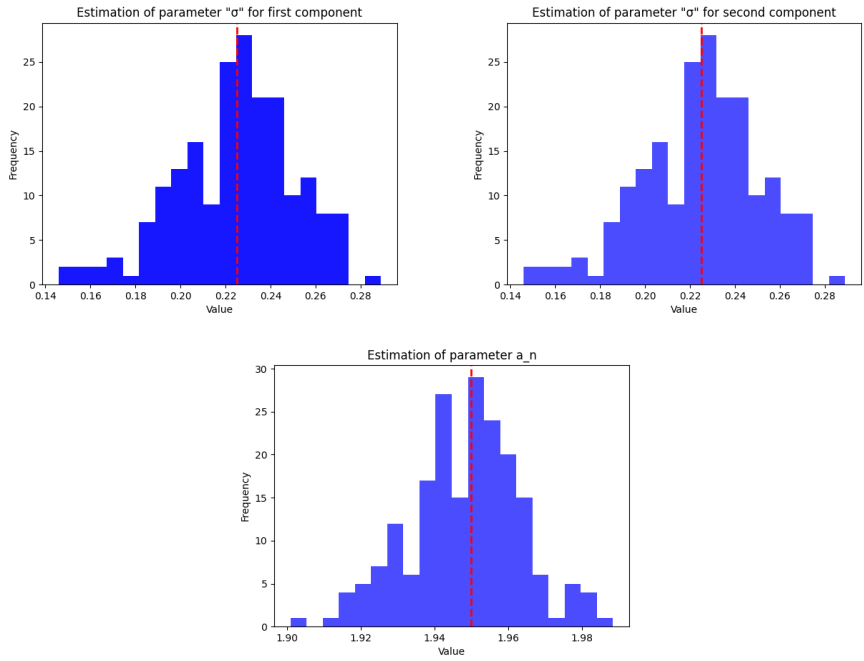

Figure 5.8: Histograms of parameter estimates of 200 paths of length 1000 of a bivariate graph supOU process with long memory. The true values are indicated by dotted lines.

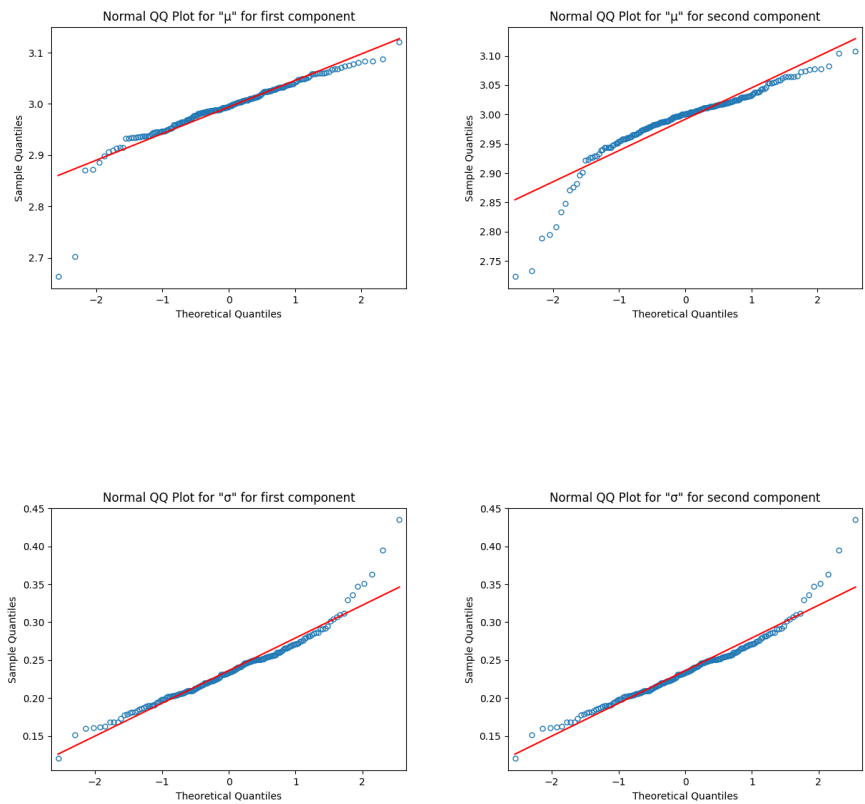

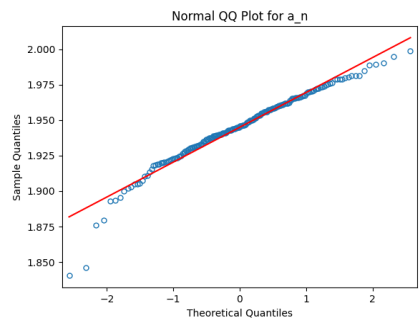

Figure 5.9: Normal QQ-plots of parameter estimates of 200 paths of length 500 of a bivariate Graph supOU model

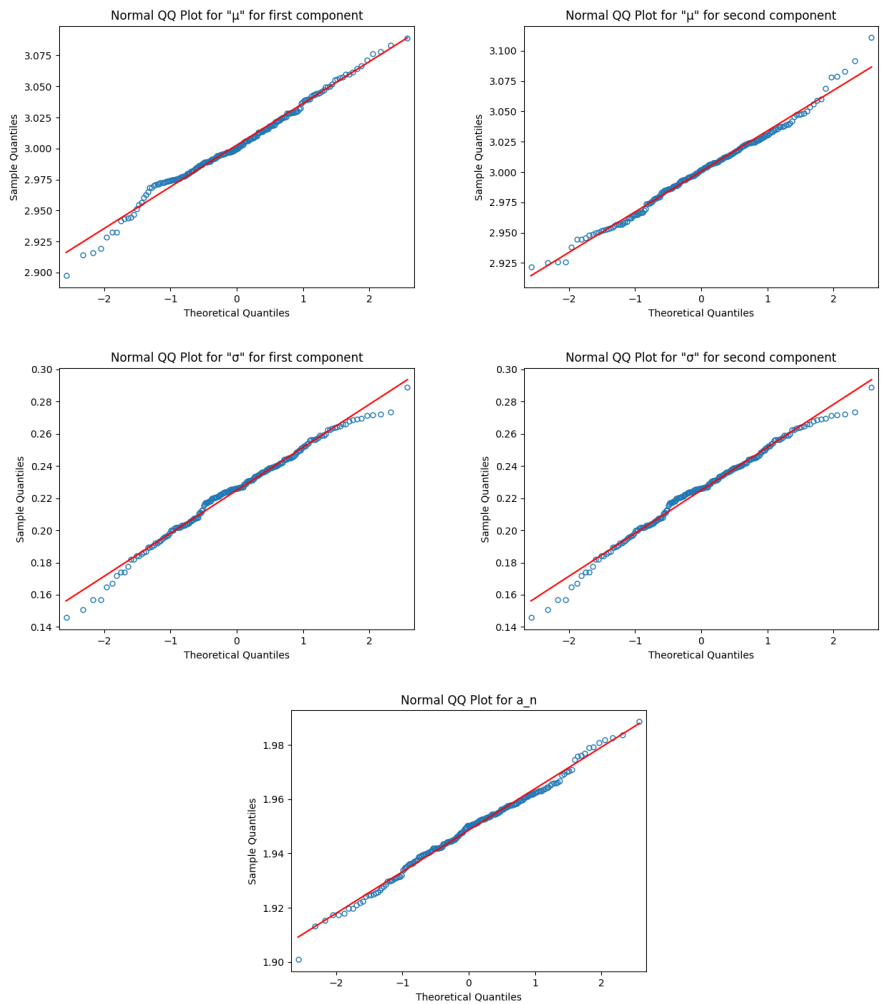

Figure 5.10: Normal QQ-plots of parameter estimates of 200 paths of length 1000 of a bivariate Graph supOU model

## 5.6 Conclusion

In this project, we established an OU model on a dynamic graph with a possibility of long(er) memory. We proved the consistency and Central Limit Theorem for the GMM estimator which excludes the case of true long memory. We developed a GMM estimation method for multivariate Graph supOU processes for long memory effects. The estimators in the simulation study are mostly remarkable indicating asymptotic normality. In the future work [MV24], we explore the effects of the network on the estimators for higher number of nodes.

As mentioned in the introduction of this chapter, the simulation study is performed using Python and the codes are available on GitHub. The implementation of the algorithm has not been optimised for the speed yet. Since the codes are very time consuming, they are only implemented for 200 paths and at most 1000 observations each. Since we observed that the estimation provides better results as we increase the number of observations from 500 to 1000. In the future work, we would optimise these codes to increase the number of samples in order to obtain even better estimates in the Monte Carlo study.

# Chapter 6

# Conclusion

## 6.1 Summary of Thesis Achievements and Further Directions

In this thesis, we have provided a setup for defining various quantum systems on infinite dimensional setting involving unbounded operators. We show that finite speed of propagation holds for the multiparticle interaction with unbounded operators. The construction of Dirichlet form is provided for multiple models along with the analysis of the corresponding dissipative dynamics. Some of these models display no spectral gap property and decay to equilibrium algebraically. One of the significant open problems involves proving the Poincaré and Logarithmic Sobolev inequalities for infinite dimensional systems and achieving uniform control over the associated constants for large dimensional systems.

Additionally, we also presented many noncommutative representations of Lie algebras. The systematic way to write the noncommutative representations of free nilpotent Lie algebras is provided. It might be possible to have similar systematic approaches to write the representations for the solvable Lie algebras or other general nilpotent Lie algebras. We also discuss some models involving the Serre-Chevalley relations which are foundational in the theory Lie algebras in combination with the creation and annihilation operators.

The models and the operators related to nilpotent Lie algebras can be a good starting point to lead to some interesting results in quantum stochastic calculus which is briefly discussed in Section 2.8.

Additionally, the representations of Lie Algebras in Chapter 4 can be utilised to construct non commutative analogue of hypocoercive generators of the form

$$L = A^*A + B$$

introduced in [Vil09] and analyse their dissipative dynamics.

In the last part of the thesis, we provide an extension of a stochastic model involving OU type relation between the nodes of a dynamic network. This extension, called the multivariate Graph supOU process, accommodates the possibility of obtaining long(er) memory in the process which can have many applications. We prove the consistency and asymptotic normality of the moment estimator, although our proof only covers the short memory case. It is an interesting open question to extend the asymptotic theory of the moment estimator for true long memory. Additionally, there is a possibility to explore the development of a similar model for the multivariate supOU stochastic volatility model, as described in Definition 2.7 of [STW15], to incorporate stochastic volatility.

At the intersection of the two areas of quantum generalisations and statistical modelling, it would be interesting to extend the model developed in the last chapter for the quantum setup by utilising the construction of quantum Ornstein–Uhlenbeck process and literature on quantum Lévy processes [Fra04]. The simulations could potentially be performed using the recently developed quantum machine learning techniques [Bia+17].